\documentclass[groupedaddress,aps,pra,reprint,longbibliography]{revtex4-1}
\usepackage{amssymb,amsfonts,amsmath}
\usepackage{graphicx} 
\usepackage{bm}        
\usepackage{float}
\usepackage{subfigure}
\usepackage{times}
\usepackage{notoccite}
\usepackage{lipsum}
\usepackage[mathscr]{euscript} 
\usepackage{longtable}
\usepackage{tabularx}

\DeclareSymbolFont{rsfs}{U}{rsfs}{m}{n}
\DeclareSymbolFontAlphabet{\mathscrsfs}{rsfs}

\newcommand{\X}[0]{$X^2\Sigma_g^+$}
\newcommand{\basis}[1]{\left|\phi_{#1}\right\rangle}
\newcommand{\eigstate}[1]{\left|\psi_{#1}\right\rangle}
\newcommand{\NF}[1]{\mathscrsfs{#1}} 

\begin{document}

\title{From megahertz to terahertz qubits encoded in molecular ions: theoretical analysis of dipole-forbidden spectroscopic transitions in N$\mathbf{_2^+}$}

\author{Kaveh Najafian}
\author{Ziv Meir}
\author{Stefan Willitsch}
\email[To whom correspondence should be addressed: ]{stefan.willitsch@unibas.ch}

\affiliation{Department of Chemistry, University of Basel, Klingelbergstrasse 80, 4056 Basel, Switzerland}

\date{\today}

\begin{abstract}
Recent advances in quantum technologies have enabled the precise control of single trapped molecules on the quantum level. 
%
Exploring the scope of these new technologies, we studied theoretically the implementation of qubits and clock transitions in the spin, rotational, and vibrational degrees of freedom of molecular nitrogen ions including the effects of magnetic fields. The relevant spectroscopic transitions span six orders of magnitude in frequency illustrating the versatility of the molecular spectrum for encoding quantum information. 
%
We identified two types of magnetically insensitive qubits with very low (``stretched''-state qubits) or even zero (``magic'' magnetic-field qubits) linear Zeeman shifts.
The corresponding spectroscopic transitions are predicted to shift by as little as a few mHz for an amplitude of magnetic-field fluctuations on the order of a few mG translating into Zeeman-limited coherence times of tens of minutes encoded in the rotations and vibrations of the molecule.
%
We also found that the Q(0) line of the fundamental vibrational transition is magnetic-dipole allowed by interaction with the first excited electronic state of the molecule. The Q(0) transitions, which benefit from small systematic shifts for clock operation and high sensitivity to a possible variation in the proton-to-electron mass ratio, were so far not considered in single-photon spectra. 
%
Finally, we explored possibilities to coherently control the nuclear-spin configuration of N$_2^+$ through the magnetically enhanced mixing of nuclear-spin states.
%
\end{abstract}

\maketitle

\section{Introduction}
Over the past two decades, a range of different methods have been devised for the generation of cold trapped molecules in the gas phase. These include molecular-beam slowing and trapping \cite{meerakker12a, segev19b}, direct laser cooling \cite{anderegg18a,mccarron18a,caldwell19a}, assembly from ultracold atoms \cite{moses17}, and sympathetic cooling \cite{molhave00a,tong10a}.
In this context, experiments in which single molecular ions are co-trapped with single atomic ions \cite{wolf16a,chou17a,sinhal20,chou20a,lin19,Najafian20a} show excellent prospects for achieving the long-standing goal of gaining full control over the quantum state and dynamics of single isolated molecules. In these experiments, a quantum-logic approach \cite{schmidt05a} is pursued in which the co-trapped atomic ion is used to detect the state of the molecular ion. Coherent Rabi and Ramsey spectroscopy \cite{chou17a,chou20a}, quantum-non demolition state detection \cite{wolf16a,sinhal20,Najafian20a} and atom-molecule entanglement \cite{lin19} have recently been demonstrated.

A full control over the quantum states of cold and trapped molecules will enable improved experiments in the realm of precision spectroscopy. Applications range from precisely validating existing physical theories such as quantum electrodynamics \cite{semeria20a,alighanbari18a,biesheuvel16a}, testing fundamental concepts \cite{safronova18a,demille17a} such as a possible time variation of physical constants \cite{schiller05a,flambaum07a} and the putative existence of new forces of nature \cite{salumbides13a}, benchmarking molecular-structure theory \cite{chou20a, hoelsch19a}, performing controlled chemical reactions \cite{sikorsky18a,dorfler19b}, to implementing new time standards based on narrow rovibrational molecular transitions in the mid-infrared spectral domain \cite{germann14a,karr14a,schiller14a}.

While spectroscopy can be performed in a destructive fashion \cite{willitsch11a, germann14a}, the newly developed methods for non-destructive detection and coherent manipulation of molecular ions promise an increase of several orders of magnitude in the experimental duty cycle \cite{sinhal20,meir19a}. This increase will result in a markedly improved spectroscopic sensitivity and, therefore, precision. Another exciting aspect of this technology is the implementation of molecular qubits which can be used for applications in quantum computation \cite{demille02a}, simulation \cite{blackmore18a}, metrology \cite{manovitz19a}, and communication \cite{kimble08a}.

Here, we studied theoretically the implementation of molecular qubits and their prospective application for spectroscopic precision measurements in the homonuclear $^{14}$N$_2^+$ molecular ion. We chose this molecule due to its prospects for investigating a possible time variation of the electron-to-proton mass ratio \cite{kajita14a} and for serving as a mid-infrared (MIR) frequency standard \cite{kajita15a, karr14a, schiller14a}. These applications are enabled by the lack of a permanent dipole moment of the molecule such that rovibrational transitions within the same electronic state are electric-dipole forbidden. These transitions only become allowed in higher order and thus exhibit very narrow natural linewidths \cite{germann14a, germann16d} and low to vanishing susceptibility to external perturbations such as blackbody radiation and stray electric fields \cite{kajita14a, kajita15a}. These qualities also make N$_2^+$ an excellent system for encoding qubits in its rovibrational state manifold in which radiative lifetimes of excited states are estimated to be on the order of months to years \cite{germann16d}. 

While electric perturbations are inherently small in N$_2^+$ (see discussion in Ref. \cite{kajita14a} and in Appendix \ref{sec:electric_shifts}), the molecular states are strongly coupled to external magnetic fields due to the doublet electron-spin character of the molecule \cite{kajita15a}. Finite magnetic fields are present in a typical experimental apparatus, especially in ion-trapping experiments in which they are a perquisite for operation. Moreover, an external magnetic field is used to lift the degeneracy of Zeeman states and to define the quantization axis of qubits realized in atomic systems. Therefore, there is a need for a comprehensive theoretical analysis of the influence of external magnetic fields on the rovibrational states of N$_2^+$. 

Here, we expanded the theory on the hyperfine structure of N$_2^+$ in Ref. \cite{berrahmansour91a} to include the Zeeman effect. We numerically diagonalized the effective molecular Hamiltonian of N$_2^+$ in the electronic ground-state, \X{}, including the interaction with magnetic fields. From the energy-level structure thus derived, we analyzed several classes of spectroscopic transitions from the radio (MHz) to mid-infrared (THz) domains. The different types of transitions (Zeeman, hyperfine-structure, fine-structure, rotational and vibrational) are discussed with respect to their applications as qubits and in precision spectroscopy. 

Magnetic-field insensitive transitions are important since magnetic-field fluctuations are amongst the dominant effects causing decoherence of qubit superpositions. The use of magnetic-field-insensitive transitions for molecular qubits can dramatically increase their coherence time \cite{wang17a}. We identified ``magic'' transitions \cite{langer05a} for which the relative Zeeman shift between the energy levels involved cancels to first order at a experimentally practicable magnetic-field strength of a few Gauss. These transitions allow for magnetic-field-limited coherence times of tens of minutes in rotational and vibrational qubits at realistic levels of magnetic field noise without the need for magnetic shielding or active magnetic-noise cancellation. We also identified transitions in which the linear Zeeman shift is only on the order of 10 Hz/G irrespective of the magnetic-field strength. The latter are transitions between "stretched" states of different ro-vibrational manifolds in the electronic ground state for which the contribution of the electron spin to the Zeeman shift largely cancels \cite{kajita14a,caldwell20a}. These ``streched'' magnetic-insensitive transitions are unique to molecular qubits.  

Previous experimental and theoretical works on N$_2^+$ analyzed the S(0) rotational component of the fundamental vibrational transition \cite{germann14a,kajita15a}, i.e, the transition from the vibrational and rotational ground state to first vibrationally and second rotationally excited state. This transition is single-photon allowed by electric-quadrupole (E2) selection rules. The corresponding Q(0) transition, i.e, the pure vibrational transition with no excitation of the rotation, was predicted to exhibit superior properties for clock and precision-spectroscopy applications due to smaller systematic shifts \cite{kajita14a}. Here, we show that the Q(0) transitions, which were previously considered to be forbidden in single-photon excitation in the present system \cite{kajita15a}, are actually magnetic dipole (M1) allowed through the anisotropy of the interaction of the electron spin with the magnetic field. This is enabled by a mixing of the first excited electronic state, $A^2\Pi_u$, with the electronic ground-state, \X{}, of the nitrogen ion \cite{bruna04a}.

In addition, we identified avoided crossings of energy levels originating from two different nuclear-spin configurations with nuclear-spin quantum numbers $I=0$ and $I=2$. The avoided crossings occur at low, experimentally accessible magnetic-field strengths of a few tens of Gauss. Around these avoided crossings, the molecular eigenstates have a mixed character of the $I=0$ and $I=2$ spin states. This magnetically enhanced nuclear-spin mixing opens up opportunities for transmuting molecular-spin states on demand by coherent two-photon processes, e.g, stimulated Raman pumping \cite{gaubatz1990}, through the highly mixed states around the avoided crossings. 

Finally, we found that for some transitions, M1 coupling dominates the spectrum while for others E2 coupling prevails due to selection rules forbidding M1 coupling. We also found that hyperfine mixing terms in the Hamiltonian allow for otherwise forbidden transitions which significantly changes the spectra compared to zeroth-order expectations.  

\section{Theory}
\subsection{Basis states}
The molecular nitrogen ion, N$_2^+$, in the electronic ground state, \X{}, is adequately described within the Hund's case ($b_{\beta_J}$) angular momentum coupling scheme \cite{frosch52a} given by,
\begin{eqnarray}
\mathbf{N + S = J}, \\
\mathbf{J + I = F}. 
\end{eqnarray}
Here, $\mathbf{N}$ is the rotational angular momentum in a $\Sigma$ electronic state, $\mathbf{J}$ is the angular momentum resulting from the coupling between the electron spin $\mathbf{S}$ and the rotation, and $\mathbf{F}$ is the total angular momentum including the nuclear spin $\mathbf{I}$. The Hund's case ($b_{\beta_J}$) basis describing this coupling scheme is denoted by,
\begin{equation}\label{eq:basis}
    |\phi_i\rangle = |v, N, S, J, I, F, m\rangle.  
\end{equation}
Here, $v$ is the vibrational, $N$ the rotational, $S$ the electron-spin, $J$ the fine-structure (spin-rotation), $I$ the nuclear-spin and $F$ the hyper-fine quantum number. We denote the projection of the total angular momentum $\mathbf{F}$ on the axis of the external magnetic field by $m$, and $i$ is a compound index for all quantum numbers. We used an effective Hamiltonian approach \cite{brown03a,brown78a} in which global perturbations from other electronic and vibrational states are absorbed in the molecular constants. Therefore, we omit the electronic index of the basis states.

Since each $^{14}$N atom has a nuclear spin of $1$, the total nuclear spin, $I$, of the $^{14}$N$_2^+$ molecule can take the values of $I = 0, 1, 2$. This gives rise to different nuclear-spin-symmetry isomers with even (odd) $I$ denoted as ortho (para). In N$_2^+$, even (odd) values of $I$ allow for only even (odd) rotational quantum numbers $N$ due to the total permutation symmetry of the molecular wavefunction imposed by the generalized Pauli principle. While our results are applicable for both spin isomers of N$_2^+$, in this manuscript, we mainly focus on the ortho nuclear-spin isomer with $I = 0, 2$ which is associated with the rotational ground state of particular interest in experiments.

\subsection{Effective Hamiltonian}

We considered the following effective Hamiltonian for the electronic ground state, \X, of N$_2^+$ \cite{berrahmansour91a,balasubramanian94,brown03a},
\begin{equation}\label{eq:H}
\NF{H} = \NF{H}_{vib} + \NF{H}_{rot} + \NF{H}_{fs} + \NF{H}_{hfs} + \NF{H}_{z}.
\end{equation}
The first three terms describing the vibrational, rotational and fine structure are diagonal in the Hund's case (b) basis (Eq. \ref{eq:basis}). Their matrix elements are given by $\NF{H}_{vib,ii}=G_v$, $\NF{H}_{rot,ii}=B_vN(N+1)-D_v(N(N+1))^2$, and  $\NF{H}_{fs,ii}=\gamma_{v,N}(J(J+1)-N(N+1)-S(S+1))/2$. Here, the subscript $v$ indicates that the molecular constants are effective values for a given vibrational and Born-Oppenheimer electronic state, $G_v$ is the vibrational energy,  $B_v$ is the rotational constant, $D_v$ is the centrifugal-distortion constant and $\gamma_{v,N}$ is the electron spin-rotation coupling constant which includes a centrifugal correction term $\gamma_{v,N}=\gamma_v+\gamma_{D_v} N(N+1)$ \cite{berrahmansour91a}. The relevant spectroscopic constants are listed in Table \ref{tab:constants}. Note that our notation of the constants differs in places from the one found in the literature \cite{berrahmansour91a} to render it unambiguous in the present context.

\begin{figure}
	\centering
	\includegraphics[width=\linewidth,trim={9cm 2cm 10.5cm 0cm},clip]{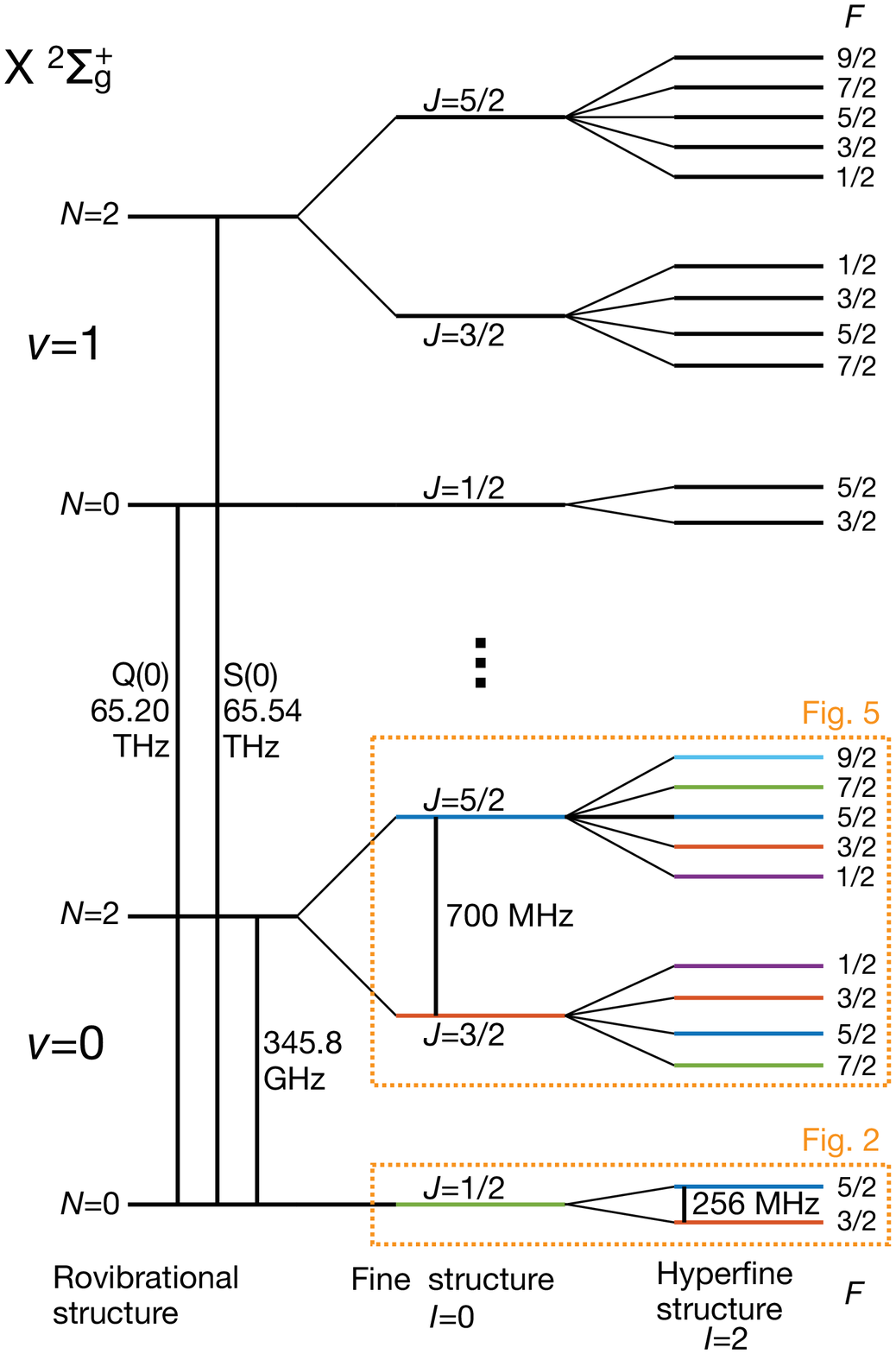}
 	\caption{Partial schematic of the field-free energy levels of N$_2^+$ in the electronic ground state, \X{} (not to scale). The states are labeled using the Hund's case ($b_{\beta_J}$) basis (Eq. \ref{eq:basis}). The dashed boxes indicate the level subspaces shown in Figs. \ref{fig:EN0} and \ref{fig:EN2} where the relevant Zeeman manifolds are displayed. The color coding of the levels is identical with the one used in those figures.}
	\label{fig:scheme}
\end{figure}

The effective hyperfine-interaction Hamiltonian takes the form \cite{berrahmansour91a},
\begin{equation}
\NF{H}_{hfs} = \NF{H}_{b_F} + \NF{H}_{t} + \NF{H}_{eqQ} + \NF{H}_{c_I}.
\end{equation}
Here, $\NF{H}_{b_F}$ represents the Fermi-contact interaction which has off-diagonal matrix elements in the $J$ quantum number, $\NF{H}_{t}$ is the dipolar hyperfine interaction with off-diagonal matrix elements in $N$ and $J$, $\NF{H}_{eqQ}$ is the electric-quadrupole hyperfine interaction with off-diagonal matrix elements in $N$, $J$ and $I$ and $\NF{H}_{c_I}$ is the magnetic nuclear spin-rotation interaction which mixes states with different $J$ quantum numbers. All matrix elements are given in Appendix \ref{sec:hamiltonian}, and the effective coupling constants are given in Table \ref{tab:constants}. A schematic of the resulting energy levels is shown in Fig. \ref{fig:scheme}.  

In the \X{} ground state of N$_2^+$, the effective Zeeman Hamiltonian, $\NF{H}_z$, neglecting relativistic and radiative corrections \cite{karr08}, has four first-order contributions corresponding to the interaction of the magnetic field $\mathbf{B}$ with the magnetic moments of the electron spin, rotation and nuclear spin \cite{brown78a,brown03a,ma09a,chen06a},
\begin{eqnarray}
\NF{H}_z = g_s\mu_B T^1_{p=0}(\mathbf{B})T^1_{p=0}(\mathbf{S})  
\\ \nonumber 
- g_r\mu_B T^1_{p=0}(\mathbf{B})T^1_{p=0}(\mathbf{N})
\\ \nonumber
- g_n\mu_N T^1_{p=0}(\mathbf{B})T^1_{p=0}(\mathbf{I})
\\ \nonumber 
+g_l\mu_B T^1_{p=0}(\mathbf{B})\sum\limits_{q=\pm1} \NF{D}^1_{p=0,q}(\omega)^* T^1_{q}(\mathbf{S}).
\label{eq:Hz}
\end{eqnarray}
Here, $g_s$, $g_r$ and $g_n$ are the $g$-factors for the spin, rotation and nuclear spin and $\mu_B$ ($\mu_N)$ is the Bohr (nuclear) magneton. The last term in $\NF{H}_z$ represents the anisotropic correction to the electron-spin Zeeman interaction and $g_l$ is the corresponding effective g-factor. $T^1_{p}$ denotes a spherical tensor operator of rank 1 in the space-fixed (subscript $p$) coordinate system, $\NF{D}^1_{pq}(\omega)$ is a Wigner rotation-matrix element, and the subscript $q$ denotes spherical tensor components in the molecule-fixed coordinate system. The $p=0$ component of the space-fixed coordinate system is taken to be aligned with the direction of the magnetic field, $T^1_{p=0}(\mathbf{B})=B_Z\hat{Z}$.
The rotational and anisotropic spin $g$-factors, $g_r$ and $g_l$, show a non-negligible dependence on the vibrational state (see Table \ref{tab:constants}). Diagonal terms in the interaction of the magnetic field with the electronic orbital angular momentum ($\mathbf{L}$) vanish in a $\Sigma$ state and terms of higher order in the magnetic field ($\propto \mathbf{B}^2$) \cite{schiff39,brown03a} are neglected in our analysis.

The interaction of the magnetic field with the electron spin mixes states with different $J$ and $F$ quantum numbers. The matrix-elements are given by,
\begin{widetext}
\begin{eqnarray}
\label{eq:M1S}
 \langle  N', S', J', I', F', m'| T^1_{p=0}(\mathbf{B}) T^1_{p=0}(\mathbf{S}) | N, S, J, I, F, m\rangle = 
 \\ \nonumber
 B_Z \delta_{N N'}\delta_{S S'}\delta_{I I'} (-1)^{F'+F-m'+2J'+N+S+I}
 \\ \nonumber
 \times 
 \sqrt{(2F'+1)(2F+1)(2J'+1)(2J+1)S(S+1)(2S+1)} 
 \\ \nonumber
 \times
 \begin{pmatrix}
    F' & 1 &  F \\ -m' & p & m
 \end{pmatrix}
 \begin{Bmatrix}
    J' & F' &  I \\ F & J & 1
 \end{Bmatrix}
 \begin{Bmatrix}
    S & J' & N \\ J & S & 1
 \end{Bmatrix}.
\end{eqnarray}
\end{widetext}
The same type of mixing occurs also for the interaction with the rotational magnetic moment, 
\begin{widetext}
\begin{eqnarray}
\label{eq:M1N}
 \langle N', S', J', I', F', m'| T^1_{p=0}(\mathbf{B}) T^1_{p=0}(\mathbf{N}) | N, S, J, I, F, m\rangle = 
 \\ \nonumber
 B_Z \delta_{N N'}\delta_{S S'}\delta_{I I'} (-1)^{F'+F-m'+J'+J+N+S+I}
 \\ \nonumber
 \times 
 \sqrt{(2F'+1)(2F+1)(2J'+1)(2J+1)N(N+1)(2N+1)} 
 \\ \nonumber
 \times
 \begin{pmatrix}
    F' & 1 &  F \\ -m' & p & m
 \end{pmatrix}
 \begin{Bmatrix}
    J' & F' &  I \\ F & J & 1
 \end{Bmatrix}
 \begin{Bmatrix}
    N & J' & S \\ J & N & 1
 \end{Bmatrix}.
\end{eqnarray}
\end{widetext}
Interaction with the nuclear spin only mixes states with different $F$ quantum numbers,
\begin{widetext}
\begin{eqnarray}
\label{eq:M1I}
 \langle N', S', J', I', F', m'| T^1_{p=0}(\mathbf{B})T^1_{p=0}(\mathbf{I}) |N, S, J, I, F, m\rangle = 
 \\ \nonumber
 B_Z \delta_{J J'}\delta_{N N'}\delta_{S S'}\delta_{I I'} (-1)^{2F'-m'+J'+I+1}
 \\ \nonumber
 \times 
 \sqrt{(2F'+1)(2F+1)I(I+1)(2I+1)} 
 \begin{pmatrix}
    F' & 1 &  F \\ -m' & p & m
 \end{pmatrix}
 \begin{Bmatrix}
    I & F' &  J \\ F & I & 1
 \end{Bmatrix}.
\end{eqnarray}
\end{widetext}

The matrix elements of the anisotropic correction to the electron-spin interaction in the Zeeman Hamiltonian are given by,
\begin{widetext}
\begin{eqnarray}
\label{eq:M1l}
\langle N', S', J', I', F', m'| T^1_{p=0}(\mathbf{B}) \sum\limits_{q = \pm1}\NF{D}_{{p=0},q}^{1}(\omega)^*T^1_{q}(\mathbf{S}) | N, S, J, I, F, m\rangle=
\\ \nonumber
B_Z \delta_{SS'}
\delta_{II'}
(-1)^{F'-m' + F + J' + I' +1 + N'}
\\ \nonumber
\times
\sqrt{(2F'+1)(2F+1)(2J'+1)(2J+1)(2N'+1)(2N+1)S(S+1)(2S+1)}
\begin{pmatrix}
    F' & 1 & F \\ -m' & p & m
\end{pmatrix}
\begin{Bmatrix}
    J' & F' & I \\ F & J & 1
\end{Bmatrix}
\\ \nonumber
\times  
2 \sum\limits_{k = 0, 2} (2k+1)
\begin{pmatrix}
        1 & 1 & k \\ -1 & 1 & 0
    \end{pmatrix}
\begin{pmatrix}
        N' & k & N \\ 0 & 0 & 0
\end{pmatrix}
\begin{Bmatrix}
    J' & J & 1 \\
    N' & N & k \\
    S' & S & 1
\end{Bmatrix}.
\end{eqnarray}
\end{widetext}
This interaction mixes different $F$,$J$ and $N$ quantum numbers.

The complete Hamiltonian given in Eq. \ref{eq:H} was diagonalized numerically by solving $\NF{H}(B_Z)\eigstate{k}=E_k(B_Z)\eigstate{k}$ in the Hund's case ($b_{\beta_J}$) basis (Eq. \ref{eq:basis}) to obtain the energies, $E_k(B_Z)$, and mixing-coefficients, $c^k_i(B_Z)$, 
\begin{equation}\label{eq:eigen}
    \eigstate{k}=\sum_i c^k_i(B_Z) \basis{i}, 
\end{equation}
as function of the external magnetic value, $B_Z$.
A basis set of 2 vibrational ($v=0,1$), 3 rotational ($N=0,2,4$), 2 nuclear ($I=0,2$) and all resulting fine, hyperfine and Zeeman states was used yielding a total of 360 states. 

\begin{table}[h]
\begin{center}
\caption{Spectroscopic constants of $^{14}$N$_2^+$ in the $v=0$ and $v=1$ vibrational states of the electronic ground state, \X{}, used to calculate the energy levels. The numbers in parentheses are uncertainties given in the literature (references in square brackets). The values of the effective coupling constants of the electric-quadrupole hyperfine interaction, $eqQ_{v=0}$, and the magnetic nuclear spin-rotation hyperfine interaction, $c_{I{v=0}}$, in the vibrational ground state, $v=0$, are not reported in the literature. It was assumed that they are equal to the values reported for the first excited vibrational state, $v=1$.}
\begin{tabular}{l  l  l}
\hline
\hline 
& $v$ = 0 & $v$ = 1 \\ \hline
$G_v-G_0$ (cm$^{-1}$) & 0 & 2174.746(1) \cite{michaud00a} \\
$B_v$ (cm$^{-1}$) & 1.9223897(53) \cite{wu07a} & 1.90330(2) \cite{collet98a} \\
$D_v$ ($\times 10^6$ cm$^{-1}$) & 5.9748(50) \cite{wu07a} & 5.904(21) \cite{collet98a} \\
$\gamma_v$ (MHz) & 280.25(45) \cite{scholl98a} & 276.92253(13) \cite{berrahmansour91a} \\
$\gamma_{Dv}$ (kHz) & 0 & -0.39790(23) \cite{berrahmansour91a} \\
$b_{Fv}$ (MHz) & 102.4(1.1) \cite{scholl98a} & 100.6040(15) \cite{berrahmansour91a} \\
$t_v$ (MHz) & 23.3(1.0) \cite{scholl98a} & 28.1946(13) \cite{berrahmansour91a} \\
$t_{Dv}$ (Hz) & 0 \cite{scholl98a} & -73.5(2.7) \cite{berrahmansour91a} \\
$eqQ_v$ (MHz) & -- & 0.7079(60) \cite{berrahmansour91a} \\
$c_{Iv}$ (kHz) & -- & 11.32(85) \cite{berrahmansour91a} \\
$g_s\mu_B$ (MHz/G) & 2.8025 \cite{CODATA2018, kajita15a} & 2.8025 \cite{CODATA2018, kajita15a} \\
$g_r\mu_B$ (Hz/G)  & 50.107 \cite{kajita15a} & 49.547 \cite{kajita15a} \\
$g_n\mu_N$ (Hz/G)  & 307.92 \cite{kajita15a} & 307.92 \cite{kajita15a} \\
$g_l\mu_B$ (Hz/G)  & -3793 \cite{bruna04a} & -3821 \cite{bruna04a} \\
\hline
\hline
\end{tabular}
\label{tab:constants}
\end{center}
\end{table}

\subsection{Transition moments}
In homonuclear diatomic molecules, transitions within the same Born-Oppenheimer electronic state are electric-dipole (E1) forbidden due to the permutation symmetry of the two nuclei. We therefore derived general expressions for magnetic-dipole (M1) and electric-quadrupole (E2) transitions and calculated their strengths under the influence of an external magnetic field.

In the basis set of Eq. \ref{eq:basis}, the square of the transition moment $S_{kl}$ between different Zeeman levels can be separated into an angular ($A$) and a radial ($R$) part as \cite{papouvsek89a},
\begin{eqnarray}\label{eq:S12}
\\ \nonumber
S_{kl} = \sum_p\Big|\langle\psi_k|{T}^{u}_p(\hat{\mu})|\psi_l\rangle\Big|^2  
\\ \nonumber
= \sum_p\Big|\sum_{i,j} c^{k*}_j c^l_i \langle\phi_j|{T}^{u}_p(\hat{\mu})|\phi_i\rangle\Big|^2
\\ \nonumber
= \sum_p \Big| \sum_{i,j} c^{k*}_j c^l_i A(...,F_j,m_j,F_i,m_i,p)R(v_j,v_i)\Big|^2.
\end{eqnarray}
Here, $\eigstate{k(l)}$ is the upper (lower) state of the transition and ${T}^{u}_p(\mathbf{\hat{\mu}})$ is the transition operator in spherical tensor notation. For M1 transitions and E2 transitions, $u = 1$ and 2, respectively. The quantum number $p=-u,...,u$ represents the polarization of the radiation in the space-fixed frame with respect to the quantization axis defined by the direction of the static magnetic field. The sum over the different polarizations in Eq. \ref{eq:S12} yields a polarization-independent transition moment. 
\\

\subsection{Magnetic-dipole transitions within the same vibrational state}
For magnetic-dipole transitions, the operators that couple to the radiation have the same form as the Zeeman Hamiltonian for coupling with an external magnetic field given in Eq. \ref{eq:Hz} with the substitution $B_Z \rightarrow \mathbf{B}(t)$ \cite{brown03a}. Therefore, the angular part of the transition moment Eq. \ref{eq:S12} for M1 transitions can be obtained from the matrix elements Eqs. \ref{eq:M1S},\ref{eq:M1N}, \ref{eq:M1I} and \ref{eq:M1l} where $p$ is now the polarization index of the magnetic-field of the radiation, $\mathbf{B}(t)$. Transitions induced by isotropic and anisotropic interaction with the electron spin and interaction with the rotation and nuclear spin are denoted by M1$_S$, M1$_{aS}$, M1$_N$ and M1$_I$. From the angular part of the transition moment, the following selection rules can be derived,
\begin{equation}
\begin{split}
\forall \: \textrm{M1}:&\Delta m=0,\pm1,\:\Delta F=0,\pm1, \\ 
        &\Delta I=0,\:\Delta S=0,
\end{split}
\end{equation}
\begin{align}\label{eq:selectionrules}
\textrm{M1}_S&: S\neq0,\:\Delta N=0, \\
\textrm{M1}_{aS}&: S\neq0,\:\Delta N=0,2, \\
\textrm{M1}_N&: N\neq0,\:\Delta N=0, \\
\textrm{M1}_I&: I\neq0,\:\Delta N=0,\:\Delta J=0.
\end{align}

For transitions within the same vibrational state, $\Delta v = 0$, the radial part of the transition moment is given by the expectation value of the magnetic moment, ${R(v,v)} \equiv g $, where the values of the $g$-factors are determined by the underlying interaction (Table \ref{tab:constants}). 

\subsection{Electric-quadrupole transitions within the same vibrational state}
For E2 transitions, the coupling operator is ${T}^2_p(Q_{n \Lambda})$ in spherical tensor notation where $Q_{n \Lambda}$ is the electric quadrupole moment in a specific electronic state \cite{germann16c}. The matrix elements for the E2 transition moments are given by \cite{germann16c},
\begin{widetext}
\begin{eqnarray}
\label{eq:E2}
    \langle v', N', S', J', I', F', m'| T^2_p(\mathbf{\hat{Q}}) | v, N, S, J, I, F, m\rangle =
    \delta_{S S'}\delta_{I I'}(-1)^{S+I+J+J'+F+F'-m'}
    \\ \nonumber
    \times
    \sqrt{(2N+1)(2N'+1)(2J+1)(2J'+1)(2F+1)(2F'+1)}
    \\ \nonumber
    \times
    \begin{pmatrix}
        N' & 2 & N \\ 0 & 0 & 0
    \end{pmatrix}
    \begin{pmatrix}
        F' & 2 & F \\ -m' & p & m
    \end{pmatrix}
    \begin{Bmatrix}
        N' & J' & S \\ J & N & 2
    \end{Bmatrix}
    \begin{Bmatrix}
        J' & F' & I \\ F & J & 2
    \end{Bmatrix}
    |R(v',v)|.
\end{eqnarray}
\end{widetext}

From the angular part of the transition moment, the following selection rules can be derived for E2 transitions,
\begin{equation}
\begin{split}
    \textrm{E2}:&\Delta m=0,\pm1,\pm2, \: \Delta N = 0, \pm2, \: \Delta F=0,\pm1,\pm2 \\
    & \Delta I=0, \: \Delta S =0.    
\end{split}
\end{equation}
In addition, $N=0\rightarrow0$ transitions are not allowed within a $\Sigma$ electronic state.

For transitions within the same vibrational level $\Delta v = 0$, the radial part of the transition moment is given by the permanent electric quadrupole moment, $R(v,v)=Q_v=1.86~ea_0^2$ \cite{bruna04a} for low vibrational states. 

\subsection{Vibrational transitions}
The transition strength between different vibrational levels was estimated by expanding the radial part of the transition moment to first order around the equilibrium bond length ($R_e$) \cite{papouvsek89a},
\begin{equation}\label{eq:Radial}
    R(v',v) \approx  \mu_p \langle v' | v \rangle + \frac{d \mu}{d R}\Bigr|_{\substack{(R= R_e)}} \langle v' |R - R_e| v \rangle.
\end{equation}
Here, $\mu_p = \mu(R)|_{\substack{(R= R_e)}}$ is the permanent (electric quadrupole or magnetic dipole) moment and ${d\mu}/{dR}|_{\substack{(R= R_e)}}$ is its derivative as a function of internuclear distance evaluated at the equilibrium bond length. 

From Eq. \ref{eq:Radial}, it seems that the first term only contributes to transitions within the same vibrational manifold, since $\langle v'|v\rangle=\delta_{vv'}$. However, rovibrational mixing, which is not explicitly apparent in the effective Hamiltonian approach taken here, introduces a non-zero overlap between different vibrational states \cite{balasubramanian94}. Therefore, the first term in Eq. \ref{eq:Radial} allows for vibrational transitions according to Eq. \ref{eq:S12}. 

The second term in Eq. \ref{eq:Radial} introduces vibrational transitions through the change in the transition moment with internuclear distance. The vibrational matrix element for the fundamental vibrational transition within the harmonic approximation is given by,
\begin{equation}
    \langle v'=1 |R - R_e| v=0 \rangle \approx R_e\sqrt{\frac{B_e}{\omega_e}}. \label{eq:fund}
\end{equation}
Here, $R_e=2.13~a_0$ \cite{bruna04a} is the equilibrium bond length, and $\omega_e \approx 2207$ cm$^{-1}$ \cite{michaud00a} and $B_e \approx 1.93$ cm$^{-1}$ \cite{scholl98a} are the harmonic vibration frequency and the equilibrium rotational constant (which both need to be inserted in the same units in Eq. \ref{eq:fund}). 

For low rotational states, we found that the strongest (M1$_S$) transitions caused by the first term in Eq. \ref{eq:Radial} are 4-5 orders of magnitude weaker than those originating from the second term for E2 and M1$_{aS}$ coupling. Transitions due to vibrational mixing are therefore neglected in the following. The reader is referred to Appendix \ref{sec:rovibmixing} for further details
\footnote{While for rovibrational E2 transitions it is well established to calculate the transition moment via the second term in Eq. \ref{eq:Radial} \cite{karl67a,germann14a,goldman07a}, the situation is less clear for rovibrational transitions of M1 type. Such transitions were first reported in Ref. \cite{dang-nhu90a} for the $^3\Sigma^-_g$ ground electronic state of the O$_2$ molecule. In Ref. \cite{balasubramanian94}, two types of mechanisms were elaborated to rationalize the M1 rovibrational transition intensities observed in Ref. \cite{dang-nhu90a}. One is rovibrational mixing and the other is due to coupling of different Born-Oppenheimer states. The former, analyzed in Appendix \ref{sec:rovibmixing}, was found to be small in the present case, the latter is included in the anisotropic electron spin Zeeman coupling which was found to give the dominant contribution the rovibrational M1 transition intensities studied here.} .

The couplings can be estimated from the change in the relevant $g$-factors with the effective bond length upon vibrational excitation given in Table \ref{tab:constants} yielding $\Delta g_r / \Delta R \approx 4\cdot10^{-5}~\mu_B/a_0$ and $\Delta g_l / \Delta R \approx 2\cdot10^{-3}~\mu_B/a_0$. The difference in averaged bond lengths, $\Delta R$, between $v = 0$ and $v = 1$ is estimated from the relation of the rotational constant to the equilibrium positions, $B_{v=1}/B_{v=0} = R_{v=0}^2/R_{v=1}^2$, such that $\Delta R\approx0.01~a_0$. 

For E2 transitions, the change in the electric quadrupole moment with the internuclear distance is given by $dQ/dR = 2.63~ea_0$ \cite{bruna04a}.


\subsection{Einstein $A$ coefficients}

The relative importance of M1 and E2 transitions to the spectra was assessed by comparing their Einstein $A$ coefficients. For M1 transitions \cite{drake06}, one obtains
\begin{equation}
\label{eq:AM1}
A_{kl}^{M1} = \frac{16\pi^3\mu_0}{3h\lambda_{kl}^3} S^{M1}_{kl},
\end{equation}
while for E2 transitions \cite{drake06}, the $A$ coefficient is given by,
\begin{equation}
\label{eq:AE2}
A_{kl}^{E2} = \frac{16\pi^5}{15 h \varepsilon_0 \lambda_{kl}^5} S^{E2}_{kl}.
\end{equation}
Here, $\lambda_{kl}$ is the transition wavelength, $\mu_0$ is the vacuum permeability, $\varepsilon_0$ is the vacuum permittivity and $h$ is the Planck constant. All values are in SI units. Since $S_{kl}$ in Eq. \ref{eq:AM1} and \ref{eq:AE2} is the square of the polarization-independent transition moment that was defined in Eq. \ref{eq:S12}, the Einstein $A$ coefficients slightly differ from their regular definitions as they explicitly depend on the Zeeman levels. 

\section{Results and discussion}

\subsection{Hyperfine and Zeeman qubits in the rotational ground state, N=0}
\begin{figure}
	\centering
	\includegraphics[width=\linewidth,trim={0cm 0cm 0cm 0cm},clip]{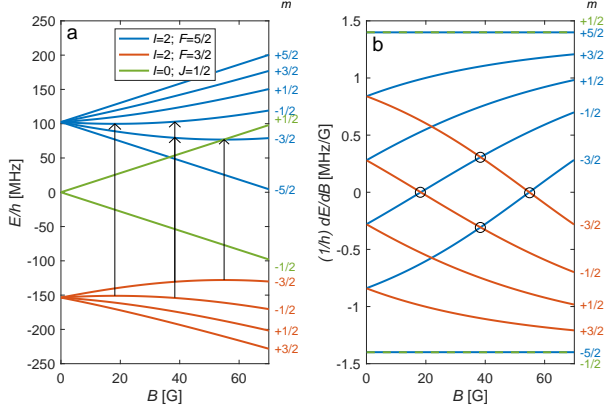}
 	\caption{a) Energies, $E$, of hyperfine-Zeeman levels in the rovibronic ground-state ($v=0$, $N=0$) manifold of N$_2^+$ as a function of the external-magnetic-field strength, $B$. b) Derivatives of the energies in a) with respect to the magnetic field. Circles indicate positions of ``magic'' magnetic-field values at which the transition energy between two levels is independent of the magnetic field to first order. The corresponding M1$_S$ transitions are indicated by the arrows in a). All of these are allowed by the selection rules.}
	\label{fig:EN0}
\end{figure}

The hyperfine-Zeeman energy levels of the rovibronic ground-state manifold, \X{}($v=0$, $N=0$), of N$_2^+$ as function of the strength of an external magnetic field are displayed in Fig. \ref{fig:EN0}a. 

\begin{figure}
	\centering
	\includegraphics[width=\linewidth,trim={0cm 5cm 0cm 0cm},clip]{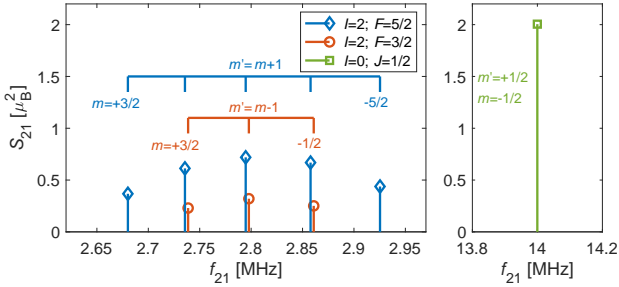}
 	\caption{Strengths, $S_{21}$, of M1$_S$ transitions, $m \rightarrow m'$, between Zeeman levels within the hyperfine manifolds of the rovibronic ground state, \X{}($v=0$, $N=0$), of the $I=0$ (green) and $I = 2$ (blue and red) nuclear-spin species of N$_2^+$ as a function of transition frequency, $f_{21}$. The abscissa indicates the transition frequencies at a magnetic field value of 5~G. The color code is the same as in Fig. \ref{fig:EN0}.}
	\label{fig:SN0_Zeeman}
\end{figure}

For the $I=0$ isomer, the situation is similar to the ground state of bosonic alkaline-earth ions (e.g., $^{88}$Sr$^+$) which are also used as qubits \cite{keselman11}. The total angular momentum $J=1/2$ results in two Zeeman levels which are separated by $(g_s+2/3g_l) \mu_B\approx2.8$ MHz/G (green traces in Fig. \ref{fig:EN0}). All terms in the Zeeman Hamiltonian are zero except for the isotropic and anisotropic electron-spin terms. Thus, the situation is formally identical (apart from negligible mixing terms to higher rotational states) to the atomic $^2$S$_{1/2}$ case. Transitions between the two Zeeman levels can be driven by M1$_S$ coupling (green stick in Fig. \ref{fig:SN0_Zeeman}). 

For the $I=2$ isomer, the hyperfine interaction splits the rovibronic ground state into two hyperfine manifolds with total angular-momentum quantum numbers $F=3/2$ and $F=5/2$ (red and blue traces in Fig. \ref{fig:EN0}). The relatively small splitting of $5/2b_{F,v=0}\approx256$ MHz (for $B$=0 G), together with the strong magnetic coupling, M1$_S$, leads to a deviation of the Zeeman splittings from the weak coupling regime (linear Zeeman effect) to the intermediate coupling regime already at relatively low magnetic fields of few tens of Gauss. The full decoupling of the spin and orbital angular momenta (Paschen-Back regime) occurs already at magnetic fields of a few hundreds of Gauss.

As a consequence, Zeeman transitions within each hyperfine manifold are not equally spaced (Fig. \ref{fig:SN0_Zeeman}, blue and red bars). The unequal spacing can be used to address Zeeman transitions individually and to allow for optical pumping and state readout as was demonstrated with polar CaH$^+$ molecules \cite{chou17a}. The transitions are dominated by M1$_S$ coupling (see Fig. \ref{fig:SN0_Zeeman} for the transition strengths). M1 transitions arising from the anisotropic-spin, rotational and nuclear-spin terms were found to be 3-5 orders of magnitude weaker due to the difference in magnitude between $g_s$ and $g_r$, $g_n$, and $g_l$.

\begin{figure}
	\centering
	\includegraphics[width=\linewidth,trim={0cm 0cm 0cm 0cm},clip]{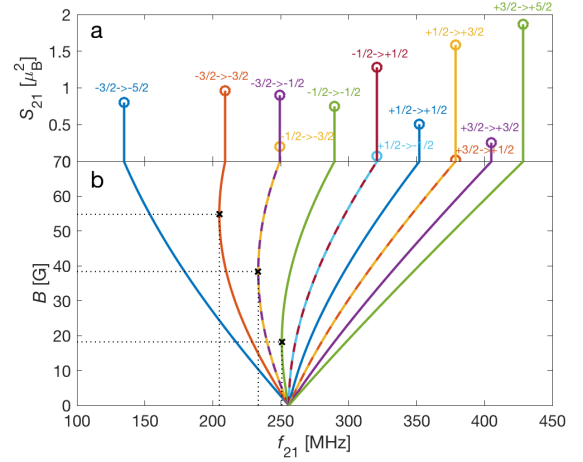}
 	\caption{a) Strengths, $S_{21}$, of hyperfine-Zeeman M1$_S$ transitions, $|F=3/2\rangle\rightarrow|F'=5/2\rangle$, within the rovibronic ground state, \X{}($v=0$, $N=0$), of the $I=2$ nuclear-spin species of N$_2^+$. The abscissa indicates the transition frequencies, $f_{21}$, at a magnetic-field strength of 70 G. The Zeeman components, $m\rightarrow m'$, of each transition are indicated. b) Dependence of the transition frequencies, $f_{21}$, on the magnetic field, $B$. Dotted lines indicate ``magic'' values of the magnetic field in which the transition frequency is insensitive to changes of the magnetic field to first order.}
	\label{fig:EN0_B_HF}
\end{figure}

Transition between the two hyperfine manifolds, $|F = 3/2 \rangle \rightarrow |F' = 5/2\rangle $, are also allowed by M1$_S$ coupling. These transitions are commonly used as long-lived qubits in atomic ions \cite{langer05a}. Here, we identified transitions in which the dependence of the energy levels on the magnetic field is equal for both the lower and upper states for specific values of the magnetic field (see arrows in Fig. \ref{fig:EN0}a, circles in Fig. \ref{fig:EN0}b and dotted lines in Fig. \ref{fig:EN0_B_HF}b). This equal dependency results in an insensitivity of the transitions to magnetic field fluctuations to first order. Insensitive transitions at ``magic'' magnetic fields are used in atomic systems \cite{langer05a,harty14} to encode qubits with improved coherence times and to circumvent the need for magnetic shielding. Due to the small hyperfine splittings in N$_2^+$, the ``magic'' magnetic field occurs at small and easily accessible values. The second-order Zeeman susceptibility of the transitions around the ``magic'' values is $\sim16$~mHz/mG$^2$ (for all the hyperfine ``magic'' transitions in Fig. \ref{fig:EN0_B_HF}b) from which we estimated a shift of as low as $\Delta E/h = \Delta f \approx 16$~mHz in the transition frequencies for a magnetic-field fluctuation of 1 mG. Thus, these transitions are ideally suited for encoding qubits with magnetic-field-limited coherence times of up to $1/\Delta f \approx 60$ s \cite{saleh07} as well as for applications in precision spectroscopy and in clocks. Typical strengths for these hyperfine transitions are given in Fig. \ref{fig:EN0_B_HF}a.

\subsection{Hyperfine and Zeeman qubits in the rotationally excited state, N=2}
\begin{figure}
	\centering
	\includegraphics[width=\linewidth,trim={0cm 0cm 0cm 0cm},clip]{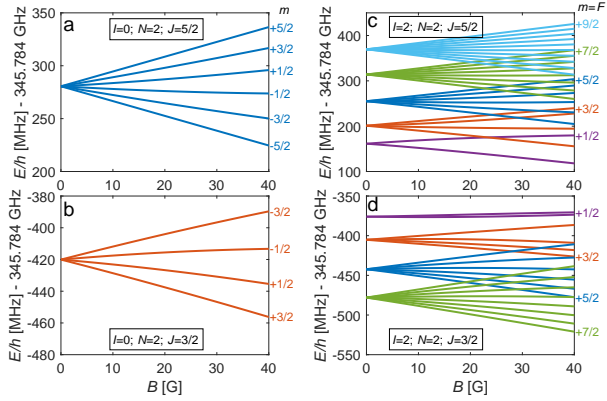}
 	\caption{Energies, $E$, of the hyperfine-Zeeman levels of the $v=0$, $N=2$ state as function of magnetic-field strength, $B$, for a),b) the $I=0$ and c), d) the $I=2$ nuclear-spin species. The spin-rotation quantum number is $J=5/2$ in panels a) and c), and $J=3/2$ in panels b) and d). Color code for the different hyperfine states in c), d): $F=1/2$ in purple, $F=3/2$ in red, $F=5/2$ in blue, $F=7/2$ in green and $F=9/2$ in light blue. In panels a) and b), all Zeeman quantum numbers are shown while in panels c) and d) only the level with the highest value of the Zeeman quantum number ($m=F$) is indicated.}
	\label{fig:EN2}
\end{figure}

The energy levels of the second rotationally excited state in the vibronic ground state, \X{}~$(v=0, N=2)$, of the ortho nuclear-spin isomer, are displayed in Fig. \ref{fig:EN2}. The rotational excitation shifts the spectrum by $6B_{v=0}-36D_{v=0}\approx345.784$ GHz compared to the rotationless case. The spin-rotation coupling, $H_{fs}$, splits the levels into $J=3/2$ and $J=5/2$ manifolds which are separated by $5/2\gamma_{v=0}\approx700$ MHz (for $B$=0~G). For the $I=0$ species, this coupling generates an energy-level structure which is qualitatively similar to the that of the $I=2$ configuration within $N=0$ (Fig. \ref{fig:EN0}). However, because of the large spin-rotation splitting, the deviation from a linear Zeeman effect occurs at higher magnetic fields compared to the situation in Fig. \ref{fig:EN0}. For instance, the first ``magic'' magnetic field for the $\left|J=3/2,m=-1/2\right\rangle\rightarrow\left|J'=5/2,m'=-1/2\right\rangle$ transition occurs at $\sim$49 G compared to $\sim$18 G for the $\left|F=3/2,m=-1/2\right\rangle\rightarrow\left|F'=5/2,m'=-1/2\right\rangle$ transition in the $N=0$, $I=2$ state. 

\begin{figure}
	\centering
	\includegraphics[width=\linewidth,trim={0cm 6.5cm 0cm 0cm},clip]{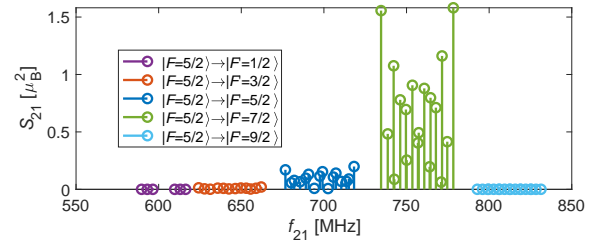}
 	\caption{Strengths, $S_{21}$, of transitions between hyperfine components of the spin-rotation manifolds in the \X{}~($v=0$, $N=2$) state of N$_2^+$ as a function of the transition frequency ,$f_{21}$, at a magnetic-field strength of 10~G. The transitions shown are of the form $|N=2,J=3/2,I=2,F=5/2\rangle \rightarrow |N'=2,J'=5/2,I'=2,F'\rangle$.}
	\label{fig:SN2F32toFAll}
\end{figure}

For the $I=2$ nuclear-spin species in the $N=2$ rotational state, the levels are further split by the hyperfine interaction which is dominated by the Fermi-contact ($H_{b_F}$) and dipolar ($H_{t}$) terms. Thus, the energy levels split into $F=9/2,...,1/2$ and $F=7/2,...,1/2$ for the $J=5/2$ (Fig. \ref{fig:EN2}c) and $J=3/2$ Fig. \ref{fig:EN2}d) spin-rotation manifolds, respectively. M1$_S$ coupling is again dominant. For spin-rotation transitions, we found that $\Delta F = 1$ components are prevalent, as can be seen in the spectrum displayed in Fig. \ref{fig:SN2F32toFAll}. ``Magic'' transitions can be found at magnetic fields as low as few Gauss (e.g., the $|F=5/2,m=+1/2\rangle \rightarrow |F'=7/2,m'=-1/2\rangle$ at $\sim$756.3 MHz and $B = 1.55$ G with second-order Zeeman-shifts as low as $\sim 8$ mHz/mG$^2$ (see Appendix \ref{sec:magic_transitions} for a partial list of the strongest ``magic'' transitions below 70~G). 

\begin{figure}
	\centering
	\includegraphics[width=\linewidth,trim={0cm 0cm 0cm 0cm},clip]{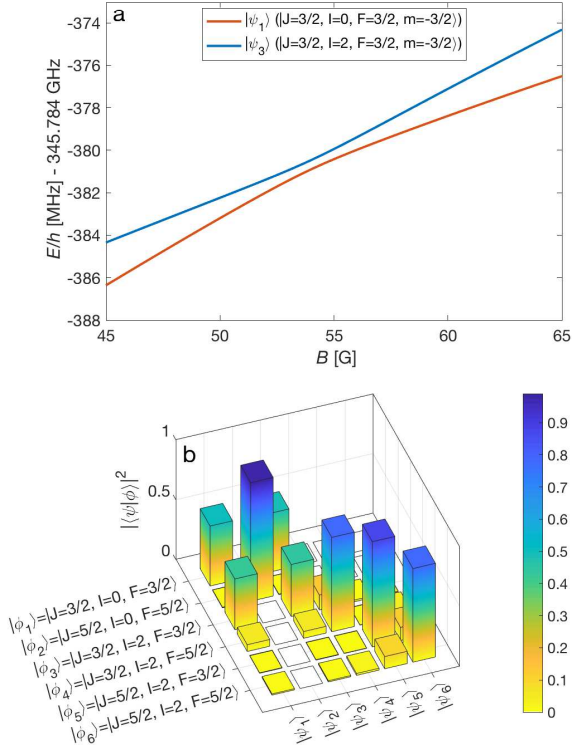}
 	\caption{a) Adiabatic level energies, $E$, as a function of the magnetic field strength, $B$, showing an avoided crossing between two states originating from two different nuclear-spin states ($I=0,2$) in the \X{}(v=0, N=2) level. The states indicated in the legend represent the dominant contributions at zero magnetic field. b) Overlap of the eigenstates (Eq. \ref{eq:eigen}) with the basis vectors (Eq. \ref{eq:basis}), $|\langle\phi_i|\psi_k\rangle|^2=|c_i^k|^2$, at a magnetic field of 54~G. Here, $|\langle\phi_i|\psi_k\rangle|^2\approx\delta_{ik}$ for zero magnetic field.}
	\label{fig:avoided}
\end{figure}

An interesting effect in N$_2^+$ as exemplified here with the $N=2$ manifold is the coupling between nuclear-spin states through the electric-quadrupole hyperfine interaction, $H_{eqQ}$, which mixes levels with even (or odd) total nuclear spin $I$. In $^{14}$N$_2^+$, there is only a single para nuclear-spin state with $I=1$ such that only the ortho species with $I=0,2$ exhibit this coupling. This interaction results in avoided crossings of energy levels originating from the different ortho spin states. As an example, Fig. \ref{fig:avoided}a shows such an avoided crossing between the $|F=3/2,m=-3/2\rangle$ states originating from the $I=0$ (red) and $I=2$ (blue) species. This avoided crossing occurs at a relatively low magnetic field of $\sim$54 G. Around the crossing point, the levels exhibit a strong mixing of the $I=0$ and $I=2$ basis states (see Fig. \ref{fig:avoided}b). This magnetically enhanced nuclear-spin mixing is interesting as it opens up possibilities to manipulate the nuclear-spin configuration of the molecule on demand (see Fig. \ref{fig:stirap} and the accompanying discussion further below).

\subsection{Rotational qubits}\label{sec:rot}

We now consider transitions from the rotational ground state $N=0$ to the second excited rotational state $N'=2$ at frequencies around $\sim$345 GHz. The sensitivity of these transitions to the proton-to-electron mass ratio \cite{hanneke16} renders them interesting for testing a possible time variation of these fundamental constants as they are within the reach of stabilized THz sources \cite{schiller08a}. 

\begin{figure}
	\centering
	\includegraphics[width=\linewidth,trim={0cm 0cm 0cm 0cm},clip]{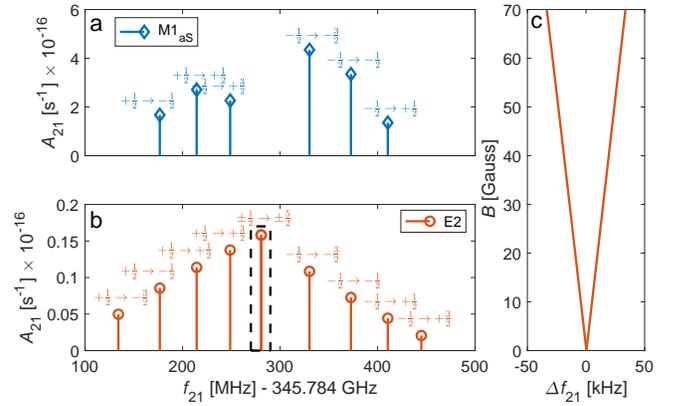}
 	\caption{Spectrum of Zeeman components of the spin-rotational transition  $|N=0,J=1/2\rangle \rightarrow |N'=2,J'=5/2\rangle$ for the $I=0$ isomer. The intensities of the transitions are given in the form of Einstein $A$ coefficients (Eqs. \ref{eq:AE2},\ref{eq:AM1}) for comparison of a) M1$_{aS}$ coupling with b) E2 coupling. The magnetic field was assumed to be 70~G. Labels indicate the Zeeman components of the transitions. c) Magnification of the dashed rectangle in b) showing the stretched transitions, $|J=1/2,m=\pm1/2\rangle \rightarrow |J'=5/2,m'=\pm5/2\rangle$, which show a very small dependence on the magnetic field and are only separated by 66.5~kHz at 70~G. The frequency axis in c) is referenced to the field-free line positions.}  
	\label{fig:RotQubitI0}
\end{figure}

In general, M1$_S$ transition selection rules do not permit a change of rotational quantum numbers by $\Delta N=2$, but this mechanism must still be considered due to mixing of rotational states. In addition, the anisotropy of the electron-spin $g$-factor tensor allows for $\Delta N=2$ transitions through M1$_{aS}$ coupling. We also consider electric-quadrupole (E2) transitions which also permit such a change in the rotational quantum number. In addition, E2 transitions permit changes in the angular-momentum-projection quantum number $\Delta m=\pm2,\pm1,0$. Thus, magnetic-dipole and electric-quadrupole rotational spectra will show different signatures as illustrated in Fig. \ref{fig:RotQubitI0}a,b. For the $|J=1/2\rangle \rightarrow |J'=3/2\rangle$ transitions, M1$_{aS}$ coupling was found to be $\sim$3 orders of magnitude stronger than the E2 coupling while for the $|J=1/2\rangle \rightarrow |J'=5/2\rangle$, it was only found to be about 1 order of magnitude stronger (Fig. \ref{fig:RotQubitI0}a,b).  

The $\Delta m = \pm2$ and $\Delta J =  2$ lines are allowed for E2 coupling opening up opportunities to exploit transitions between ``stretched'' states, e.g., $\left|J=1/2,m=\pm1/2\right\rangle \rightarrow \left|J'=5/2,m'=\pm5/2\right\rangle$ in the $I=0$ nuclear-spin isomer and $\left|F=5/2,m=\pm5/2\right\rangle \rightarrow \left|F'=9/2,m'=\pm9/2\right\rangle$ in the $I=2$ nuclear-spin isomer. These transitions show a very small linear dependence on the magnetic field due to cancellation of the major contribution from the isotropic Zeeman Hamiltonian (Eq. \ref{eq:M1S}) in the ground and excited ``stretched'' states. The remaining susceptibility of these levels to magnetic field is attributed to the rotational dependence of the anisotropic (Eq. \ref{eq:M1l}) and rotational (Eq. \ref{eq:M1N}) Zeeman Hamiltonians. The isotropic term (Eq. \ref{eq:M1S}) still has a small effect due to mixing of the rotational states. Thus, precise measurements of the magnetic dependence of these transitions can be used for an accurate determination of the anisotropic electron-spin and rotational $g$-factors. 

The ``stretched'' transitions depend linearly on the magnetic field in the range considered here (up to 70 G) as can be seen in Fig. \ref{fig:RotQubitI0}c. The frequencies of these transitions will change by $\Delta E/h=\Delta f \approx $475 mHz for magnetic field fluctuations of 1 mG. Therefore, they can be exploited for encoding THz qubits with coherence times of up to $1/\Delta f \approx 2$ s \cite{saleh07} and for precision THz spectroscopy. The rotational spectrum of the $I = 2$ nuclear-spin species also exhibits ``magic'' magnetic-field insensitive transitions with second-order shifts as low as $\sim 3$ mHz/mG$^2$ (Appendix \ref{sec:magic_transitions}). Magnetic field fluctuations on the order of $\sim 1$ mG still permit qubits with Zeeman-limited coherence times of up to $1/\Delta f \approx 5$ min.

\begin{figure}
	\centering
	\includegraphics[width=\linewidth,trim={0cm 0cm 0cm 0cm},clip]{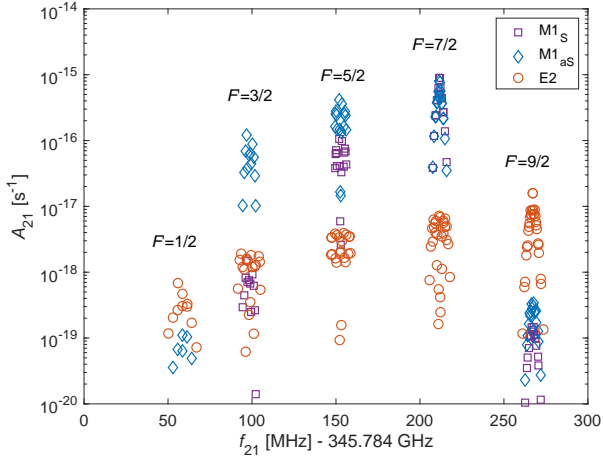}
  	\caption{Einstein $A$ coefficients of hyperfine-Zeeman components of the spin-rotational transition $|N=0,J=1/2,F=5/2\rangle \rightarrow |N'=2,J'=5/2,F'\rangle$ of the $I=2$ nuclear-spin species. Transitions due to M1$_S$, M1$_{aS}$ and E2 are indicated by purple squares, blue diamonds and red circles. The magnetic field was set to 5 G.}
	\label{fig:RotQubitI2}
\end{figure}

In Fig. \ref{fig:RotQubitI2}, the hyperfine components of the transition $|N=0,J=1/2,F=5/2\rangle \rightarrow |N'=2,J'=5/2,F'\rangle$ in the $I=2$ nuclear-spin state due to M1 and E2 coupling are shown. The M1$_S$ rotational transitions are allowed by rotational mixing induced by the dipolar hyperfine interaction, $H_t$. The strongest M1$_S$ lines are on-par with the strongest M1$_{aS}$ lines and are up to two order of magnitude stronger than the E2 lines. However, in some cases the strengths for both types of transitions are similar and in other cases, only E2 transitions are allowed due to quadrupole selections rules. Thus, one should consider both types of transitions when analyzing the molecular spectrum. To directly compare the strength of both types of couplings with Eq. \ref{eq:S12}, we calculated the relevant Einstein $A$ coefficients using Eq. \ref{eq:AM1} and Eq. \ref{eq:AE2}. 

\subsection{Rovibrational qubits}

\begin{figure}
    \centering
    \includegraphics[width=\linewidth,trim={0cm 0cm 0cm 0cm},clip]{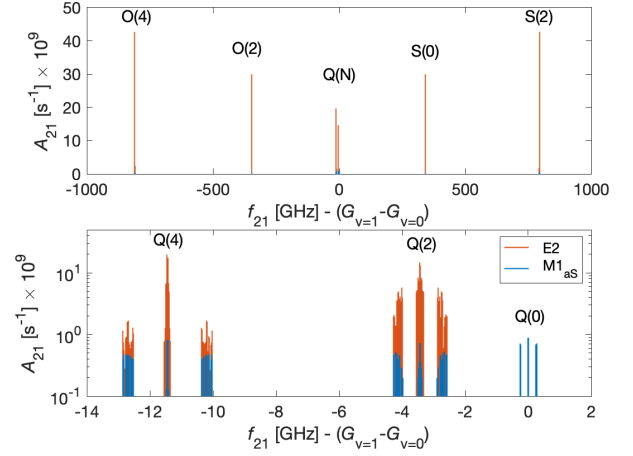}
    \caption{Einstein $A$ coefficients of E2 (red lines) and M1$_{aS}$ (blue lines) transitions of the O($N$=2,4), Q($N$=0,2,4) and S($N$=0,2) branches of the fundamental vibrational transition $|v=0,N\rangle \rightarrow |v'=1,N'\rangle$. The magnetic field was assumed to be 5 G. The lower panel shows a magnification of the spectrum in the region of the Q branch.}
    \label{fig:VibOverview}
\end{figure}

Dipole-forbidden rovibrational lines in N$_2^+$ were first observed by Germann et al. \cite{germann14a}. Vibrational transitions are promising for tests of a possible temporal variation of the proton-to-electron mass ratio because of their sensitivity to these constants. Also, they benefit from higher transition frequencies than rotational lines and thus allow for a better relative precision \cite{kajita14a}. Fig. \ref{fig:VibOverview} shows the O($N$), Q($N$) and S($N$) branches of the fundamental vibrational spectrum, i.e., transitions with $|v=0,N\rangle \rightarrow |v'=1,N'=N+ (-2,0,+2)\rangle$ and $N=0,2,4$ and for the two ortho nuclear-spin species.

Transitions within the Q(0) manifold, $\left|v=0,N=0\right\rangle \rightarrow \left|v'=1,N'=0\right\rangle$, are usually considered to be forbidden for single photon excitation within a $\Sigma$ electronic state \cite{kajita14a}. E2 selection rules forbid transitions from $N=0$ to $N'=0$. However, the anisotropic electron-spin interaction (Eq. \ref{eq:M1l}) permits $N=0$ to $N'=0$ transitions and it varies considerably with the internuclear distance (Eq. \ref{eq:Radial}). This leads to the appearance of Q(0) lines in the spectrum which, to the best of our knowledge, were so far not considered for the present vibrational spectrum.

\begin{figure}
    \centering
    \includegraphics[width=\linewidth,trim={0cm 0cm 0cm 0cm},clip]{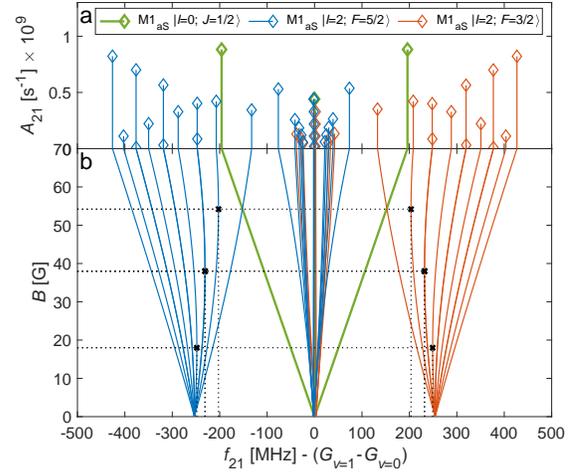}
    \caption{a) Einstein $A$ coefficients  b) and line positions as function of magnetic-field strength, $B$, of hyperfine-Zeeman components of the Q(0) line of the fundamental vibrational transition $|v=0,N=0\rangle \rightarrow |v'=1,N'=0\rangle$. Black crosses and dotted lines indicate positions of ``magic'' magnetic-field transitions.}
    \label{fig:VibQubitQ0}
\end{figure}

In Fig. \ref{fig:VibQubitQ0}, components of the Q(0) transition, i.e. $|v=0,N=0,J=1/2\rangle \rightarrow |v'=1,N'=0,J'=1/2\rangle$, of both the $I=0$ and $I=2$ species is shown. For both nuclear-spin configurations, the ``stretched'' transitions, i.e $|J=1/2,m=\pm1/2\rangle \rightarrow |J'=1/2,m'=\pm1/2\rangle$ and $|F=5/2,m=\pm5/2\rangle \rightarrow |F'=5/2,m'=\pm5/2\rangle$ are allowed by M1$_{aS}$ coupling and show very small linear Zeeman shifts of $\Delta g_l/3\approx 9.3$ mHz/mG. This dependency is $\sim 50$ times smaller than for transitions between ``stretched'' states in the S(0) ($|N=0\rangle \rightarrow |N'=2\rangle$) manifold. Precise measurements of the magnetic dependence of these transitions constitute a direct measurement of the anisotropy of the electron-spin $g$-factor tensor.

The Q(0) spectrum also exhibits ``magic`` transitions for the $I=2$ species (indicated by black crosses in Fig. \ref{fig:VibQubitQ0}) at relatively low magnetic fields of a few 10~G. The second-order Zeeman susceptibility of these transitions is $\sim16$ mHz/mG$^2$. Note that there are no ``magic'' transitions for the $I=0$ nuclear-spin configuration when driving a transition from the rotational ground state, $N=0$. This is due to the linear Zeeman shifts of the rotational ground state at the magnetic field values considered here (see Fig. \ref{fig:EN0}a green lines). 

We now turn to discuss S(0) transitions, i.e. $\left|v=0,N=0\right\rangle \rightarrow \left|v'=1,N'=2\right\rangle$. The S(0) spectrum is predicted to be $\sim30$ times stronger (see Fig. \ref{fig:VibOverview} red lines) than the the Q(0) spectrum due to E2 transitions (Eq. \ref{eq:Radial}). The second largest contribution to the S(0) spectrum is due to M1$_{aS}$ coupling (Fig. \ref{fig:VibOverview} blue lines). All other coupling mechanisms were found to be more than 5 orders of magnitude smaller. Q(2) transitions, $\left|v=0,N=2\right\rangle \rightarrow \left|v'=1,N'=2\right\rangle$, are also dominated by E2 coupling.

The S(0) spectrum is predicted to exhibit ``magic'' transitions at low magnetic fields of a few Gauss and with second-order Zeeman susceptibilities as low as $\sim$ 1 mHz/mG$^2$ (see Appendix \ref{sec:magic_transitions}). With magnetic field fluctuations on the order of $\sim 1$ mG, they can be used for encoding vibrational qubits with coherence times of up to $\approx 15$ min. This corresponds to a relative Zeeman shift of $\Delta E / E \approx 1 \times 10^{-17}$ without any active or passive magnetic field stabilization. 
The S(0) spectrum also features ``streched'' transitions that have a low linear Zeeman shift of $\sim$480 mHz/mG. 

The S(0) transitions at 4.574 $\mu$m with $A \approx 3 \times 10^{-8}$~Hz can be driven using commercial quantum-cascade lasers as demonstrated in Ref. \cite{germann14a}. With typical values for the laser power of 100 mW and a $1/e$ beam radius of 50 $\mu$m at the position of the molecule, Rabi frequencies \cite{leibfried03a} of $\Omega \sim(2\pi)$10 kHz are estimated yielding $\pi$-pulse times of $t_\pi\sim$50 $\mu$s thus enabling an efficient coherent manipulation of the rovibrational levels of the molecule. For the Q(0) transitions, $A \approx 4 \times 10^{-10}$ Hz. With the same laser parameters, $\Omega \sim(2\pi)$0.5~kHz and $t_\pi\sim$1~ms are estimated.

\begin{figure}
    \centering
    \includegraphics[width=\linewidth,trim={0cm 0cm 0cm 0cm},clip]{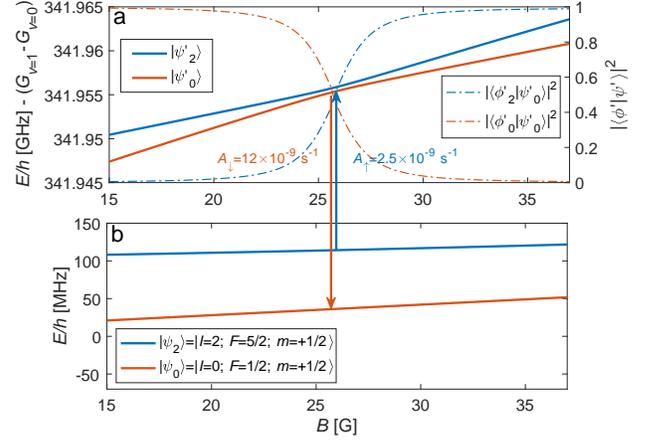}
    \caption{a) Adiabatic level energies, $E$, as a function of the magnetic field, $B$, showing an avoided crossing between two states originating from $|I'=2,F'=3/2,m'=-3/2\rangle$ ($|\psi'_2\rangle$, blue line) and $|I'=0,J'=3/2,m'=-3/2\rangle$ ($|\psi'_0\rangle$, red line) in the $v'=1,N'=2$ rovibrational-state manifold. The dashed-dotted lines indicate the degree of mixing, $\left|\langle\phi'_j|\psi'_0\rangle\right|^2$, where $|\phi'_j\rangle$ are the basis states ($j=0,2$) describing $|\psi'_j\rangle$ at low magnetic fields. At a magnetic-field strength of $\sim$25.8 G, a 50-50 mixture of the $I=0$ and $I=2$ states is predicted. b) Levels of the $I=2$ (blue) and the $I=0$ (red) nuclear-spin configurations in the rovibrational ground state are coupled by a resonant two-photon process (arrows). The coupling is performed through S(0) transitions to one of the highly mixed nuclear-spin states. The coupling strengths to the $|\psi'_{0}\rangle$ state in the form of the Einstein $A$ coefficients for $\Delta m=-2$ transitions are given in the plot.}
    \label{fig:stirap}
\end{figure}

The $|v'=1,N'=2\rangle$ rotational manifold of the first excited vibrational state exhibits avoided crossings between levels of the two ortho-nuclear-spin species as illustrated in Fig. \ref{fig:stirap}. The mixing is again induced by the quadrupole hyperfine interaction, $H_{eqQ}$. At a magnetic field of $\sim$25.8 G, the states labeled $|\psi'_{0,2}\rangle$ in Fig. \ref{fig:stirap} are composed of a 50-50 mixture of the $|I'=2,F'=3/2,m'=-3/2\rangle$ and $|I'=0,J'=3/2,m'=-3/2\rangle$ basis states. This opens up the possibility of coupling two distinct molecular states of different nuclear-spin character, for instance the $|\psi_2\rangle=|I=2,F=5/2,m=1/2\rangle$ and $|\psi_0\rangle=|I=0,J=1/2,m=1/2\rangle$ in the rovibrational ground state, $v=0,N=0$. These states show negligible nuclear-spin mixing. Fig. \ref{fig:stirap} illustrates how these two states in the vibrational ground state can be interconverted by excitation and deexcitation to the mixed states $|\psi'_{0,2}\rangle$ in $v=1$. Alternatively, interconversion of the nuclear-spin states can be achieved by populating one of the mixed state in $v=1$ and appropriately tuning the magnetic field across the crossing region.   

\section{Summary and conclusions}
In this paper, we presented a theoretical study of dipole-forbidden spectroscopic transitions in N$_2^+$ considering the complete hyperfine, rovibrational and Zeeman level structure. We identified magnetic-field insensitive transitions which are promising for encoding qubits because of their excellent coherence properties and for clock operation because of their weak dependence on magnetic fields. We calculated the strengths of magnetic-dipole and electric-quadrupole allowed spectroscopic transitions showing the importance of both in the radiofrequency, microwave and infrared spectrum of the nitrogen molecular ion. We found that pure vibrational transitions, Q(0), are allowed by M1 coupling induced by the anisotropic spin-magnetic field interaction. These transitions, which benefit from the lowest systematic shifts for clock operation, were not considered in the single-photon spectrum of N$_2^+$ so far. Finally, we showed that the electric-quadrupole hyperfine interaction causes avoided crossings between states of the two ortho nuclear-spin configurations of nitrogen. This magnetically enhanced nuclear-spin mixing opens the possibility to coherently transmute the the nuclear-spin state on demand.   


It is instructive to make a quantitative comparison between the magnetic insensitivity of clock transitions embedded in N$_2^+$ to other clock systems, e.g., Al$^+$ quantum-logic clocks which currently exhibit among the lowest systematic uncertainties \cite{brewer19}. The Al$^+$ clock is based on the $^1$S$_0 \leftrightarrow ^3$P$_0$ electronic transition which is first-order magnetically sensitive due to the nuclear spin, $I=5/2$, of $^{27}$Al$^+$. By averaging two stretched Zeeman transitions, the first-order shift is canceled, and the clock only depends on second-order Zeeman shifts \cite{brewer2019b}. This averaging technique is not an option for qubit applications. The second-order sensitivity of Al$^+$, $\sim7.2\times10^{-4}$ mHz/mG$^2$, is five orders of magnitude smaller than the sensitivity of N$_2^+$ ``magic'' transitions analysed here. However, since the Al$^+$ clock works at a finite magnetic field of 1.2 G \cite{brewer19}, the clock transition acquires an effective first-order sensitivity of $\sim1.7$ mHz/mG. This first-order sensitivity is five times smaller than that of the Q(0) ``streched'' transitions in N$_2^+$. However, for the ``magic'' transitions in N$_2^+$, which have a vanishing first-order sensitivity, the magnetic-field sensitivity breaks even with the Al$^+$ clock transition at a fluctuating magnetic field value of $\sim0.1$ mG which is a typical value for a system with actively stabilized magnetic field. Below this value, the N$_2^+$ ``magic'' transitions are less sensitive than the Al$^+$ clock while above they are more sensitive to magnetic field fluctuations.

\section*{Acknowledgements}

We thank Prof. Timothy Steimle (Arizona State University) for his critical reading of the manuscript and for his comments. This work has been supported by the Swiss National Science Foundation as part of the National Centre of Competence in Research, Quantum Science and Technology (NCCR-QSIT), grant nr. CRSII5\_183579, and by the University of Basel.


\begin{thebibliography}{78}%
\makeatletter
\providecommand \@ifxundefined [1]{%
 \@ifx{#1\undefined}
}%
\providecommand \@ifnum [1]{%
 \ifnum #1\expandafter \@firstoftwo
 \else \expandafter \@secondoftwo
 \fi
}%
\providecommand \@ifx [1]{%
 \ifx #1\expandafter \@firstoftwo
 \else \expandafter \@secondoftwo
 \fi
}%
\providecommand \natexlab [1]{#1}%
\providecommand \enquote  [1]{``#1''}%
\providecommand \bibnamefont  [1]{#1}%
\providecommand \bibfnamefont [1]{#1}%
\providecommand \citenamefont [1]{#1}%
\providecommand \href@noop [0]{\@secondoftwo}%
\providecommand \href [0]{\begingroup \@sanitize@url \@href}%
\providecommand \@href[1]{\@@startlink{#1}\@@href}%
\providecommand \@@href[1]{\endgroup#1\@@endlink}%
\providecommand \@sanitize@url [0]{\catcode `\\12\catcode `\$12\catcode
  `\&12\catcode `\#12\catcode `\^12\catcode `\_12\catcode `\%12\relax}%
\providecommand \@@startlink[1]{}%
\providecommand \@@endlink[0]{}%
\providecommand \url  [0]{\begingroup\@sanitize@url \@url }%
\providecommand \@url [1]{\endgroup\@href {#1}{\urlprefix }}%
\providecommand \urlprefix  [0]{URL }%
\providecommand \Eprint [0]{\href }%
\providecommand \doibase [0]{http://dx.doi.org/}%
\providecommand \selectlanguage [0]{\@gobble}%
\providecommand \bibinfo  [0]{\@secondoftwo}%
\providecommand \bibfield  [0]{\@secondoftwo}%
\providecommand \translation [1]{[#1]}%
\providecommand \BibitemOpen [0]{}%
\providecommand \bibitemStop [0]{}%
\providecommand \bibitemNoStop [0]{.\EOS\space}%
\providecommand \EOS [0]{\spacefactor3000\relax}%
\providecommand \BibitemShut  [1]{\csname bibitem#1\endcsname}%
\let\auto@bib@innerbib\@empty
\bibitem [{\citenamefont {{\mbox{van de} Meerakker}}\ \emph
  {et~al.}(2012)\citenamefont {{\mbox{van de} Meerakker}}, \citenamefont
  {Bethlem}, \citenamefont {Vanhaecke},\ and\ \citenamefont
  {Meijer}}]{meerakker12a}%
  \BibitemOpen
  \bibfield  {author} {\bibinfo {author} {\bibfnamefont {S.~Y.~T.}\
  \bibnamefont {{\mbox{van de} Meerakker}}}, \bibinfo {author} {\bibfnamefont
  {H.~L.}\ \bibnamefont {Bethlem}}, \bibinfo {author} {\bibfnamefont
  {N.}~\bibnamefont {Vanhaecke}}, \ and\ \bibinfo {author} {\bibfnamefont
  {G.}~\bibnamefont {Meijer}},\ }\href@noop {} {\bibfield  {journal} {\bibinfo
  {journal} {Chem. Rev.}\ }\textbf {\bibinfo {volume} {112}},\ \bibinfo {pages}
  {4828} (\bibinfo {year} {2012})}\BibitemShut {NoStop}%
\bibitem [{\citenamefont {Segev}\ \emph {et~al.}(2019)\citenamefont {Segev},
  \citenamefont {Pitzer}, \citenamefont {Karpov}, \citenamefont {Akerman},
  \citenamefont {Narevicius},\ and\ \citenamefont {Narevicius}}]{segev19b}%
  \BibitemOpen
  \bibfield  {author} {\bibinfo {author} {\bibfnamefont {Y.}~\bibnamefont
  {Segev}}, \bibinfo {author} {\bibfnamefont {M.}~\bibnamefont {Pitzer}},
  \bibinfo {author} {\bibfnamefont {M.}~\bibnamefont {Karpov}}, \bibinfo
  {author} {\bibfnamefont {N.}~\bibnamefont {Akerman}}, \bibinfo {author}
  {\bibfnamefont {J.}~\bibnamefont {Narevicius}}, \ and\ \bibinfo {author}
  {\bibfnamefont {E.}~\bibnamefont {Narevicius}},\ }\href@noop {} {\bibfield
  {journal} {\bibinfo  {journal} {Nature}\ }\textbf {\bibinfo {volume} {572}},\
  \bibinfo {pages} {189} (\bibinfo {year} {2019})}\BibitemShut {NoStop}%
\bibitem [{\citenamefont {Anderegg}\ \emph {et~al.}(2018)\citenamefont
  {Anderegg}, \citenamefont {Augenbraun}, \citenamefont {Bao}, \citenamefont
  {Burchesky}, \citenamefont {Cheuk}, \citenamefont {Ketterle},\ and\
  \citenamefont {Doyle}}]{anderegg18a}%
  \BibitemOpen
  \bibfield  {author} {\bibinfo {author} {\bibfnamefont {L.}~\bibnamefont
  {Anderegg}}, \bibinfo {author} {\bibfnamefont {B.~L.}\ \bibnamefont
  {Augenbraun}}, \bibinfo {author} {\bibfnamefont {Y.}~\bibnamefont {Bao}},
  \bibinfo {author} {\bibfnamefont {S.}~\bibnamefont {Burchesky}}, \bibinfo
  {author} {\bibfnamefont {L.~W.}\ \bibnamefont {Cheuk}}, \bibinfo {author}
  {\bibfnamefont {W.}~\bibnamefont {Ketterle}}, \ and\ \bibinfo {author}
  {\bibfnamefont {J.~M.}\ \bibnamefont {Doyle}},\ }\href@noop {} {\bibfield
  {journal} {\bibinfo  {journal} {Nat. Phys.}\ }\textbf {\bibinfo {volume}
  {14}},\ \bibinfo {pages} {890} (\bibinfo {year} {2018})}\BibitemShut
  {NoStop}%
\bibitem [{\citenamefont {McCarron}\ \emph {et~al.}(2018)\citenamefont
  {McCarron}, \citenamefont {Steinecker}, \citenamefont {Zhu},\ and\
  \citenamefont {DeMille}}]{mccarron18a}%
  \BibitemOpen
  \bibfield  {author} {\bibinfo {author} {\bibfnamefont {D.~J.}\ \bibnamefont
  {McCarron}}, \bibinfo {author} {\bibfnamefont {M.~H.}\ \bibnamefont
  {Steinecker}}, \bibinfo {author} {\bibfnamefont {Y.}~\bibnamefont {Zhu}}, \
  and\ \bibinfo {author} {\bibfnamefont {D.}~\bibnamefont {DeMille}},\
  }\href@noop {} {\bibfield  {journal} {\bibinfo  {journal} {Phys. Rev. Lett.}\
  }\textbf {\bibinfo {volume} {121}},\ \bibinfo {pages} {013202} (\bibinfo
  {year} {2018})}\BibitemShut {NoStop}%
\bibitem [{\citenamefont {Caldwell}\ \emph {et~al.}(2019)\citenamefont
  {Caldwell}, \citenamefont {Devlin}, \citenamefont {Williams}, \citenamefont
  {Fitch}, \citenamefont {Hinds}, \citenamefont {Sauer},\ and\ \citenamefont
  {Tarbutt}}]{caldwell19a}%
  \BibitemOpen
  \bibfield  {author} {\bibinfo {author} {\bibfnamefont {L.}~\bibnamefont
  {Caldwell}}, \bibinfo {author} {\bibfnamefont {J.}~\bibnamefont {Devlin}},
  \bibinfo {author} {\bibfnamefont {H.}~\bibnamefont {Williams}}, \bibinfo
  {author} {\bibfnamefont {N.}~\bibnamefont {Fitch}}, \bibinfo {author}
  {\bibfnamefont {E.}~\bibnamefont {Hinds}}, \bibinfo {author} {\bibfnamefont
  {B.}~\bibnamefont {Sauer}}, \ and\ \bibinfo {author} {\bibfnamefont
  {M.}~\bibnamefont {Tarbutt}},\ }\href@noop {} {\bibfield  {journal} {\bibinfo
   {journal} {Phys. Rev. Lett.}\ }\textbf {\bibinfo {volume} {123}},\ \bibinfo
  {pages} {033202} (\bibinfo {year} {2019})}\BibitemShut {NoStop}%
\bibitem [{\citenamefont {Moses}\ \emph {et~al.}(2017)\citenamefont {Moses},
  \citenamefont {Covey}, \citenamefont {Miecnikowski}, \citenamefont {Jin},\
  and\ \citenamefont {Ye}}]{moses17}%
  \BibitemOpen
  \bibfield  {author} {\bibinfo {author} {\bibfnamefont {S.~A.}\ \bibnamefont
  {Moses}}, \bibinfo {author} {\bibfnamefont {J.~P.}\ \bibnamefont {Covey}},
  \bibinfo {author} {\bibfnamefont {M.~T.}\ \bibnamefont {Miecnikowski}},
  \bibinfo {author} {\bibfnamefont {D.~S.}\ \bibnamefont {Jin}}, \ and\
  \bibinfo {author} {\bibfnamefont {J.}~\bibnamefont {Ye}},\ }\href@noop {}
  {\bibfield  {journal} {\bibinfo  {journal} {Nat. Phys.}\ }\textbf {\bibinfo
  {volume} {13}},\ \bibinfo {pages} {13} (\bibinfo {year} {2017})}\BibitemShut
  {NoStop}%
\bibitem [{\citenamefont {M{\o}lhave}\ and\ \citenamefont
  {Drewsen}(2000)}]{molhave00a}%
  \BibitemOpen
  \bibfield  {author} {\bibinfo {author} {\bibfnamefont {K.}~\bibnamefont
  {M{\o}lhave}}\ and\ \bibinfo {author} {\bibfnamefont {M.}~\bibnamefont
  {Drewsen}},\ }\href@noop {} {\bibfield  {journal} {\bibinfo  {journal}
  {{Phys. Rev. A}}\ }\textbf {\bibinfo {volume} {62}},\ \bibinfo {pages}
  {011401} (\bibinfo {year} {2000})}\BibitemShut {NoStop}%
\bibitem [{\citenamefont {Tong}\ \emph {et~al.}(2010)\citenamefont {Tong},
  \citenamefont {Winney},\ and\ \citenamefont {Willitsch}}]{tong10a}%
  \BibitemOpen
  \bibfield  {author} {\bibinfo {author} {\bibfnamefont {X.}~\bibnamefont
  {Tong}}, \bibinfo {author} {\bibfnamefont {A.~H.}\ \bibnamefont {Winney}}, \
  and\ \bibinfo {author} {\bibfnamefont {S.}~\bibnamefont {Willitsch}},\
  }\href@noop {} {\bibfield  {journal} {\bibinfo  {journal} {{Phys. Rev.
  Lett.}}\ }\textbf {\bibinfo {volume} {105}},\ \bibinfo {pages} {143001}
  (\bibinfo {year} {2010})}\BibitemShut {NoStop}%
\bibitem [{\citenamefont {Wolf}\ \emph {et~al.}(2016)\citenamefont {Wolf},
  \citenamefont {Wan}, \citenamefont {Heip}, \citenamefont {Gebert},
  \citenamefont {Shi},\ and\ \citenamefont {Schmidt}}]{wolf16a}%
  \BibitemOpen
  \bibfield  {author} {\bibinfo {author} {\bibfnamefont {F.}~\bibnamefont
  {Wolf}}, \bibinfo {author} {\bibfnamefont {Y.}~\bibnamefont {Wan}}, \bibinfo
  {author} {\bibfnamefont {J.~C.}\ \bibnamefont {Heip}}, \bibinfo {author}
  {\bibfnamefont {F.}~\bibnamefont {Gebert}}, \bibinfo {author} {\bibfnamefont
  {C.}~\bibnamefont {Shi}}, \ and\ \bibinfo {author} {\bibfnamefont {P.~O.}\
  \bibnamefont {Schmidt}},\ }\href@noop {} {\bibfield  {journal} {\bibinfo
  {journal} {Nature}\ }\textbf {\bibinfo {volume} {530}},\ \bibinfo {pages}
  {457} (\bibinfo {year} {2016})}\BibitemShut {NoStop}%
\bibitem [{\citenamefont {Chou}\ \emph {et~al.}(2017)\citenamefont {Chou},
  \citenamefont {Kurz}, \citenamefont {Hume}, \citenamefont {Plessow},
  \citenamefont {Leibrandt},\ and\ \citenamefont {Leibfried}}]{chou17a}%
  \BibitemOpen
  \bibfield  {author} {\bibinfo {author} {\bibfnamefont {C.~W.}\ \bibnamefont
  {Chou}}, \bibinfo {author} {\bibfnamefont {C.}~\bibnamefont {Kurz}}, \bibinfo
  {author} {\bibfnamefont {D.~B.}\ \bibnamefont {Hume}}, \bibinfo {author}
  {\bibfnamefont {P.~N.}\ \bibnamefont {Plessow}}, \bibinfo {author}
  {\bibfnamefont {D.~R.}\ \bibnamefont {Leibrandt}}, \ and\ \bibinfo {author}
  {\bibfnamefont {D.}~\bibnamefont {Leibfried}},\ }\href@noop {} {\bibfield
  {journal} {\bibinfo  {journal} {Nature}\ }\textbf {\bibinfo {volume} {545}},\
  \bibinfo {pages} {203} (\bibinfo {year} {2017})}\BibitemShut {NoStop}%
\bibitem [{\citenamefont {Sinhal}\ \emph {et~al.}(2020)\citenamefont {Sinhal},
  \citenamefont {Meir}, \citenamefont {Najafian}, \citenamefont {Hegi},\ and\
  \citenamefont {Willitsch}}]{sinhal20}%
  \BibitemOpen
  \bibfield  {author} {\bibinfo {author} {\bibfnamefont {M.}~\bibnamefont
  {Sinhal}}, \bibinfo {author} {\bibfnamefont {Z.}~\bibnamefont {Meir}},
  \bibinfo {author} {\bibfnamefont {K.}~\bibnamefont {Najafian}}, \bibinfo
  {author} {\bibfnamefont {G.}~\bibnamefont {Hegi}}, \ and\ \bibinfo {author}
  {\bibfnamefont {S.}~\bibnamefont {Willitsch}},\ }\href@noop {} {\bibfield
  {journal} {\bibinfo  {journal} {Science}\ }\textbf {\bibinfo {volume}
  {367}},\ \bibinfo {pages} {1213} (\bibinfo {year} {2020})}\BibitemShut
  {NoStop}%
\bibitem [{\citenamefont {Chou}\ \emph {et~al.}(2020)\citenamefont {Chou},
  \citenamefont {Collopy}, \citenamefont {Kurz}, \citenamefont {Lin},
  \citenamefont {Harding}, \citenamefont {Plessow}, \citenamefont {Fortier},
  \citenamefont {Diddams}, \citenamefont {Leibfried},\ and\ \citenamefont
  {Leibrandt}}]{chou20a}%
  \BibitemOpen
  \bibfield  {author} {\bibinfo {author} {\bibfnamefont {C.}~\bibnamefont
  {Chou}}, \bibinfo {author} {\bibfnamefont {A.}~\bibnamefont {Collopy}},
  \bibinfo {author} {\bibfnamefont {C.}~\bibnamefont {Kurz}}, \bibinfo {author}
  {\bibfnamefont {Y.}~\bibnamefont {Lin}}, \bibinfo {author} {\bibfnamefont
  {M.}~\bibnamefont {Harding}}, \bibinfo {author} {\bibfnamefont
  {P.}~\bibnamefont {Plessow}}, \bibinfo {author} {\bibfnamefont
  {T.}~\bibnamefont {Fortier}}, \bibinfo {author} {\bibfnamefont
  {S.}~\bibnamefont {Diddams}}, \bibinfo {author} {\bibfnamefont
  {D.}~\bibnamefont {Leibfried}}, \ and\ \bibinfo {author} {\bibfnamefont
  {D.}~\bibnamefont {Leibrandt}},\ }\href@noop {} {\bibfield  {journal}
  {\bibinfo  {journal} {Science}\ }\textbf {\bibinfo {volume} {367}},\ \bibinfo
  {pages} {1458} (\bibinfo {year} {2020})}\BibitemShut {NoStop}%
\bibitem [{\citenamefont {Lin}\ \emph {et~al.}(2020)\citenamefont {Lin},
  \citenamefont {Leibrandt}, \citenamefont {Leibfried},\ and\ \citenamefont
  {Chou}}]{lin19}%
  \BibitemOpen
  \bibfield  {author} {\bibinfo {author} {\bibfnamefont {Y.}~\bibnamefont
  {Lin}}, \bibinfo {author} {\bibfnamefont {D.~R.}\ \bibnamefont {Leibrandt}},
  \bibinfo {author} {\bibfnamefont {D.}~\bibnamefont {Leibfried}}, \ and\
  \bibinfo {author} {\bibfnamefont {C.-w.}\ \bibnamefont {Chou}},\ }\href@noop
  {} {\bibfield  {journal} {\bibinfo  {journal} {Nature}\ }\textbf {\bibinfo
  {volume} {581}},\ \bibinfo {pages} {273} (\bibinfo {year}
  {2020})}\BibitemShut {NoStop}%
\bibitem [{\citenamefont {Najafian}\ \emph {et~al.}(2020)\citenamefont
  {Najafian}, \citenamefont {Meir}, \citenamefont {Sinhal},\ and\ \citenamefont
  {Willitsch}}]{Najafian20a}%
  \BibitemOpen
  \bibfield  {author} {\bibinfo {author} {\bibfnamefont {K.}~\bibnamefont
  {Najafian}}, \bibinfo {author} {\bibfnamefont {Z.}~\bibnamefont {Meir}},
  \bibinfo {author} {\bibfnamefont {M.}~\bibnamefont {Sinhal}}, \ and\ \bibinfo
  {author} {\bibfnamefont {S.}~\bibnamefont {Willitsch}},\ }\href@noop {}
  {\bibfield  {journal} {\bibinfo  {journal} {arxiv: 2004.05306}\ } (\bibinfo
  {year} {2020})}\BibitemShut {NoStop}%
\bibitem [{\citenamefont {Schmidt}\ \emph {et~al.}(2005)\citenamefont
  {Schmidt}, \citenamefont {Rosenband}, \citenamefont {Langer}, \citenamefont
  {Itano}, \citenamefont {Bergquist},\ and\ \citenamefont
  {Wineland}}]{schmidt05a}%
  \BibitemOpen
  \bibfield  {author} {\bibinfo {author} {\bibfnamefont {P.~O.}\ \bibnamefont
  {Schmidt}}, \bibinfo {author} {\bibfnamefont {T.}~\bibnamefont {Rosenband}},
  \bibinfo {author} {\bibfnamefont {C.}~\bibnamefont {Langer}}, \bibinfo
  {author} {\bibfnamefont {W.~M.}\ \bibnamefont {Itano}}, \bibinfo {author}
  {\bibfnamefont {J.~C.}\ \bibnamefont {Bergquist}}, \ and\ \bibinfo {author}
  {\bibfnamefont {D.~J.}\ \bibnamefont {Wineland}},\ }\href@noop {} {\bibfield
  {journal} {\bibinfo  {journal} {Science}\ }\textbf {\bibinfo {volume}
  {309}},\ \bibinfo {pages} {749} (\bibinfo {year} {2005})}\BibitemShut
  {NoStop}%
\bibitem [{\citenamefont {Semeria}\ \emph {et~al.}(2020)\citenamefont
  {Semeria}, \citenamefont {Jansen}, \citenamefont {Camenisch}, \citenamefont
  {Mellini}, \citenamefont {Schmutz},\ and\ \citenamefont
  {Merkt}}]{semeria20a}%
  \BibitemOpen
  \bibfield  {author} {\bibinfo {author} {\bibfnamefont {L.}~\bibnamefont
  {Semeria}}, \bibinfo {author} {\bibfnamefont {P.}~\bibnamefont {Jansen}},
  \bibinfo {author} {\bibfnamefont {G.-M.}\ \bibnamefont {Camenisch}}, \bibinfo
  {author} {\bibfnamefont {F.}~\bibnamefont {Mellini}}, \bibinfo {author}
  {\bibfnamefont {H.}~\bibnamefont {Schmutz}}, \ and\ \bibinfo {author}
  {\bibfnamefont {F.}~\bibnamefont {Merkt}},\ }\href@noop {} {\bibfield
  {journal} {\bibinfo  {journal} {Phys. Rev. Lett.}\ }\textbf {\bibinfo
  {volume} {124}},\ \bibinfo {pages} {213001} (\bibinfo {year}
  {2020})}\BibitemShut {NoStop}%
\bibitem [{\citenamefont {Alighanbari}\ \emph {et~al.}(2018)\citenamefont
  {Alighanbari}, \citenamefont {Hansen}, \citenamefont {Korobov},\ and\
  \citenamefont {Schiller}}]{alighanbari18a}%
  \BibitemOpen
  \bibfield  {author} {\bibinfo {author} {\bibfnamefont {S.}~\bibnamefont
  {Alighanbari}}, \bibinfo {author} {\bibfnamefont {M.~G.}\ \bibnamefont
  {Hansen}}, \bibinfo {author} {\bibfnamefont {V.}~\bibnamefont {Korobov}}, \
  and\ \bibinfo {author} {\bibfnamefont {S.}~\bibnamefont {Schiller}},\
  }\href@noop {} {\bibfield  {journal} {\bibinfo  {journal} {Nat. Phys.}\
  }\textbf {\bibinfo {volume} {14}},\ \bibinfo {pages} {555} (\bibinfo {year}
  {2018})}\BibitemShut {NoStop}%
\bibitem [{\citenamefont {Biesheuvel}\ \emph {et~al.}(2016)\citenamefont
  {Biesheuvel}, \citenamefont {Karr}, \citenamefont {Hilico}, \citenamefont
  {Eikema}, \citenamefont {Ubachs},\ and\ \citenamefont
  {Koelemeij}}]{biesheuvel16a}%
  \BibitemOpen
  \bibfield  {author} {\bibinfo {author} {\bibfnamefont {J.}~\bibnamefont
  {Biesheuvel}}, \bibinfo {author} {\bibfnamefont {J.~P.}\ \bibnamefont
  {Karr}}, \bibinfo {author} {\bibfnamefont {L.}~\bibnamefont {Hilico}},
  \bibinfo {author} {\bibfnamefont {K.~S.~E.}\ \bibnamefont {Eikema}}, \bibinfo
  {author} {\bibfnamefont {W.}~\bibnamefont {Ubachs}}, \ and\ \bibinfo {author}
  {\bibfnamefont {J.~C.~J.}\ \bibnamefont {Koelemeij}},\ }\href@noop {}
  {\bibfield  {journal} {\bibinfo  {journal} {Nat. Commun.}\ }\textbf {\bibinfo
  {volume} {7}},\ \bibinfo {pages} {10385} (\bibinfo {year}
  {2016})}\BibitemShut {NoStop}%
\bibitem [{\citenamefont {Safronova}\ \emph {et~al.}(2018)\citenamefont
  {Safronova}, \citenamefont {Budker}, \citenamefont {DeMille}, \citenamefont
  {Kimball}, \citenamefont {Derevianko},\ and\ \citenamefont
  {Clark}}]{safronova18a}%
  \BibitemOpen
  \bibfield  {author} {\bibinfo {author} {\bibfnamefont {M.~S.}\ \bibnamefont
  {Safronova}}, \bibinfo {author} {\bibfnamefont {D.}~\bibnamefont {Budker}},
  \bibinfo {author} {\bibfnamefont {D.}~\bibnamefont {DeMille}}, \bibinfo
  {author} {\bibfnamefont {D.~F.~J.}\ \bibnamefont {Kimball}}, \bibinfo
  {author} {\bibfnamefont {A.}~\bibnamefont {Derevianko}}, \ and\ \bibinfo
  {author} {\bibfnamefont {C.~W.}\ \bibnamefont {Clark}},\ }\href@noop {}
  {\bibfield  {journal} {\bibinfo  {journal} {Rev. Mod. Phys.}\ }\textbf
  {\bibinfo {volume} {90}},\ \bibinfo {pages} {025008} (\bibinfo {year}
  {2018})}\BibitemShut {NoStop}%
\bibitem [{\citenamefont {DeMille}\ \emph {et~al.}(2017)\citenamefont
  {DeMille}, \citenamefont {Doyle},\ and\ \citenamefont
  {Sushkov}}]{demille17a}%
  \BibitemOpen
  \bibfield  {author} {\bibinfo {author} {\bibfnamefont {D.}~\bibnamefont
  {DeMille}}, \bibinfo {author} {\bibfnamefont {J.~M.}\ \bibnamefont {Doyle}},
  \ and\ \bibinfo {author} {\bibfnamefont {A.~O.}\ \bibnamefont {Sushkov}},\
  }\href@noop {} {\bibfield  {journal} {\bibinfo  {journal} {Science}\ }\textbf
  {\bibinfo {volume} {357}},\ \bibinfo {pages} {990} (\bibinfo {year}
  {2017})}\BibitemShut {NoStop}%
\bibitem [{\citenamefont {Schiller}\ and\ \citenamefont
  {Korobov}(2005)}]{schiller05a}%
  \BibitemOpen
  \bibfield  {author} {\bibinfo {author} {\bibfnamefont {S.}~\bibnamefont
  {Schiller}}\ and\ \bibinfo {author} {\bibfnamefont {V.}~\bibnamefont
  {Korobov}},\ }\href {2505} {\bibfield  {journal} {\bibinfo  {journal} {{Phys.
  Rev. A}}\ }\textbf {\bibinfo {volume} {71}},\ \bibinfo {pages} {032505}
  (\bibinfo {year} {2005})}\BibitemShut {NoStop}%
\bibitem [{\citenamefont {Flambaum}\ and\ \citenamefont
  {Kozlov}(2007)}]{flambaum07a}%
  \BibitemOpen
  \bibfield  {author} {\bibinfo {author} {\bibfnamefont {V.}~\bibnamefont
  {Flambaum}}\ and\ \bibinfo {author} {\bibfnamefont {M.}~\bibnamefont
  {Kozlov}},\ }\href@noop {} {\bibfield  {journal} {\bibinfo  {journal} {Phys.
  Rev. Lett.}\ }\textbf {\bibinfo {volume} {99}},\ \bibinfo {pages} {150801}
  (\bibinfo {year} {2007})}\BibitemShut {NoStop}%
\bibitem [{\citenamefont {Salumbides}\ \emph {et~al.}(2013)\citenamefont
  {Salumbides}, \citenamefont {Koelemeij}, \citenamefont {Komasa},
  \citenamefont {Pachucki}, \citenamefont {Eikema},\ and\ \citenamefont
  {Ubachs}}]{salumbides13a}%
  \BibitemOpen
  \bibfield  {author} {\bibinfo {author} {\bibfnamefont {E.~J.}\ \bibnamefont
  {Salumbides}}, \bibinfo {author} {\bibfnamefont {J.~C.~J.}\ \bibnamefont
  {Koelemeij}}, \bibinfo {author} {\bibfnamefont {J.}~\bibnamefont {Komasa}},
  \bibinfo {author} {\bibfnamefont {K.}~\bibnamefont {Pachucki}}, \bibinfo
  {author} {\bibfnamefont {K.~S.~E.}\ \bibnamefont {Eikema}}, \ and\ \bibinfo
  {author} {\bibfnamefont {W.}~\bibnamefont {Ubachs}},\ }\href@noop {}
  {\bibfield  {journal} {\bibinfo  {journal} {{Phys. Rev. D}}\ }\textbf
  {\bibinfo {volume} {{87}}},\ \bibinfo {pages} {{112008}} (\bibinfo {year}
  {{2013}})}\BibitemShut {NoStop}%
\bibitem [{\citenamefont {H\"olsch}\ \emph {et~al.}(2019)\citenamefont
  {H\"olsch}, \citenamefont {Beyer}, \citenamefont {Salumbides}, \citenamefont
  {Eikema}, \citenamefont {Ubachs}, \citenamefont {\mbox{Ch.} Jungen},\ and\
  \citenamefont {Merkt}}]{hoelsch19a}%
  \BibitemOpen
  \bibfield  {author} {\bibinfo {author} {\bibfnamefont {N.}~\bibnamefont
  {H\"olsch}}, \bibinfo {author} {\bibfnamefont {M.}~\bibnamefont {Beyer}},
  \bibinfo {author} {\bibfnamefont {E.~J.}\ \bibnamefont {Salumbides}},
  \bibinfo {author} {\bibfnamefont {K.~S.~E.}\ \bibnamefont {Eikema}}, \bibinfo
  {author} {\bibfnamefont {W.}~\bibnamefont {Ubachs}}, \bibinfo {author}
  {\bibnamefont {\mbox{Ch.} Jungen}}, \ and\ \bibinfo {author} {\bibfnamefont
  {F.}~\bibnamefont {Merkt}},\ }\href@noop {} {\bibfield  {journal} {\bibinfo
  {journal} {Phys. Rev. Lett.}\ }\textbf {\bibinfo {volume} {122}},\ \bibinfo
  {pages} {103002} (\bibinfo {year} {2019})}\BibitemShut {NoStop}%
\bibitem [{\citenamefont {Sikorsky}\ \emph {et~al.}(2018)\citenamefont
  {Sikorsky}, \citenamefont {Meir}, \citenamefont {Ben-shlomi}, \citenamefont
  {Akerman},\ and\ \citenamefont {Ozeri}}]{sikorsky18a}%
  \BibitemOpen
  \bibfield  {author} {\bibinfo {author} {\bibfnamefont {T.}~\bibnamefont
  {Sikorsky}}, \bibinfo {author} {\bibfnamefont {Z.}~\bibnamefont {Meir}},
  \bibinfo {author} {\bibfnamefont {R.}~\bibnamefont {Ben-shlomi}}, \bibinfo
  {author} {\bibfnamefont {N.}~\bibnamefont {Akerman}}, \ and\ \bibinfo
  {author} {\bibfnamefont {R.}~\bibnamefont {Ozeri}},\ }\href@noop {}
  {\bibfield  {journal} {\bibinfo  {journal} {Nat. Commun.}\ }\textbf {\bibinfo
  {volume} {9}},\ \bibinfo {pages} {920} (\bibinfo {year} {2018})}\BibitemShut
  {NoStop}%
\bibitem [{\citenamefont {D{\"o}rfler}\ \emph {et~al.}(2019)\citenamefont
  {D{\"o}rfler}, \citenamefont {Eberle}, \citenamefont {Koner}, \citenamefont
  {Tomza}, \citenamefont {Meuwly},\ and\ \citenamefont
  {Willitsch}}]{dorfler19b}%
  \BibitemOpen
  \bibfield  {author} {\bibinfo {author} {\bibfnamefont {A.~D.}\ \bibnamefont
  {D{\"o}rfler}}, \bibinfo {author} {\bibfnamefont {P.}~\bibnamefont {Eberle}},
  \bibinfo {author} {\bibfnamefont {D.}~\bibnamefont {Koner}}, \bibinfo
  {author} {\bibfnamefont {M.}~\bibnamefont {Tomza}}, \bibinfo {author}
  {\bibfnamefont {M.}~\bibnamefont {Meuwly}}, \ and\ \bibinfo {author}
  {\bibfnamefont {S.}~\bibnamefont {Willitsch}},\ }\href@noop {} {\bibfield
  {journal} {\bibinfo  {journal} {Nat. Commun.}\ }\textbf {\bibinfo {volume}
  {10}},\ \bibinfo {pages} {5429} (\bibinfo {year} {2019})}\BibitemShut
  {NoStop}%
\bibitem [{\citenamefont {Germann}\ \emph {et~al.}(2014)\citenamefont
  {Germann}, \citenamefont {Tong},\ and\ \citenamefont
  {Willitsch}}]{germann14a}%
  \BibitemOpen
  \bibfield  {author} {\bibinfo {author} {\bibfnamefont {M.}~\bibnamefont
  {Germann}}, \bibinfo {author} {\bibfnamefont {X.}~\bibnamefont {Tong}}, \
  and\ \bibinfo {author} {\bibfnamefont {S.}~\bibnamefont {Willitsch}},\
  }\href@noop {} {\bibfield  {journal} {\bibinfo  {journal} {Nat. Phys.}\
  }\textbf {\bibinfo {volume} {10}},\ \bibinfo {pages} {820} (\bibinfo {year}
  {2014})}\BibitemShut {NoStop}%
\bibitem [{\citenamefont {\mbox{J.-Ph.} Karr}(2014)}]{karr14a}%
  \BibitemOpen
  \bibfield  {author} {\bibinfo {author} {\bibnamefont {\mbox{J.-Ph.} Karr}},\
  }\href@noop {} {\bibfield  {journal} {\bibinfo  {journal} {{J. Mol.
  Spectrosc.}}\ }\textbf {\bibinfo {volume} {300}},\ \bibinfo {pages} {37}
  (\bibinfo {year} {2014})}\BibitemShut {NoStop}%
\bibitem [{\citenamefont {Schiller}\ \emph {et~al.}(2014)\citenamefont
  {Schiller}, \citenamefont {Bakalov},\ and\ \citenamefont
  {Korobov}}]{schiller14a}%
  \BibitemOpen
  \bibfield  {author} {\bibinfo {author} {\bibfnamefont {S.}~\bibnamefont
  {Schiller}}, \bibinfo {author} {\bibfnamefont {D.}~\bibnamefont {Bakalov}}, \
  and\ \bibinfo {author} {\bibfnamefont {V.~I.}\ \bibnamefont {Korobov}},\
  }\href@noop {} {\bibfield  {journal} {\bibinfo  {journal} {{Phys. Rev.
  Lett.}}\ }\textbf {\bibinfo {volume} {113}},\ \bibinfo {pages} {023004}
  (\bibinfo {year} {2014})}\BibitemShut {NoStop}%
\bibitem [{\citenamefont {Willitsch}(2011)}]{willitsch11a}%
  \BibitemOpen
  \bibfield  {author} {\bibinfo {author} {\bibfnamefont {S.}~\bibnamefont
  {Willitsch}},\ }in\ \href@noop {} {\emph {\bibinfo {booktitle} {{Handbook of
  High-Resolution Spectroscopy}}}},\ Vol.~\bibinfo {volume} {3},\ \bibinfo
  {editor} {edited by\ \bibinfo {editor} {\bibfnamefont {M.}~\bibnamefont
  {Quack}}\ and\ \bibinfo {editor} {\bibfnamefont {F.}~\bibnamefont {Merkt}}}\
  (\bibinfo  {publisher} {John Wiley \& Sons},\ \bibinfo {year} {2011})\ p.\
  \bibinfo {pages} {1691}\BibitemShut {NoStop}%
\bibitem [{\citenamefont {Meir}\ \emph {et~al.}(2019)\citenamefont {Meir},
  \citenamefont {Hegi}, \citenamefont {Najafian}, \citenamefont {Sinhal},\ and\
  \citenamefont {Willitsch}}]{meir19a}%
  \BibitemOpen
  \bibfield  {author} {\bibinfo {author} {\bibfnamefont {Z.}~\bibnamefont
  {Meir}}, \bibinfo {author} {\bibfnamefont {G.}~\bibnamefont {Hegi}}, \bibinfo
  {author} {\bibfnamefont {K.}~\bibnamefont {Najafian}}, \bibinfo {author}
  {\bibfnamefont {M.}~\bibnamefont {Sinhal}}, \ and\ \bibinfo {author}
  {\bibfnamefont {S.}~\bibnamefont {Willitsch}},\ }\href@noop {} {\bibfield
  {journal} {\bibinfo  {journal} {Faraday Discuss.}\ }\textbf {\bibinfo
  {volume} {217}},\ \bibinfo {pages} {561} (\bibinfo {year}
  {2019})}\BibitemShut {NoStop}%
\bibitem [{\citenamefont {DeMille}(2002)}]{demille02a}%
  \BibitemOpen
  \bibfield  {author} {\bibinfo {author} {\bibfnamefont {D.}~\bibnamefont
  {DeMille}},\ }\href@noop {} {\bibfield  {journal} {\bibinfo  {journal}
  {{Phys. Rev. Lett.}}\ }\textbf {\bibinfo {volume} {88}},\ \bibinfo {pages}
  {067901} (\bibinfo {year} {2002})}\BibitemShut {NoStop}%
\bibitem [{\citenamefont {Blackmore}\ \emph {et~al.}(2018)\citenamefont
  {Blackmore}, \citenamefont {Caldwell}, \citenamefont {Gregory}, \citenamefont
  {Bridge}, \citenamefont {Sawant}, \citenamefont {Aldegunde}, \citenamefont
  {Mur-Petit}, \citenamefont {Jaksch}, \citenamefont {Hutson}, \citenamefont
  {Sauer}, \citenamefont {Tarbutt},\ and\ \citenamefont
  {Cornish}}]{blackmore18a}%
  \BibitemOpen
  \bibfield  {author} {\bibinfo {author} {\bibfnamefont {J.~A.}\ \bibnamefont
  {Blackmore}}, \bibinfo {author} {\bibfnamefont {L.}~\bibnamefont {Caldwell}},
  \bibinfo {author} {\bibfnamefont {P.~D.}\ \bibnamefont {Gregory}}, \bibinfo
  {author} {\bibfnamefont {E.~M.}\ \bibnamefont {Bridge}}, \bibinfo {author}
  {\bibfnamefont {R.}~\bibnamefont {Sawant}}, \bibinfo {author} {\bibfnamefont
  {J.}~\bibnamefont {Aldegunde}}, \bibinfo {author} {\bibfnamefont
  {J.}~\bibnamefont {Mur-Petit}}, \bibinfo {author} {\bibfnamefont
  {D.}~\bibnamefont {Jaksch}}, \bibinfo {author} {\bibfnamefont {J.~M.}\
  \bibnamefont {Hutson}}, \bibinfo {author} {\bibfnamefont {B.~E.}\
  \bibnamefont {Sauer}}, \bibinfo {author} {\bibfnamefont {M.~R.}\ \bibnamefont
  {Tarbutt}}, \ and\ \bibinfo {author} {\bibfnamefont {S.~L.}\ \bibnamefont
  {Cornish}},\ }\href@noop {} {\bibfield  {journal} {\bibinfo  {journal}
  {Quantum Sci. Technol.}\ }\textbf {\bibinfo {volume} {4}},\ \bibinfo {pages}
  {014010} (\bibinfo {year} {2018})}\BibitemShut {NoStop}%
\bibitem [{\citenamefont {Manovitz}\ \emph {et~al.}(2019)\citenamefont
  {Manovitz}, \citenamefont {Shaniv}, \citenamefont {Shapira}, \citenamefont
  {Ozeri},\ and\ \citenamefont {Akerman}}]{manovitz19a}%
  \BibitemOpen
  \bibfield  {author} {\bibinfo {author} {\bibfnamefont {T.}~\bibnamefont
  {Manovitz}}, \bibinfo {author} {\bibfnamefont {R.}~\bibnamefont {Shaniv}},
  \bibinfo {author} {\bibfnamefont {Y.}~\bibnamefont {Shapira}}, \bibinfo
  {author} {\bibfnamefont {R.}~\bibnamefont {Ozeri}}, \ and\ \bibinfo {author}
  {\bibfnamefont {N.}~\bibnamefont {Akerman}},\ }\href@noop {} {\bibfield
  {journal} {\bibinfo  {journal} {Phys. Rev. Lett.}\ }\textbf {\bibinfo
  {volume} {123}},\ \bibinfo {pages} {203001} (\bibinfo {year}
  {2019})}\BibitemShut {NoStop}%
\bibitem [{\citenamefont {Kimble}(2008)}]{kimble08a}%
  \BibitemOpen
  \bibfield  {author} {\bibinfo {author} {\bibfnamefont {H.~J.}\ \bibnamefont
  {Kimble}},\ }\href@noop {} {\bibfield  {journal} {\bibinfo  {journal}
  {Nature}\ }\textbf {\bibinfo {volume} {453}},\ \bibinfo {pages} {1023}
  (\bibinfo {year} {2008})}\BibitemShut {NoStop}%
\bibitem [{\citenamefont {Kajita}\ \emph {et~al.}(2014)\citenamefont {Kajita},
  \citenamefont {Gopakumar}, \citenamefont {Abe}, \citenamefont {Hada},\ and\
  \citenamefont {Keller}}]{kajita14a}%
  \BibitemOpen
  \bibfield  {author} {\bibinfo {author} {\bibfnamefont {M.}~\bibnamefont
  {Kajita}}, \bibinfo {author} {\bibfnamefont {G.}~\bibnamefont {Gopakumar}},
  \bibinfo {author} {\bibfnamefont {M.}~\bibnamefont {Abe}}, \bibinfo {author}
  {\bibfnamefont {M.}~\bibnamefont {Hada}}, \ and\ \bibinfo {author}
  {\bibfnamefont {M.}~\bibnamefont {Keller}},\ }\href@noop {} {\bibfield
  {journal} {\bibinfo  {journal} {{Phys. Rev. A}}\ }\textbf {\bibinfo {volume}
  {89}},\ \bibinfo {pages} {032509} (\bibinfo {year} {2014})}\BibitemShut
  {NoStop}%
\bibitem [{\citenamefont {Kajita}(2015)}]{kajita15a}%
  \BibitemOpen
  \bibfield  {author} {\bibinfo {author} {\bibfnamefont {M.}~\bibnamefont
  {Kajita}},\ }\href@noop {} {\bibfield  {journal} {\bibinfo  {journal} {{Phys.
  Rev. A}}\ }\textbf {\bibinfo {volume} {92}},\ \bibinfo {pages} {043423}
  (\bibinfo {year} {2015})}\BibitemShut {NoStop}%
\bibitem [{\citenamefont {Germann}(2016)}]{germann16d}%
  \BibitemOpen
  \bibfield  {author} {\bibinfo {author} {\bibfnamefont {M.}~\bibnamefont
  {Germann}},\ }\href@noop {} {Ph.D. thesis},\ \bibinfo  {school} {University
  of Basel} (\bibinfo {year} {2016})\BibitemShut {NoStop}%
\bibitem [{\citenamefont {Mansour}\ \emph {et~al.}(1991)\citenamefont
  {Mansour}, \citenamefont {Kurtz}, \citenamefont {Steimle}, \citenamefont
  {Goodman}, \citenamefont {Young}, \citenamefont {Scholl}, \citenamefont
  {Rosner},\ and\ \citenamefont {Holt}}]{berrahmansour91a}%
  \BibitemOpen
  \bibfield  {author} {\bibinfo {author} {\bibfnamefont {N.~B.}\ \bibnamefont
  {Mansour}}, \bibinfo {author} {\bibfnamefont {C.}~\bibnamefont {Kurtz}},
  \bibinfo {author} {\bibfnamefont {T.}~\bibnamefont {Steimle}}, \bibinfo
  {author} {\bibfnamefont {G.}~\bibnamefont {Goodman}}, \bibinfo {author}
  {\bibfnamefont {L.}~\bibnamefont {Young}}, \bibinfo {author} {\bibfnamefont
  {T.}~\bibnamefont {Scholl}}, \bibinfo {author} {\bibfnamefont
  {S.}~\bibnamefont {Rosner}}, \ and\ \bibinfo {author} {\bibfnamefont
  {R.}~\bibnamefont {Holt}},\ }\href@noop {} {\bibfield  {journal} {\bibinfo
  {journal} {Phys. Rev. A}\ }\textbf {\bibinfo {volume} {44}},\ \bibinfo
  {pages} {4418} (\bibinfo {year} {1991})}\BibitemShut {NoStop}%
\bibitem [{\citenamefont {Wang}\ \emph {et~al.}(2017)\citenamefont {Wang},
  \citenamefont {Um}, \citenamefont {Zhang}, \citenamefont {An}, \citenamefont
  {Lyu}, \citenamefont {Zhang}, \citenamefont {Duan}, \citenamefont {Yum},\
  and\ \citenamefont {Kim}}]{wang17a}%
  \BibitemOpen
  \bibfield  {author} {\bibinfo {author} {\bibfnamefont {Y.}~\bibnamefont
  {Wang}}, \bibinfo {author} {\bibfnamefont {M.}~\bibnamefont {Um}}, \bibinfo
  {author} {\bibfnamefont {J.}~\bibnamefont {Zhang}}, \bibinfo {author}
  {\bibfnamefont {S.}~\bibnamefont {An}}, \bibinfo {author} {\bibfnamefont
  {M.}~\bibnamefont {Lyu}}, \bibinfo {author} {\bibfnamefont {J.-N.}\
  \bibnamefont {Zhang}}, \bibinfo {author} {\bibfnamefont {L.-M.}\ \bibnamefont
  {Duan}}, \bibinfo {author} {\bibfnamefont {D.}~\bibnamefont {Yum}}, \ and\
  \bibinfo {author} {\bibfnamefont {K.}~\bibnamefont {Kim}},\ }\href@noop {}
  {\bibfield  {journal} {\bibinfo  {journal} {Nat. Photonics}\ }\textbf
  {\bibinfo {volume} {11}},\ \bibinfo {pages} {646} (\bibinfo {year}
  {2017})}\BibitemShut {NoStop}%
\bibitem [{\citenamefont {Langer}\ \emph {et~al.}(2005)\citenamefont {Langer},
  \citenamefont {Ozeri}, \citenamefont {Jost}, \citenamefont {Chiaverini},
  \citenamefont {DeMarco}, \citenamefont {Ben-Kish}, \citenamefont {Blakestad},
  \citenamefont {Britton}, \citenamefont {Hume}, \citenamefont {Itano},
  \citenamefont {Leibfried}, \citenamefont {Reichle}, \citenamefont
  {Rosenband}, \citenamefont {Schaetz}, \citenamefont {Schmidt},\ and\
  \citenamefont {Wineland}}]{langer05a}%
  \BibitemOpen
  \bibfield  {author} {\bibinfo {author} {\bibfnamefont {C.}~\bibnamefont
  {Langer}}, \bibinfo {author} {\bibfnamefont {R.}~\bibnamefont {Ozeri}},
  \bibinfo {author} {\bibfnamefont {J.~D.}\ \bibnamefont {Jost}}, \bibinfo
  {author} {\bibfnamefont {J.}~\bibnamefont {Chiaverini}}, \bibinfo {author}
  {\bibfnamefont {B.}~\bibnamefont {DeMarco}}, \bibinfo {author} {\bibfnamefont
  {A.}~\bibnamefont {Ben-Kish}}, \bibinfo {author} {\bibfnamefont {R.~B.}\
  \bibnamefont {Blakestad}}, \bibinfo {author} {\bibfnamefont {J.}~\bibnamefont
  {Britton}}, \bibinfo {author} {\bibfnamefont {D.~B.}\ \bibnamefont {Hume}},
  \bibinfo {author} {\bibfnamefont {W.~M.}\ \bibnamefont {Itano}}, \bibinfo
  {author} {\bibfnamefont {D.}~\bibnamefont {Leibfried}}, \bibinfo {author}
  {\bibfnamefont {R.}~\bibnamefont {Reichle}}, \bibinfo {author} {\bibfnamefont
  {T.}~\bibnamefont {Rosenband}}, \bibinfo {author} {\bibfnamefont
  {T.}~\bibnamefont {Schaetz}}, \bibinfo {author} {\bibfnamefont {P.~O.}\
  \bibnamefont {Schmidt}}, \ and\ \bibinfo {author} {\bibfnamefont {D.~J.}\
  \bibnamefont {Wineland}},\ }\href@noop {} {\bibfield  {journal} {\bibinfo
  {journal} {Phys. Rev. Lett.}\ }\textbf {\bibinfo {volume} {95}},\ \bibinfo
  {pages} {060502} (\bibinfo {year} {2005})}\BibitemShut {NoStop}%
\bibitem [{\citenamefont {Caldwell}\ \emph {et~al.}(2020)\citenamefont
  {Caldwell}, \citenamefont {Williams}, \citenamefont {Fitch}, \citenamefont
  {Aldegunde}, \citenamefont {Hutson}, \citenamefont {Sauer},\ and\
  \citenamefont {Tarbutt}}]{caldwell20a}%
  \BibitemOpen
  \bibfield  {author} {\bibinfo {author} {\bibfnamefont {L.}~\bibnamefont
  {Caldwell}}, \bibinfo {author} {\bibfnamefont {H.}~\bibnamefont {Williams}},
  \bibinfo {author} {\bibfnamefont {N.}~\bibnamefont {Fitch}}, \bibinfo
  {author} {\bibfnamefont {J.}~\bibnamefont {Aldegunde}}, \bibinfo {author}
  {\bibfnamefont {J.~M.}\ \bibnamefont {Hutson}}, \bibinfo {author}
  {\bibfnamefont {B.}~\bibnamefont {Sauer}}, \ and\ \bibinfo {author}
  {\bibfnamefont {M.}~\bibnamefont {Tarbutt}},\ }\href@noop {} {\bibfield
  {journal} {\bibinfo  {journal} {Phys. Rev. Lett.}\ }\textbf {\bibinfo
  {volume} {124}},\ \bibinfo {pages} {063001} (\bibinfo {year}
  {2020})}\BibitemShut {NoStop}%
\bibitem [{\citenamefont {Bruna}\ and\ \citenamefont {Grein}(2004)}]{bruna04a}%
  \BibitemOpen
  \bibfield  {author} {\bibinfo {author} {\bibfnamefont {P.~J.}\ \bibnamefont
  {Bruna}}\ and\ \bibinfo {author} {\bibfnamefont {F.}~\bibnamefont {Grein}},\
  }\href@noop {} {\bibfield  {journal} {\bibinfo  {journal} {J. Mol.
  Spectrosc.}\ }\textbf {\bibinfo {volume} {227}},\ \bibinfo {pages} {67}
  (\bibinfo {year} {2004})}\BibitemShut {NoStop}%
\bibitem [{\citenamefont {Gaubatz}\ \emph {et~al.}(1990)\citenamefont
  {Gaubatz}, \citenamefont {Rudecki}, \citenamefont {Schiemann},\ and\
  \citenamefont {Bergmann}}]{gaubatz1990}%
  \BibitemOpen
  \bibfield  {author} {\bibinfo {author} {\bibfnamefont {U.}~\bibnamefont
  {Gaubatz}}, \bibinfo {author} {\bibfnamefont {P.}~\bibnamefont {Rudecki}},
  \bibinfo {author} {\bibfnamefont {S.}~\bibnamefont {Schiemann}}, \ and\
  \bibinfo {author} {\bibfnamefont {K.}~\bibnamefont {Bergmann}},\ }\href@noop
  {} {\bibfield  {journal} {\bibinfo  {journal} {J. Chem. Phys.}\ }\textbf
  {\bibinfo {volume} {92}},\ \bibinfo {pages} {5363} (\bibinfo {year}
  {1990})}\BibitemShut {NoStop}%
\bibitem [{\citenamefont {Frosch}\ and\ \citenamefont
  {Foley}(1952)}]{frosch52a}%
  \BibitemOpen
  \bibfield  {author} {\bibinfo {author} {\bibfnamefont {R.~A.}\ \bibnamefont
  {Frosch}}\ and\ \bibinfo {author} {\bibfnamefont {H.~M.}\ \bibnamefont
  {Foley}},\ }\href@noop {} {\bibfield  {journal} {\bibinfo  {journal} {{Phys.
  Rev.}}\ }\textbf {\bibinfo {volume} {88}},\ \bibinfo {pages} {1337} (\bibinfo
  {year} {1952})}\BibitemShut {NoStop}%
\bibitem [{\citenamefont {Brown}\ and\ \citenamefont
  {Carrington}(2003)}]{brown03a}%
  \BibitemOpen
  \bibfield  {author} {\bibinfo {author} {\bibfnamefont {J.~M.}\ \bibnamefont
  {Brown}}\ and\ \bibinfo {author} {\bibfnamefont {A.}~\bibnamefont
  {Carrington}},\ }\href@noop {} {\emph {\bibinfo {title} {{Rotational
  Spectroscopy of Diatomic Molecules}}}}\ (\bibinfo  {publisher} {Cambridge
  University Press},\ \bibinfo {year} {2003})\BibitemShut {NoStop}%
\bibitem [{\citenamefont {Brown}\ \emph {et~al.}(1978)\citenamefont {Brown},
  \citenamefont {Kaise}, \citenamefont {Kerr},\ and\ \citenamefont
  {Milton}}]{brown78a}%
  \BibitemOpen
  \bibfield  {author} {\bibinfo {author} {\bibfnamefont {J.}~\bibnamefont
  {Brown}}, \bibinfo {author} {\bibfnamefont {M.}~\bibnamefont {Kaise}},
  \bibinfo {author} {\bibfnamefont {C.}~\bibnamefont {Kerr}}, \ and\ \bibinfo
  {author} {\bibfnamefont {D.}~\bibnamefont {Milton}},\ }\href@noop {}
  {\bibfield  {journal} {\bibinfo  {journal} {Mol. Phys.}\ }\textbf {\bibinfo
  {volume} {36}},\ \bibinfo {pages} {553} (\bibinfo {year} {1978})}\BibitemShut
  {NoStop}%
\bibitem [{\citenamefont {Balasubramanian}\ \emph {et~al.}(1994)\citenamefont
  {Balasubramanian}, \citenamefont {Bellary},\ and\ \citenamefont
  {Rao}}]{balasubramanian94}%
  \BibitemOpen
  \bibfield  {author} {\bibinfo {author} {\bibfnamefont {T.}~\bibnamefont
  {Balasubramanian}}, \bibinfo {author} {\bibfnamefont {V.}~\bibnamefont
  {Bellary}}, \ and\ \bibinfo {author} {\bibfnamefont {K.~N.}\ \bibnamefont
  {Rao}},\ }\href@noop {} {\bibfield  {journal} {\bibinfo  {journal} {Can. J.
  Phys.}\ }\textbf {\bibinfo {volume} {72}},\ \bibinfo {pages} {971} (\bibinfo
  {year} {1994})}\BibitemShut {NoStop}%
\bibitem [{\citenamefont {Karr}\ \emph {et~al.}(2008)\citenamefont {Karr},
  \citenamefont {Korobov},\ and\ \citenamefont {Hilico}}]{karr08}%
  \BibitemOpen
  \bibfield  {author} {\bibinfo {author} {\bibfnamefont {J.-P.}\ \bibnamefont
  {Karr}}, \bibinfo {author} {\bibfnamefont {V.~I.}\ \bibnamefont {Korobov}}, \
  and\ \bibinfo {author} {\bibfnamefont {L.}~\bibnamefont {Hilico}},\
  }\href@noop {} {\bibfield  {journal} {\bibinfo  {journal} {Phys. Rev. A}\
  }\textbf {\bibinfo {volume} {77}},\ \bibinfo {pages} {062507} (\bibinfo
  {year} {2008})}\BibitemShut {NoStop}%
\bibitem [{\citenamefont {Ma}\ \emph {et~al.}(2009)\citenamefont {Ma},
  \citenamefont {Butler}, \citenamefont {Brown}, \citenamefont {Linton},\ and\
  \citenamefont {Steimle}}]{ma09a}%
  \BibitemOpen
  \bibfield  {author} {\bibinfo {author} {\bibfnamefont {T.}~\bibnamefont
  {Ma}}, \bibinfo {author} {\bibfnamefont {C.}~\bibnamefont {Butler}}, \bibinfo
  {author} {\bibfnamefont {J.~M.}\ \bibnamefont {Brown}}, \bibinfo {author}
  {\bibfnamefont {C.}~\bibnamefont {Linton}}, \ and\ \bibinfo {author}
  {\bibfnamefont {T.~C.}\ \bibnamefont {Steimle}},\ }\href@noop {} {\bibfield
  {journal} {\bibinfo  {journal} {J. Phys. Chem. A}\ }\textbf {\bibinfo
  {volume} {113}},\ \bibinfo {pages} {8038} (\bibinfo {year}
  {2009})}\BibitemShut {NoStop}%
\bibitem [{\citenamefont {Chen}\ \emph {et~al.}(2006)\citenamefont {Chen},
  \citenamefont {Gengler}, \citenamefont {Steimle},\ and\ \citenamefont
  {Brown}}]{chen06a}%
  \BibitemOpen
  \bibfield  {author} {\bibinfo {author} {\bibfnamefont {J.}~\bibnamefont
  {Chen}}, \bibinfo {author} {\bibfnamefont {J.}~\bibnamefont {Gengler}},
  \bibinfo {author} {\bibfnamefont {T.}~\bibnamefont {Steimle}}, \ and\
  \bibinfo {author} {\bibfnamefont {J.~M.}\ \bibnamefont {Brown}},\ }\href@noop
  {} {\bibfield  {journal} {\bibinfo  {journal} {Phys. Rev. A}\ }\textbf
  {\bibinfo {volume} {73}},\ \bibinfo {pages} {012502} (\bibinfo {year}
  {2006})}\BibitemShut {NoStop}%
\bibitem [{\citenamefont {Schiff}\ and\ \citenamefont
  {Snyder}(1939)}]{schiff39}%
  \BibitemOpen
  \bibfield  {author} {\bibinfo {author} {\bibfnamefont {L.}~\bibnamefont
  {Schiff}}\ and\ \bibinfo {author} {\bibfnamefont {H.}~\bibnamefont
  {Snyder}},\ }\href@noop {} {\bibfield  {journal} {\bibinfo  {journal} {Phys.
  Rev.}\ }\textbf {\bibinfo {volume} {55}},\ \bibinfo {pages} {59} (\bibinfo
  {year} {1939})}\BibitemShut {NoStop}%
\bibitem [{\citenamefont {Michaud}\ \emph {et~al.}(2000)\citenamefont
  {Michaud}, \citenamefont {Roux}, \citenamefont {Davis}, \citenamefont
  {Nguyen},\ and\ \citenamefont {Laux}}]{michaud00a}%
  \BibitemOpen
  \bibfield  {author} {\bibinfo {author} {\bibfnamefont {F.}~\bibnamefont
  {Michaud}}, \bibinfo {author} {\bibfnamefont {F.}~\bibnamefont {Roux}},
  \bibinfo {author} {\bibfnamefont {S.~P.}\ \bibnamefont {Davis}}, \bibinfo
  {author} {\bibfnamefont {A.-D.}\ \bibnamefont {Nguyen}}, \ and\ \bibinfo
  {author} {\bibfnamefont {C.~O.}\ \bibnamefont {Laux}},\ }\href@noop {}
  {\bibfield  {journal} {\bibinfo  {journal} {J. Mol. Spectrosc.}\ }\textbf
  {\bibinfo {volume} {203}},\ \bibinfo {pages} {1} (\bibinfo {year}
  {2000})}\BibitemShut {NoStop}%
\bibitem [{\citenamefont {Wu}\ \emph {et~al.}(2007)\citenamefont {Wu},
  \citenamefont {Ben}, \citenamefont {Li}, \citenamefont {Zheng}, \citenamefont
  {Chen},\ and\ \citenamefont {Yang}}]{wu07a}%
  \BibitemOpen
  \bibfield  {author} {\bibinfo {author} {\bibfnamefont {Y.-D.}\ \bibnamefont
  {Wu}}, \bibinfo {author} {\bibfnamefont {J.-W.}\ \bibnamefont {Ben}},
  \bibinfo {author} {\bibfnamefont {B.}~\bibnamefont {Li}}, \bibinfo {author}
  {\bibfnamefont {L.-J.}\ \bibnamefont {Zheng}}, \bibinfo {author}
  {\bibfnamefont {Y.-Q.}\ \bibnamefont {Chen}}, \ and\ \bibinfo {author}
  {\bibfnamefont {X.-H.}\ \bibnamefont {Yang}},\ }\href@noop {} {\bibfield
  {journal} {\bibinfo  {journal} {Chinese J. Chem. Phys.}\ }\textbf {\bibinfo
  {volume} {20}},\ \bibinfo {pages} {285} (\bibinfo {year} {2007})}\BibitemShut
  {NoStop}%
\bibitem [{\citenamefont {Collet}\ \emph {et~al.}(1998)\citenamefont {Collet},
  \citenamefont {Destombes}, \citenamefont {Bachir},\ and\ \citenamefont
  {Huet}}]{collet98a}%
  \BibitemOpen
  \bibfield  {author} {\bibinfo {author} {\bibfnamefont {D.}~\bibnamefont
  {Collet}}, \bibinfo {author} {\bibfnamefont {J.-L.}\ \bibnamefont
  {Destombes}}, \bibinfo {author} {\bibfnamefont {I.~H.}\ \bibnamefont
  {Bachir}}, \ and\ \bibinfo {author} {\bibfnamefont {T.}~\bibnamefont
  {Huet}},\ }\href@noop {} {\bibfield  {journal} {\bibinfo  {journal} {Chem.
  Phys. Lett.}\ }\textbf {\bibinfo {volume} {286}},\ \bibinfo {pages} {311}
  (\bibinfo {year} {1998})}\BibitemShut {NoStop}%
\bibitem [{\citenamefont {Scholl}\ \emph {et~al.}(1998)\citenamefont {Scholl},
  \citenamefont {Holt},\ and\ \citenamefont {Rosner}}]{scholl98a}%
  \BibitemOpen
  \bibfield  {author} {\bibinfo {author} {\bibfnamefont {T.~J.}\ \bibnamefont
  {Scholl}}, \bibinfo {author} {\bibfnamefont {R.~A.}\ \bibnamefont {Holt}}, \
  and\ \bibinfo {author} {\bibfnamefont {S.~D.}\ \bibnamefont {Rosner}},\
  }\href@noop {} {\bibfield  {journal} {\bibinfo  {journal} {J. Mol.
  Spectrosc.}\ }\textbf {\bibinfo {volume} {192}},\ \bibinfo {pages} {424}
  (\bibinfo {year} {1998})}\BibitemShut {NoStop}%
\bibitem [{\citenamefont {Tiesinga}\ \emph {et~al.}(2020)\citenamefont
  {Tiesinga}, \citenamefont {Mohr}, \citenamefont {Newell},\ and\ \citenamefont
  {Taylor}}]{CODATA2018}%
  \BibitemOpen
  \bibfield  {author} {\bibinfo {author} {\bibfnamefont {E.}~\bibnamefont
  {Tiesinga}}, \bibinfo {author} {\bibfnamefont {P.~J.}\ \bibnamefont {Mohr}},
  \bibinfo {author} {\bibfnamefont {D.~B.}\ \bibnamefont {Newell}}, \ and\
  \bibinfo {author} {\bibfnamefont {B.~N.}\ \bibnamefont {Taylor}},\
  }\href@noop {} {\enquote {\bibinfo {title} {The 2018 codata recommended
  values of the fundamental physical constants (web version 8.1)},}\ }
  (\bibinfo {year} {2020})\BibitemShut {NoStop}%
\bibitem [{\citenamefont {Papou{\v{s}}ek}(1989)}]{papouvsek89a}%
  \BibitemOpen
  \bibfield  {author} {\bibinfo {author} {\bibfnamefont {D.}~\bibnamefont
  {Papou{\v{s}}ek}},\ }\href@noop {} {\bibfield  {journal} {\bibinfo  {journal}
  {Collect. Czech. Chem. C.}\ }\textbf {\bibinfo {volume} {54}},\ \bibinfo
  {pages} {2555} (\bibinfo {year} {1989})}\BibitemShut {NoStop}%
\bibitem [{\citenamefont {Germann}\ and\ \citenamefont
  {Willitsch}(2016)}]{germann16c}%
  \BibitemOpen
  \bibfield  {author} {\bibinfo {author} {\bibfnamefont {M.}~\bibnamefont
  {Germann}}\ and\ \bibinfo {author} {\bibfnamefont {S.}~\bibnamefont
  {Willitsch}},\ }\href@noop {} {\bibfield  {journal} {\bibinfo  {journal}
  {Mol. Phys.}\ }\textbf {\bibinfo {volume} {114}},\ \bibinfo {pages} {769}
  (\bibinfo {year} {2016})}\BibitemShut {NoStop}%
\bibitem [{Note1()}]{Note1}%
  \BibitemOpen
  \bibinfo {note} {While for rovibrational E2 transitions it is well
  established to calculate the transition moment via the second term in Eq.
  \ref {eq:Radial} \cite {karl67a,germann14a,goldman07a}, the situation is less
  clear for rovibrational transitions of M1 type. Such transitions were first
  reported in Ref. \cite {dang-nhu90a} for the $^3\Sigma ^-_g$ ground
  electronic state of the O$_2$ molecule. In Ref. \cite {balasubramanian94},
  two types of mechanisms were elaborated to rationalize the M1 rovibrational
  transition intensities observed in Ref. \cite {dang-nhu90a}. One is
  rovibrational mixing and the other is due to coupling of different
  Born-Oppenheimer states. The former, analyzed in Appendix \ref
  {sec:rovibmixing}, was found to be small in the present case, the latter is
  included in the anisotropic electron spin Zeeman coupling which was found to
  give the dominant contribution the rovibrational M1 transition intensities
  studied here.}\BibitemShut {Stop}%
\bibitem [{\citenamefont {Drake}(2006)}]{drake06}%
  \BibitemOpen
  \bibfield  {author} {\bibinfo {author} {\bibfnamefont {G.~W.~F.}\
  \bibnamefont {Drake}},\ }\href@noop {} {\emph {\bibinfo {title} {{Springer
  Handbook of Atomic, Molecular and Optical Physics}}}}\ (\bibinfo  {publisher}
  {Springer},\ \bibinfo {year} {2006})\ p.\ \bibinfo {pages} {192}\BibitemShut
  {NoStop}%
\bibitem [{\citenamefont {Keselman}\ \emph {et~al.}(2011)\citenamefont
  {Keselman}, \citenamefont {Glickman}, \citenamefont {Akerman}, \citenamefont
  {Kotler},\ and\ \citenamefont {Ozeri}}]{keselman11}%
  \BibitemOpen
  \bibfield  {author} {\bibinfo {author} {\bibfnamefont {A.}~\bibnamefont
  {Keselman}}, \bibinfo {author} {\bibfnamefont {Y.}~\bibnamefont {Glickman}},
  \bibinfo {author} {\bibfnamefont {N.}~\bibnamefont {Akerman}}, \bibinfo
  {author} {\bibfnamefont {S.}~\bibnamefont {Kotler}}, \ and\ \bibinfo {author}
  {\bibfnamefont {R.}~\bibnamefont {Ozeri}},\ }\href@noop {} {\bibfield
  {journal} {\bibinfo  {journal} {New J. Phys.}\ }\textbf {\bibinfo {volume}
  {13}},\ \bibinfo {pages} {073027} (\bibinfo {year} {2011})}\BibitemShut
  {NoStop}%
\bibitem [{\citenamefont {Harty}\ \emph {et~al.}(2014)\citenamefont {Harty},
  \citenamefont {Allcock}, \citenamefont {Ballance}, \citenamefont {Guidoni},\
  and\ \citenamefont {Janacek}}]{harty14}%
  \BibitemOpen
  \bibfield  {author} {\bibinfo {author} {\bibfnamefont {T.~P.}\ \bibnamefont
  {Harty}}, \bibinfo {author} {\bibfnamefont {D.~T.~C.}\ \bibnamefont
  {Allcock}}, \bibinfo {author} {\bibfnamefont {C.~J.}\ \bibnamefont
  {Ballance}}, \bibinfo {author} {\bibfnamefont {L.}~\bibnamefont {Guidoni}}, \
  and\ \bibinfo {author} {\bibfnamefont {H.~A.}\ \bibnamefont {Janacek}},\
  }\href@noop {} {\bibfield  {journal} {\bibinfo  {journal} {Phys. Rev. Lett.}\
  }\textbf {\bibinfo {volume} {113}},\ \bibinfo {pages} {220501} (\bibinfo
  {year} {2014})}\BibitemShut {NoStop}%
\bibitem [{\citenamefont {Saleh}\ and\ \citenamefont {Teich}(2007)}]{saleh07}%
  \BibitemOpen
  \bibfield  {author} {\bibinfo {author} {\bibfnamefont {B.~E.~A.}\
  \bibnamefont {Saleh}}\ and\ \bibinfo {author} {\bibfnamefont {M.~C.}\
  \bibnamefont {Teich}},\ }\href@noop {} {\emph {\bibinfo {title}
  {{Fundamentals of photonics, 2 Ed.}}}}\ (\bibinfo  {publisher} {Wiley Series
  in Pure and Applied Optics},\ \bibinfo {year} {2007})\BibitemShut {NoStop}%
\bibitem [{\citenamefont {Hanneke}\ \emph {et~al.}(2016)\citenamefont
  {Hanneke}, \citenamefont {Carollo},\ and\ \citenamefont {Lane}}]{hanneke16}%
  \BibitemOpen
  \bibfield  {author} {\bibinfo {author} {\bibfnamefont {D.}~\bibnamefont
  {Hanneke}}, \bibinfo {author} {\bibfnamefont {R.}~\bibnamefont {Carollo}}, \
  and\ \bibinfo {author} {\bibfnamefont {D.}~\bibnamefont {Lane}},\ }\href@noop
  {} {\bibfield  {journal} {\bibinfo  {journal} {Phys. Rev. A}\ }\textbf
  {\bibinfo {volume} {94}},\ \bibinfo {pages} {050101} (\bibinfo {year}
  {2016})}\BibitemShut {NoStop}%
\bibitem [{\citenamefont {Schiller}\ \emph {et~al.}(2008)\citenamefont
  {Schiller}, \citenamefont {Roth}, \citenamefont {Lewen}, \citenamefont
  {Ricken},\ and\ \citenamefont {Wiedner}}]{schiller08a}%
  \BibitemOpen
  \bibfield  {author} {\bibinfo {author} {\bibfnamefont {S.}~\bibnamefont
  {Schiller}}, \bibinfo {author} {\bibfnamefont {B.}~\bibnamefont {Roth}},
  \bibinfo {author} {\bibfnamefont {F.}~\bibnamefont {Lewen}}, \bibinfo
  {author} {\bibfnamefont {O.}~\bibnamefont {Ricken}}, \ and\ \bibinfo {author}
  {\bibfnamefont {M.}~\bibnamefont {Wiedner}},\ }\href@noop {} {\bibfield
  {journal} {\bibinfo  {journal} {Appl. Phys. B}\ }\textbf {\bibinfo {volume}
  {95}},\ \bibinfo {pages} {55} (\bibinfo {year} {2008})}\BibitemShut {NoStop}%
\bibitem [{\citenamefont {Leibfried}\ \emph {et~al.}(2003)\citenamefont
  {Leibfried}, \citenamefont {Blatt}, \citenamefont {Monroe},\ and\
  \citenamefont {Wineland}}]{leibfried03a}%
  \BibitemOpen
  \bibfield  {author} {\bibinfo {author} {\bibfnamefont {D.}~\bibnamefont
  {Leibfried}}, \bibinfo {author} {\bibfnamefont {R.}~\bibnamefont {Blatt}},
  \bibinfo {author} {\bibfnamefont {C.}~\bibnamefont {Monroe}}, \ and\ \bibinfo
  {author} {\bibfnamefont {D.}~\bibnamefont {Wineland}},\ }\href@noop {}
  {\bibfield  {journal} {\bibinfo  {journal} {Rev. Mod. Phys.}\ }\textbf
  {\bibinfo {volume} {75}},\ \bibinfo {pages} {281} (\bibinfo {year}
  {2003})}\BibitemShut {NoStop}%
\bibitem [{\citenamefont {Brewer}\ \emph
  {et~al.}(2019{\natexlab{a}})\citenamefont {Brewer}, \citenamefont {Chen},
  \citenamefont {Hankin}, \citenamefont {Clements}, \citenamefont {Chou},
  \citenamefont {Wineland}, \citenamefont {Hume},\ and\ \citenamefont
  {Leibrandt}}]{brewer19}%
  \BibitemOpen
  \bibfield  {author} {\bibinfo {author} {\bibfnamefont {S.~M.}\ \bibnamefont
  {Brewer}}, \bibinfo {author} {\bibfnamefont {J.-S.}\ \bibnamefont {Chen}},
  \bibinfo {author} {\bibfnamefont {A.~M.}\ \bibnamefont {Hankin}}, \bibinfo
  {author} {\bibfnamefont {E.~R.}\ \bibnamefont {Clements}}, \bibinfo {author}
  {\bibfnamefont {C.~W.}\ \bibnamefont {Chou}}, \bibinfo {author}
  {\bibfnamefont {D.~J.}\ \bibnamefont {Wineland}}, \bibinfo {author}
  {\bibfnamefont {D.~B.}\ \bibnamefont {Hume}}, \ and\ \bibinfo {author}
  {\bibfnamefont {D.~R.}\ \bibnamefont {Leibrandt}},\ }\href@noop {} {\bibfield
   {journal} {\bibinfo  {journal} {Phys. Rev. Lett.}\ }\textbf {\bibinfo
  {volume} {123}},\ \bibinfo {pages} {033201} (\bibinfo {year}
  {2019}{\natexlab{a}})}\BibitemShut {NoStop}%
\bibitem [{\citenamefont {Brewer}\ \emph
  {et~al.}(2019{\natexlab{b}})\citenamefont {Brewer}, \citenamefont {Chen},
  \citenamefont {Beloy}, \citenamefont {Hankin}, \citenamefont {Clements},
  \citenamefont {Chou}, \citenamefont {McGrew}, \citenamefont {Zhang},
  \citenamefont {Fasano}, \citenamefont {Nicolodi} \emph
  {et~al.}}]{brewer2019b}%
  \BibitemOpen
  \bibfield  {author} {\bibinfo {author} {\bibfnamefont {S.}~\bibnamefont
  {Brewer}}, \bibinfo {author} {\bibfnamefont {J.-S.}\ \bibnamefont {Chen}},
  \bibinfo {author} {\bibfnamefont {K.}~\bibnamefont {Beloy}}, \bibinfo
  {author} {\bibfnamefont {A.}~\bibnamefont {Hankin}}, \bibinfo {author}
  {\bibfnamefont {E.}~\bibnamefont {Clements}}, \bibinfo {author}
  {\bibfnamefont {C.}~\bibnamefont {Chou}}, \bibinfo {author} {\bibfnamefont
  {W.}~\bibnamefont {McGrew}}, \bibinfo {author} {\bibfnamefont
  {X.}~\bibnamefont {Zhang}}, \bibinfo {author} {\bibfnamefont
  {R.}~\bibnamefont {Fasano}}, \bibinfo {author} {\bibfnamefont
  {D.}~\bibnamefont {Nicolodi}},  \emph {et~al.},\ }\href@noop {} {\bibfield
  {journal} {\bibinfo  {journal} {Phys. Rev. A}\ }\textbf {\bibinfo {volume}
  {100}},\ \bibinfo {pages} {013409} (\bibinfo {year}
  {2019}{\natexlab{b}})}\BibitemShut {NoStop}%
\bibitem [{\citenamefont {Karl}\ and\ \citenamefont {Poll}(1967)}]{karl67a}%
  \BibitemOpen
  \bibfield  {author} {\bibinfo {author} {\bibfnamefont {G.}~\bibnamefont
  {Karl}}\ and\ \bibinfo {author} {\bibfnamefont {J.~D.}\ \bibnamefont
  {Poll}},\ }\href@noop {} {\bibfield  {journal} {\bibinfo  {journal} {J. Chem.
  Phys.}\ }\textbf {\bibinfo {volume} {46}},\ \bibinfo {pages} {2944} (\bibinfo
  {year} {1967})}\BibitemShut {NoStop}%
\bibitem [{\citenamefont {Goldman}\ \emph {et~al.}(2007)\citenamefont
  {Goldman}, \citenamefont {Tipping}, \citenamefont {Ma}, \citenamefont
  {Boone}, \citenamefont {Bernath}, \citenamefont {Demouline}, \citenamefont
  {Hase}, \citenamefont {Schneider}, \citenamefont {Hannigan}, \citenamefont
  {Coffey},\ and\ \citenamefont {Rinsland}}]{goldman07a}%
  \BibitemOpen
  \bibfield  {author} {\bibinfo {author} {\bibfnamefont {A.}~\bibnamefont
  {Goldman}}, \bibinfo {author} {\bibfnamefont {R.}~\bibnamefont {Tipping}},
  \bibinfo {author} {\bibfnamefont {Q.}~\bibnamefont {Ma}}, \bibinfo {author}
  {\bibfnamefont {C.}~\bibnamefont {Boone}}, \bibinfo {author} {\bibfnamefont
  {P.}~\bibnamefont {Bernath}}, \bibinfo {author} {\bibfnamefont
  {P.}~\bibnamefont {Demouline}}, \bibinfo {author} {\bibfnamefont
  {F.}~\bibnamefont {Hase}}, \bibinfo {author} {\bibfnamefont {M.}~\bibnamefont
  {Schneider}}, \bibinfo {author} {\bibfnamefont {J.}~\bibnamefont {Hannigan}},
  \bibinfo {author} {\bibfnamefont {M.}~\bibnamefont {Coffey}}, \ and\ \bibinfo
  {author} {\bibfnamefont {C.}~\bibnamefont {Rinsland}},\ }\href@noop {}
  {\bibfield  {journal} {\bibinfo  {journal} {J. Quant. Spectr. Rad. Transfer}\
  }\textbf {\bibinfo {volume} {103}},\ \bibinfo {pages} {168} (\bibinfo {year}
  {2007})}\BibitemShut {NoStop}%
\bibitem [{\citenamefont {Dang-Nhu}\ \emph {et~al.}(1990)\citenamefont
  {Dang-Nhu}, \citenamefont {Zander}, \citenamefont {Goldman},\ and\
  \citenamefont {Rinsland}}]{dang-nhu90a}%
  \BibitemOpen
  \bibfield  {author} {\bibinfo {author} {\bibfnamefont {M.}~\bibnamefont
  {Dang-Nhu}}, \bibinfo {author} {\bibfnamefont {R.}~\bibnamefont {Zander}},
  \bibinfo {author} {\bibfnamefont {A.}~\bibnamefont {Goldman}}, \ and\
  \bibinfo {author} {\bibfnamefont {C.~P.}\ \bibnamefont {Rinsland}},\
  }\href@noop {} {\bibfield  {journal} {\bibinfo  {journal} {J. Mol.
  Spectrosc.}\ }\textbf {\bibinfo {volume} {144}},\ \bibinfo {pages} {366}
  (\bibinfo {year} {1990})}\BibitemShut {NoStop}%
\bibitem [{\citenamefont {Lofthus}\ and\ \citenamefont
  {Krupenie}(1977)}]{lofthus77a}%
  \BibitemOpen
  \bibfield  {author} {\bibinfo {author} {\bibfnamefont {A.}~\bibnamefont
  {Lofthus}}\ and\ \bibinfo {author} {\bibfnamefont {P.~H.}\ \bibnamefont
  {Krupenie}},\ }\href@noop {} {\bibfield  {journal} {\bibinfo  {journal} {J.
  Phys. Chem. Ref. Data}\ }\textbf {\bibinfo {volume} {6}},\ \bibinfo {pages}
  {113} (\bibinfo {year} {1977})}\BibitemShut {NoStop}%
\bibitem [{\citenamefont {Riehle}(2005)}]{riehle}%
  \BibitemOpen
  \bibfield  {author} {\bibinfo {author} {\bibfnamefont {F.}~\bibnamefont
  {Riehle}},\ }\href@noop {} {\emph {\bibinfo {title} {Frequency Standards,
  Basics and Applications}}}\ (\bibinfo  {publisher} {WILEY-VCH Verlag GmbH},\
  \bibinfo {year} {2005})\BibitemShut {NoStop}%
\bibitem [{\citenamefont {Yudin}\ \emph {et~al.}(2010)\citenamefont {Yudin},
  \citenamefont {Taichenachev}, \citenamefont {Oates}, \citenamefont {Barber},
  \citenamefont {Lemke}, \citenamefont {Ludlow}, \citenamefont {Sterr},
  \citenamefont {Lisdat},\ and\ \citenamefont {Riehle}}]{yudin10}%
  \BibitemOpen
  \bibfield  {author} {\bibinfo {author} {\bibfnamefont {V.}~\bibnamefont
  {Yudin}}, \bibinfo {author} {\bibfnamefont {A.}~\bibnamefont {Taichenachev}},
  \bibinfo {author} {\bibfnamefont {C.}~\bibnamefont {Oates}}, \bibinfo
  {author} {\bibfnamefont {Z.}~\bibnamefont {Barber}}, \bibinfo {author}
  {\bibfnamefont {N.~D.}\ \bibnamefont {Lemke}}, \bibinfo {author}
  {\bibfnamefont {A.~D.}\ \bibnamefont {Ludlow}}, \bibinfo {author}
  {\bibfnamefont {U.}~\bibnamefont {Sterr}}, \bibinfo {author} {\bibfnamefont
  {C.}~\bibnamefont {Lisdat}}, \ and\ \bibinfo {author} {\bibfnamefont
  {F.}~\bibnamefont {Riehle}},\ }\href@noop {} {\bibfield  {journal} {\bibinfo
  {journal} {Phys. Rev. A}\ }\textbf {\bibinfo {volume} {82}},\ \bibinfo
  {pages} {011804} (\bibinfo {year} {2010})}\BibitemShut {NoStop}%
\bibitem [{\citenamefont {Sanner}\ \emph {et~al.}(2018)\citenamefont {Sanner},
  \citenamefont {Huntemann}, \citenamefont {Lange}, \citenamefont {Tamm},\ and\
  \citenamefont {Peik}}]{sanner18}%
  \BibitemOpen
  \bibfield  {author} {\bibinfo {author} {\bibfnamefont {C.}~\bibnamefont
  {Sanner}}, \bibinfo {author} {\bibfnamefont {N.}~\bibnamefont {Huntemann}},
  \bibinfo {author} {\bibfnamefont {R.}~\bibnamefont {Lange}}, \bibinfo
  {author} {\bibfnamefont {C.}~\bibnamefont {Tamm}}, \ and\ \bibinfo {author}
  {\bibfnamefont {E.}~\bibnamefont {Peik}},\ }\href@noop {} {\bibfield
  {journal} {\bibinfo  {journal} {Physical review letters}\ }\textbf {\bibinfo
  {volume} {120}},\ \bibinfo {pages} {053602} (\bibinfo {year}
  {2018})}\BibitemShut {NoStop}%
\bibitem [{\citenamefont {Bakalov}\ and\ \citenamefont
  {Schiller}(2014)}]{bakalov14a}%
  \BibitemOpen
  \bibfield  {author} {\bibinfo {author} {\bibfnamefont {D.}~\bibnamefont
  {Bakalov}}\ and\ \bibinfo {author} {\bibfnamefont {S.}~\bibnamefont
  {Schiller}},\ }\href {\doibase 10.1007/s00340-013-5703-z} {\bibfield
  {journal} {\bibinfo  {journal} {App. Phys. B}\ }\textbf {\bibinfo {volume}
  {114}},\ \bibinfo {pages} {213} (\bibinfo {year} {2014})}\BibitemShut
  {NoStop}%
\bibitem [{\citenamefont {Balasubramanian}\ \emph {et~al.}(1992)\citenamefont
  {Balasubramanian}, \citenamefont {Bellary}, \citenamefont {D'Cunha},\ and\
  \citenamefont {Rao}}]{balasubramanian92a}%
  \BibitemOpen
  \bibfield  {author} {\bibinfo {author} {\bibfnamefont {T.}~\bibnamefont
  {Balasubramanian}}, \bibinfo {author} {\bibfnamefont {V.}~\bibnamefont
  {Bellary}}, \bibinfo {author} {\bibfnamefont {R.}~\bibnamefont {D'Cunha}}, \
  and\ \bibinfo {author} {\bibfnamefont {K.~N.}\ \bibnamefont {Rao}},\
  }\href@noop {} {\bibfield  {journal} {\bibinfo  {journal} {J. Mol.
  Spectrosc.}\ }\textbf {\bibinfo {volume} {153}},\ \bibinfo {pages} {26}
  (\bibinfo {year} {1992})}\BibitemShut {NoStop}%
\end{thebibliography}
\bibliographystyle{apsrev4-1}

\newpage{}

\section*{Appendix}
\appendix

\section{Electric-field induced systematic shifts}
\label{sec:electric_shifts}

In the main text, we have considered the usefulness of different types of spectroscopic transitions for clock applications and as qubits in terms of their sensitivity to magnetic fields which causes the dominant systematic shifts in N$_2^+$. In addition, viable clock candidates should also have low susceptibilities to electric fields. For completeness, we consider here the AC-Stark and electric quadrupole shifts, complementing the discussion in Ref. \cite{kajita14a}.

\subsection{Stark shifts}

In a homonuclear diatomic molecule, there is no electric-dipole coupling between rovibrational states and the first-order Stark shift vanishes. There is, however, dipole coupling to excited electronic states. The AC-Stark shift of state $j$ is given by,
\begin{gather}
\label{eq:ACshift}
\Delta E_j = -\frac{1}{2}\alpha_j(\omega) E_0^2,
\end{gather}
where $E_0$ is the electric field amplitude at frequency $\omega$ and $\alpha_j(\omega)$ is the polarizability given by,
\begin{gather}
\label{eq:polari}
\alpha_j(\omega) = \sum \limits_{k} \frac{|\langle k | \mathbf{\mu} | j \rangle|^2}{\hbar} \frac{\omega_{jk}}{\omega^2 - \omega_{jk}^2}.
\end{gather}
Here, the summation runs over all states of the molecule, $k$, with non-vanishing dipole matrix element, $\langle k | \mathbf{\mu} | j \rangle$, and transition angular frequency, $\omega_{jk}$. Data on excited electronic states of N$_2^+$ can be found in the literature, e.g., Ref. \cite{lofthus77a}. The AC-Stark-shift-induced differential shift of the transition frequency between two levels $j$ and $i$ is then given by $h \Delta f_{AC} = \Delta E_j - \Delta E_i$.

Different types of AC-Stark shifts contribute to the present problem. The AC-Stark shift from the RF drive of an ion trap with frequency $\omega \approx (2\pi) 20$ MHz can be estimated in the limit $\omega \rightarrow 0$ in Eq. \eqref{eq:polari}. For vibrational transitions with $f \approx 65.2$ THz, one obtains a relative shift of $\Delta f_{AC} / (f E_0^2) = 7 \times 10^{-24}$ (m/V)$^2$. Typically, the electric field amplitude vanishes at the position of the ions in a Paul trap. However, trap imperfections can lead to a non-zero electric-field amplitude. These fields will cause a relative shift of $1.26 \times 10^{-18}$ at a field amplitude of $300$ V/m for the fundamental vibrational transition. It should be noted that such a field amplitude is excessive for an ion in a typical ion trap built for quantum-logic experiments. For other classes of transitions, e.g., hyperfine or rotational excitations, the corresponding shift is smaller as the energy spacing decreases and the cancellation of the AC-Stark shift between the upper and the lower state is more significant. 

The AC-Stark shift induced by ambient blackbody radiation can also be estimated in the limit of $\omega \rightarrow 0$ because the maximum of the thermal spectral energy density at a temperature of 300 K is situated around 31 THz which is small compared to the frequencies of electronic transitions from the vibrational ground-state. The time-averaged value of the quadratic electric-field amplitude of a 300~K radiator is $E_0^2/2 = \langle E_0^2\rangle \approx (831.9$ V/m)$^2$ \cite{riehle} yielding a relative shift of $\Delta f_{AC} / f = 1.0 \cdot 10^{-17}$ for vibrational transitions. Transitions within a vibrational state will have smaller shifts due to cancellation between the upper and lower levels. The N$_2^+$ molecular clock is therefore suitable for operation in a room-temperature environment.

AC-Stark shifts from the probe laser can be eliminated by using the Hyper-Ramsey spectroscopic method \cite{yudin10} or through a balanced Raman scheme \cite{sanner18}. In a Rabi- or Ramsey-type clock experiment, the laser power is reduced in order to minimize power broadening. In order to obtain a Rabi frequency of $\Omega \sim (2\pi) 1$ Hz on the S(0) branch of the fundamental vibrational transition, 1 nW of laser power focused to a beam radius of $50~ \mu$m are required. The intensity is thus $I = 0.26$ W/m$^2$ corresponding to an electric field of $13.9$ V/m. The AC-Stark shift obtained from Eq. \eqref{eq:ACshift} is then $\Delta f_{AC} / f = 8.9 \cdot 10^{-22}$. With a laser power more suitable for driving qubits of $\sim 100$ mW, the intensity is $I = 2.55 \cdot 10^{7}$~W/m$^2$ and an AC-Stark shift of $5.8$ Hz or $\Delta f_{AC} / f = 8.9 \times 10^{-14}$ is obtained.

\subsection{Electric-quadrupole shift}

The matrix element in Eq. \ref{eq:E2} that was used to obtain the electric quadrupole transition moment can also be used for estimating the quadrupole shift caused by static-field gradients prevalent in an ion trap. For the present purpose, the permanent moments $R(0,0) \approx 1.86$ $ea_0^2$ for $v = 0$ and $R(1,1) \approx 1.89$ $ ea_0^2$ for $v = 1$ \cite{bruna04a} were assumed in Eq. \ref{eq:E2}. Furthermore, we assumed $p = 0$ to describe a static field gradient along the trap axis. The first-order energy correction due to the quadrupole shift is then given by \cite{bakalov14a}, 
\begin{gather}
    \Delta E_j = -\frac{1}{3} \frac{dE}{dz} \langle \phi_i | \mathbf{Q} | \phi_j \rangle.
\end{gather} 
A typical field gradient, $dE/dz$, in an ion trap is $\sim 10^7$ V/m$^2$. From the matrix elements in Eq.\ref{eq:E2},  we see that the quadrupole shift vanishes for all states in the rotational ground state, $N = 0$, as is also apparent by the selection rules for E2 transitions. Further, for any state with $F = 0$ or $F = 1/2$, the shift also vanishes. For other states in $v = 0, 1$ and $N = 0, 2, 4$, the matrix element ranges between $10^{-3} - 10^{-1} R(v, v)$ corresponding to an absolute quadrupole shift between $0.013 - 1.3$ Hz.

All the Zeeman-and hyperfine transitions in $N = 0$ are therefore immune to electric-quadrupole shifts to first order. Rotational transitions of the form $N = 0 \rightarrow (N' = 2$, $F = 1/2)$, among which several ``magic'' transitions were identified (see Appendix \ref{sec:magic_transitions}), are also not affected by this shift, as is the Q(0) branch of vibrational transitions. The Q(0) transitions of the $I = 0$ nuclear-spin configuration are therefore especially suitable for clock operation because they are immune from the quadrupole shift and feature stretched-state transitions which have a low susceptibility to magnetic fields. Q(2) transitions can also be chosen with $F = F' = 1/2$ such that the quadrupole shift cancels. For S(0) transitions, the quadrupole shift vanishes in the lower states $N = 0$ and the upper state can be chosen as $F' = 1/2$. 

The differential shift of the fine-structure transitions with $\Delta J = 1$ within the $N = 2$ manifold was calculated for the ``magic'' transitions listed in Appendix \ref{sec:magic_transitions}. The differential shift ranges between $0.38 - 1.86$ Hz for the ``magic'' transitions with the exception of $|J = 3/2, I = 2, F = 7/2, M_F = -1/2 \rangle \rightarrow |J = 5/2, I = 2, F = 9/2, M_F = -3/2 \rangle$ for which an accidental cancellation leads to a vanishing shift. Therefore, suitable clock transitions with a low sensitivity to magnetic fields and vanishing quadrupole shifts were identified in every class of transitions examined in this paper.

\onecolumngrid

\section{The effective hyperfine Hamiltonian}
\label{sec:hamiltonian}

The effective hyperfine-interaction Hamiltonian takes the form,
\begin{equation}
H_{hfs} = H_{b_F} + H_{t} + H_{eqQ} + H_{c_I}.
\end{equation}
The individual terms in $H_{hfs}$ and their matrix elements have been discussed in Ref. \cite{berrahmansour91a} and are reproduced here for convenience. $H_{b_F}$ represents the Fermi-contact interaction which has off-diagonal matrix elements in $J$,
\begin{widetext}
\begin{eqnarray}
\label{eq:Hbf}
\langle v', N', S', J', I', F', m'| H_{b_F} |v, N, S, J, I, F, m\rangle = b_{Fv} \delta_{v v'}\delta_{N N'}\delta_{S S'}\delta_{I I'}\delta_{F F'}\delta_{m m'} (-1)^{F+I+J'+J+N+S+1}\\ \nonumber
\times \sqrt{I(I+1)(2I+1)S(S+1)(2S+1)(2J+1)(2J'+1)} 
\begin{Bmatrix}
I & J' &  F \\ J & I & 1
\end{Bmatrix}
\begin{Bmatrix}
S & J' & N \\ J & S & 1
\end{Bmatrix},
\end{eqnarray}
\end{widetext}
$H_{t}$ is the dipolar hyperfine interaction with off-diagonal matrix elements in $N$ and $J$, 
\begin{widetext}
\begin{eqnarray}
\label{eq:Ht}
\langle v', N', S', J', I', F', m'| H_{t} |v, N, S, J, I, F, m\rangle = t_{v,N} \delta_{v v'}\delta_{S S'}\delta_{I I'}\delta_{F F'}\delta_{m m'} (-1)^{J+I+F+N'+1}
\\ \nonumber
\times \sqrt{30I(I+1)(2I+1)S(S+1)(2S+1)(2J+1)(2J'+1)(2N+1)(2N'+1)} 
\\ \nonumber
\times 
\begin{Bmatrix}
I & J' &  F \\ J & I & 1
\end{Bmatrix}
\begin{Bmatrix}
N' & N & 2 \\ S & S & 1 \\ J' & J & 1
\end{Bmatrix}
\begin{pmatrix}
N' & 2 & N \\ 0 & 0 & 0
\end{pmatrix},
\end{eqnarray}
\end{widetext}
where $t_{v,N}=t_v+t_{Dv} N(N+1)$, $H_{eqQ}$ is the electric-quadrupole hyperfine interaction with off-diagonal matrix elements in $N$, $J$ and $I$,
\begin{widetext}
\begin{eqnarray}
\label{eq:HeqQ}
\langle v', N', S', J', I', F', m'| H_{eqQ} |v, N, S, J, I, F, m\rangle = \frac{eqQ_v}{2} \delta_{v v'}\delta_{S S'}\delta_{F F'}\delta_{m m'}
\\ \nonumber
\times 
\frac{(-1)^I+(-1)^{I'}}{2} (-1)^{F+2J+I'+2I_1+S+2N'}
\\ \nonumber
\times 
\sqrt{(2I+1)(2I'+1)(2J+1)(2J'+1)(2N+1)(2N'+1)} 
\\ \nonumber
\times 
\begin{Bmatrix}
I' & 2 &  I \\ J & F & J'
\end{Bmatrix}
\begin{Bmatrix}
I_1 & 2 & I_1 \\ I & I_1 & I'
\end{Bmatrix}
\begin{Bmatrix}
N' & 2 & N \\ J & S & J'
\end{Bmatrix}
\begin{pmatrix}
N' & 2 & N \\ 0 & 0 & 0
\end{pmatrix}
\begin{pmatrix}
I_1 & 2 & I_1 \\ -I_1 & 0 & I_1
\end{pmatrix}^{-1},
\end{eqnarray}
\end{widetext}
where $I_1=1$ is the nuclear spin of a $^{14}$N atom, and $H_{c_I}$ is the magnetic nuclear spin-rotation interaction which mixes different $J$ quantum numbers,
\begin{widetext}
\begin{eqnarray}
\label{eq:HCI}
\langle v', N', S', J', I', F', m'| H_{c_I} |v, N, S, J, I, F, m\rangle = c_I \delta_{v v'}\delta_{N N'}\delta_{S S'}\delta_{I I'}\delta_{F F'}\delta_{m m'}
(-1)^{F+I+J'+J+N+S+1}
\\ \nonumber
\times 
\sqrt{I(I+1)(2I+1)N(N+1)(2N+1)(2J+1)(2J'+1)} 
\begin{Bmatrix}
J & 1 &  J' \\ I & F & I
\end{Bmatrix}
\begin{Bmatrix}
N & 1 & N \\ J' & S & J
\end{Bmatrix}.
\end{eqnarray}
\end{widetext}
All effective coupling constants are given in Table \ref{tab:constants}. 

\section{Vibrational mixing}
\label{sec:rovibmixing}

The rovibrational mixing was estimated in the same way as in Refs. \cite{balasubramanian92a,balasubramanian94}, i.e., as a result of the dependence of the rotational constant, $B$, on the bond length, $R$, within the harmonic approximation for the vibration. The rotational constant, $B(R)\propto R^{-2}$, was expanded in a Taylor series to first order around the equilibrium internuclear distance $R= R_e$,
\begin{equation}
    B(R) = B_e \left[ 1 - 2\hat{\xi} +O(\hat{\xi}^2) \right].
    \label{eq:ber}
\end{equation}
Here, $\hat{\xi} = (\hat{R} - R_e)/R_e$ and $B_e$ is the equilibrium rotational constant. The zero-order term corresponds to the rigid-rotor rotational Hamiltonian whereas the linear term in $\hat{\xi}$ causes the rovibrational interaction.

Inserting Eq. \ref{eq:ber} in the rotational Hamiltonian, $B(R)\hat{N}^2,$ the rigid-rotor Hamiltonian, $B_e\hat{N}^2$, and the rotation-vibration coupling Hamiltonian to first order,
\begin{equation}
    \NF{H}_{ro-vib}=-2 B_e \hat{\xi}\hat{N}^2.
\end{equation}
are obtained.
$\hat{\xi}$ is further expressed in terms of creation ($\hat{a}^\dagger$) and annihilation ($\hat{a}$) operators of an harmonic oscillator,
\begin{equation}
    \hat{\xi} = \sqrt{\frac{B_e}{\omega_e}}(\hat{a}^\dagger + \hat{a}),
\label{eq:adagger}
\end{equation}
which results in the matrix elements,
\begin{eqnarray}
\label{eq:Hrovib}
\langle v', N' | \NF{H}_{\text{ro-vib}} | v, N \rangle = -2 B_e \sqrt{\frac{B_e}{\omega_e}} \delta_{N N'}
\\ \nonumber
\times
N(N+1)\left(\sqrt{v+1}\delta_{v',v+1}+\sqrt{v}\delta_{v',v-1}\right). 
\end{eqnarray}
Here, $\omega_e$ is the harmonic vibration frequency (expressed in the same units as $B_e$).

The combined vibrational and rotational Hamiltonians, $\NF{H}_{vib} = G_v$ and $\NF{H}_{rot} =B_e\hat{N}^2$, respectively, were diagonalized including the rovibrational interaction $\NF{H}_{ro-vib}$ numerically using the $v = 0, 1, 2, 3, 4$ vibrational and $N = 0, 2, 4, 6, 8, 10$ rotational states as a basis set to obtain the mixing coefficients. In this treatment, we found $\sum\limits_{ij} c_i c_j R(v=0, v=1) \sim 5 \times 10^{-6} \mu_p$ in $N = 2$ according to Eq. \ref{eq:Radial}. For E2 transitions, this corresponds to transition moment of $\sim 10^{-5}~ea_0^2$. The second term in Eq. \ref{eq:Radial}, however, leads to a much stronger transition moment of $\sim10^{-1}~ea_0^2$. Therefore, the effect of ro-vibrational mixing in the calculation of the transition moments for low-lying rotational states can be neglected.

\onecolumngrid
\section{``Magic'' magnetic-field-insensitive transitions}
\label{sec:magic_transitions}
\centering
\begin{longtable}{l l| p{1.0cm} |p{1.5cm}|c|c}
\caption{Partial list of the strongest ``magic''  magnetic-field insensitive transitions within the hyperfine, fine, rotational and vibrational manifolds below 70~G. The labels $\left|v,N,S,J,I,F,m\right\rangle$ correspond to those basis states which exhibit the largest overlap with with the true molecular eigenstates. For each transition, the ``magic'' magnetic-field strength, $B$, the Einstein $A$ coefficient, the transition frequency, $f-f_0$, and the second-order magnetic-field dependence, $a$, of the transition are indicated. The dominant coupling mechanism (M1$_S$, M1$_{aS}$ or E2) is also listed for each type of transition. The transition frequencies are given with respect to a reference frequency, $f_0$, defined as follows: $f_0=0$ for pure hyperfine and fine-structure transitions, $f_0=B_0\times6-D_0\times6^2\approx345'784.31$ MHz for rotational transitions, $f_0=G_1-G_0\approx65'197'244.88$ MHz for Q(0) rovibrational transitions, and $f_0=G_1-G_0+B_1\times6-D_1\times6^2\approx65'539'595.50$ MHz for S(0) rovibrational transitions.}
\endfirsthead
\endhead
\endfoot
\endlastfoot
\\
Hyperfine transitions: ($I=2$) M1$_S$ & $|v=0,N=0\rangle \rightarrow |v'=0,N'=0\rangle$ & $B$ [G] & $A$ [s$^{-1}$] & $f-f_0$ [MHz] & $a$ [mHz/mG$^2$] \\
\hline
$|J=1/2, I=2, F=3/2, m=-3/2 \rangle\rightarrow$ & $|J=1/2, I=2, F=5/2, m=-3/2 \rangle$ &54.85 &8.6$\times 10^{-18}$ &204.80 &19.1\\
$|J=1/2, I=2, F=3/2, m=-1/2 \rangle\rightarrow$ & $|J=1/2, I=2, F=5/2, m=-3/2 \rangle$ &38.40 &6.1$\times 10^{-18}$ &233.51 &16.1\\
$|J=1/2, I=2, F=3/2, m=-3/2 \rangle\rightarrow$ & $|J=1/2, I=2, F=5/2, m=-1/2 \rangle$ &38.42 &6.1$\times 10^{-18}$ &233.48 &16.1\\
$|J=1/2, I=2, F=3/2, m=-1/2 \rangle\rightarrow$ & $|J=1/2, I=2, F=5/2, m=-1/2 \rangle$ &18.28 &1.6$\times 10^{-17}$ &250.83 &15.6\\
[1pt]\\
Fine-structure transitions: ($I=0,2$) M1$_S$ & $|v=0,N=2\rangle \rightarrow |v'=0,N'=2\rangle$ & $B$ [G] & $A$ [s$^{-1}$] & $f-f_0$ [MHz] & $a$ [mHz/mG$^2$] \\
\hline
$|J=3/2, I=0, F=3/2, m=-1/2 \rangle\rightarrow$ & $|J=5/2, I=0, F=5/2, m=-1/2 \rangle$ &49.20 &3.1$\times 10^{-16}$ &686.52 &8.9\\
$|J=3/2, I=2, F=3/2, m=-1/2 \rangle\rightarrow$ & $|J=5/2, I=2, F=5/2, m=-1/2 \rangle$ &17.15 &2.2$\times 10^{-16}$ &656.74 &12.0\\
$|J=3/2, I=2, F=3/2, m=1/2 \rangle\rightarrow$ & $|J=5/2, I=2, F=5/2, m=-1/2 \rangle$ &3.81 &1.0$\times 10^{-16}$ &660.16 &10.5\\
$|J=3/2, I=2, F=5/2, m=-1/2 \rangle\rightarrow$ & $|J=5/2, I=2, F=7/2, m=-3/2 \rangle$ &48.30 &1.9$\times 10^{-16}$ &738.06 &8.0\\
$|J=3/2, I=2, F=5/2, m=-3/2 \rangle\rightarrow$ & $|J=5/2, I=2, F=7/2, m=-1/2 \rangle$ &48.72 &1.6$\times 10^{-16}$ &739.40 &7.0\\
$|J=3/2, I=2, F=5/2, m=-1/2 \rangle\rightarrow$ & $|J=5/2, I=2, F=7/2, m=-1/2 \rangle$ &22.31 &3.9$\times 10^{-16}$ &752.31 &8.2\\
$|J=3/2, I=2, F=5/2, m=1/2 \rangle\rightarrow$ & $|J=5/2, I=2, F=7/2, m=-1/2 \rangle$ &1.55 &1.9$\times 10^{-16}$ &756.33 &7.9\\
$|J=3/2, I=2, F=7/2, m=-1/2 \rangle\rightarrow$ & $|J=5/2, I=2, F=9/2, m=-3/2 \rangle$ &49.84 &2.8$\times 10^{-16}$ &832.14 &6.3\\
$|J=3/2, I=2, F=7/2, m=-3/2 \rangle\rightarrow$ & $|J=5/2, I=2, F=9/2, m=-1/2 \rangle$ &49.39 &2.7$\times 10^{-16}$ &832.67 &6.2\\
$|J=3/2, I=2, F=7/2, m=-1/2 \rangle\rightarrow$ & $|J=5/2, I=2, F=9/2, m=-1/2 \rangle$ &24.06 &5.9$\times 10^{-16}$ &843.94 &6.3\\
[1pt]\\
Rotational transitions: ($I=2$) M1$_{aS}$ & $|v=0,N=0\rangle \rightarrow |v'=0,N'=2\rangle$ & $B$ [G] & $A$ [s$^{-1}$] & $f-f_0$ [MHz] & $a$ [mHz/mG$^2$] \\
\hline
$|J=1/2, I=2, F=3/2, m=-3/2 \rangle\rightarrow$ & $|J=3/2, I=2, F=1/2, m=-1/2 \rangle$ &45.57 &2.2$\times 10^{-15}$ &-244.08 &9.4\\
$|J=1/2, I=2, F=3/2, m=-1/2 \rangle\rightarrow$ & $|J=3/2, I=2, F=1/2, m=-1/2 \rangle$ &15.80 &9.3$\times 10^{-15}$ &-225.74 &12.3\\
$|J=1/2, I=2, F=3/2, m=-1/2 \rangle\rightarrow$ & $|J=3/2, I=2, F=1/2, m=1/2 \rangle$ &9.37 &2.7$\times 10^{-15}$ &-223.57 &9.3\\
$|J=1/2, I=2, F=3/2, m=-3/2 \rangle\rightarrow$ & $|J=3/2, I=2, F=3/2, m=-3/2 \rangle$ &29.85 &9.8$\times 10^{-15}$ &-257.50 &7.3\\
$|J=1/2, I=2, F=3/2, m=-1/2 \rangle\rightarrow$ & $|J=3/2, I=2, F=3/2, m=-1/2 \rangle$ &36.99 &1.2$\times 10^{-15}$ &-254.73 &3.8\\
$|J=1/2, I=2, F=5/2, m=-3/2 \rangle\rightarrow$ & $|J=3/2, I=2, F=3/2, m=-1/2 \rangle$ &38.86 &1.4$\times 10^{-15}$ &-488.22 &-12.1\\
$|J=1/2, I=2, F=5/2, m=-1/2 \rangle\rightarrow$ & $|J=3/2, I=2, F=3/2, m=-1/2 \rangle$ &15.30 &1.9$\times 10^{-15}$ &-504.42 &-14.9\\
$|J=1/2, I=2, F=3/2, m=-1/2 \rangle\rightarrow$ & $|J=3/2, I=2, F=3/2, m=1/2 \rangle$ &48.51 &7.4$\times 10^{-15}$ &-264.22 &5.3\\
$|J=1/2, I=2, F=3/2, m=1/2 \rangle\rightarrow$ & $|J=3/2, I=2, F=3/2, m=3/2 \rangle$ &17.64 &5.9$\times 10^{-15}$ &-253.71 &5.4\\
$|J=1/2, I=2, F=3/2, m=-3/2 \rangle\rightarrow$ & $|J=3/2, I=2, F=5/2, m=-3/2 \rangle$ &44.15 &7.7$\times 10^{-15}$ &-297.86 &5.9\\
$|J=1/2, I=2, F=5/2, m=-3/2 \rangle\rightarrow$ & $|J=3/2, I=2, F=5/2, m=-3/2 \rangle$ &60.07 &1.9$\times 10^{-15}$ &-501.62 &-12.6\\
$|J=1/2, I=2, F=5/2, m=-1/2 \rangle\rightarrow$ & $|J=3/2, I=2, F=5/2, m=-3/2 \rangle$ &35.38 &5.4$\times 10^{-15}$ &-531.06 &-10.9\\
$|J=1/2, I=2, F=3/2, m=-1/2 \rangle\rightarrow$ & $|J=3/2, I=2, F=5/2, m=-1/2 \rangle$ &16.71 &7.1$\times 10^{-15}$ &-289.93 &3.4\\
$|J=1/2, I=2, F=5/2, m=-3/2 \rangle\rightarrow$ & $|J=3/2, I=2, F=5/2, m=-1/2 \rangle$ &43.39 &7.8$\times 10^{-15}$ &-521.55 &-13.8\\
$|J=1/2, I=2, F=3/2, m=-1/2 \rangle\rightarrow$ & $|J=3/2, I=2, F=5/2, m=1/2 \rangle$ &60.74 &6.2$\times 10^{-15}$ &-302.52 &3.2\\
$|J=1/2, I=2, F=5/2, m=-1/2 \rangle\rightarrow$ & $|J=3/2, I=2, F=5/2, m=1/2 \rangle$ &4.82 &6.8$\times 10^{-15}$ &-544.71 &-11.7\\
$|J=1/2, I=2, F=3/2, m=1/2 \rangle\rightarrow$ & $|J=3/2, I=2, F=5/2, m=3/2 \rangle$ &30.76 &7.5$\times 10^{-15}$ &-292.22 &3.0\\
$|J=1/2, I=2, F=5/2, m=-3/2 \rangle\rightarrow$ & $|J=3/2, I=2, F=7/2, m=-3/2 \rangle$ &58.94 &1.4$\times 10^{-14}$ &-539.71 &-12.8\\
$|J=1/2, I=2, F=5/2, m=-1/2 \rangle\rightarrow$ & $|J=3/2, I=2, F=7/2, m=-3/2 \rangle$ &33.79 &9.0$\times 10^{-15}$ &-568.41 &-10.6\\
$|J=1/2, I=2, F=5/2, m=-3/2 \rangle\rightarrow$ & $|J=3/2, I=2, F=7/2, m=-1/2 \rangle$ &46.34 &6.2$\times 10^{-15}$ &-556.10 &-12.8\\
$|J=1/2, I=2, F=5/2, m=-1/2 \rangle\rightarrow$ & $|J=3/2, I=2, F=7/2, m=-1/2 \rangle$ &19.49 &1.4$\times 10^{-14}$ &-576.51 &-11.1\\
$|J=1/2, I=2, F=5/2, m=-1/2 \rangle\rightarrow$ & $|J=3/2, I=2, F=7/2, m=1/2 \rangle$ &6.12 &7.4$\times 10^{-15}$ &-580.31 &-10.9\\
[1pt]\\
Rovibrational transitions: ($I=2$): M1$_{aS}$ & $|v=0,N=0\rangle \rightarrow |v'=1,N'=0\rangle$ & $B$ [G] & $A$ [s$^{-1}$] & $f-f_0$ [MHz] & $a$ [mHz/mG$^2$] \\
\hline
$|J=1/2, I=2, F=5/2, m=-3/2 \rangle\rightarrow$ & $|J=1/2, I=2, F=3/2, m=-3/2 \rangle$ &54.37 &4.4$\times 10^{-10}$ &-202.56 &-19.3\\
$|J=1/2, I=2, F=5/2, m=-1/2 \rangle\rightarrow$ & $|J=1/2, I=2, F=3/2, m=-3/2 \rangle$ &38.05 &2.1$\times 10^{-10}$ &-230.99 &-16.3\\
$|J=1/2, I=2, F=5/2, m=-3/2 \rangle\rightarrow$ & $|J=1/2, I=2, F=3/2, m=-1/2 \rangle$ &38.10 &2.1$\times 10^{-10}$ &-231.01 &-16.2\\
$|J=1/2, I=2, F=5/2, m=-1/2 \rangle\rightarrow$ & $|J=1/2, I=2, F=3/2, m=-1/2 \rangle$ &18.12 &4.4$\times 10^{-10}$ &-248.18 &-15.8\\
$|J=1/2, I=2, F=3/2, m=-3/2 \rangle\rightarrow$ & $|J=1/2, I=2, F=5/2, m=-3/2 \rangle$ &54.37 &4.4$\times 10^{-10}$ &203.45 &19.3\\
$|J=1/2, I=2, F=3/2, m=-1/2 \rangle\rightarrow$ & $|J=1/2, I=2, F=5/2, m=-3/2 \rangle$ &38.03 &2.1$\times 10^{-10}$ &231.91 &16.3\\
$|J=1/2, I=2, F=3/2, m=-3/2 \rangle\rightarrow$ & $|J=1/2, I=2, F=5/2, m=-1/2 \rangle$ &38.12 &2.1$\times 10^{-10}$ &231.88 &16.2\\
$|J=1/2, I=2, F=3/2, m=-1/2 \rangle\rightarrow$ & $|J=1/2, I=2, F=5/2, m=-1/2 \rangle$ &18.12 &4.4$\times 10^{-10}$ &249.07 &15.8\\
[1pt] \\
Rovibrational transitions: ($I=2$): E2 & $|v=0,N=0\rangle \rightarrow |v'=1,N'=2\rangle$ & $B$ [G] & $A$ [s$^{-1}$] & $f-f_0$ [MHz] & $a$ [mHz/mG$^2$] \\
\hline 
$|J=1/2, I=2, F=5/2, m=-1/2 \rangle\rightarrow$ & $|J=3/2, I=2, F=1/2, m=-1/2 \rangle$ &30.06 &5.8$\times 10^{-9}$ &-480.97 &-5.2\\
$|J=1/2, I=2, F=5/2, m=-1/2 \rangle\rightarrow$ & $|J=3/2, I=2, F=3/2, m=-3/2 \rangle$ &47.64 &8.5$\times 10^{-9}$ &-491.42 &-8.0\\
$|J=1/2, I=2, F=5/2, m=1/2 \rangle\rightarrow$ & $|J=3/2, I=2, F=3/2, m=-3/2 \rangle$ &14.87 &3.8$\times 10^{-9}$ &-508.36 &-6.7\\
$|J=1/2, I=2, F=3/2, m=-1/2 \rangle\rightarrow$ & $|J=3/2, I=2, F=3/2, m=-1/2 \rangle$ &36.60 &1.8$\times 10^{-9}$ &-257.19 &3.9\\
$|J=1/2, I=2, F=5/2, m=-3/2 \rangle\rightarrow$ & $|J=3/2, I=2, F=3/2, m=-1/2 \rangle$ &39.07 &6.7$\times 10^{-9}$ &-490.68 &-11.5\\
$|J=1/2, I=2, F=3/2, m=-1/2 \rangle\rightarrow$ & $|J=3/2, I=2, F=3/2, m=1/2 \rangle$ &46.75 &4.2$\times 10^{-9}$ &-265.99 &5.6\\
$|J=1/2, I=2, F=5/2, m=-3/2 \rangle\rightarrow$ & $|J=3/2, I=2, F=3/2, m=1/2 \rangle$ &33.59 &3.3$\times 10^{-9}$ &-498.86 &-9.9\\
$|J=1/2, I=2, F=5/2, m=-1/2 \rangle\rightarrow$ & $|J=3/2, I=2, F=3/2, m=1/2 \rangle$ &4.28 &5.6$\times 10^{-9}$ &-509.68 &-13.1\\
$|J=1/2, I=2, F=3/2, m=-1/2 \rangle\rightarrow$ & $|J=3/2, I=2, F=3/2, m=3/2 \rangle$ &61.63 &1.1$\times 10^{-9}$ &-275.92 &4.3\\
$|J=1/2, I=2, F=3/2, m=1/2 \rangle\rightarrow$ & $|J=3/2, I=2, F=3/2, m=3/2 \rangle$ &16.85 &5.0$\times 10^{-9}$ &-255.67 &5.6\\
$|J=1/2, I=2, F=5/2, m=-1/2 \rangle\rightarrow$ & $|J=3/2, I=2, F=5/2, m=-5/2 \rangle$ &64.93 &8.9$\times 10^{-9}$ &-506.02 &-7.5\\
$|J=1/2, I=2, F=3/2, m=-3/2 \rangle\rightarrow$ & $|J=3/2, I=2, F=5/2, m=-3/2 \rangle$ &44.62 &3.2$\times 10^{-9}$ &-294.67 &6.0\\
$|J=1/2, I=2, F=5/2, m=-3/2 \rangle\rightarrow$ & $|J=3/2, I=2, F=5/2, m=-3/2 \rangle$ &59.97 &6.0$\times 10^{-9}$ &-498.49 &-12.5\\
$|J=1/2, I=2, F=5/2, m=1/2 \rangle\rightarrow$ & $|J=3/2, I=2, F=5/2, m=-3/2 \rangle$ &10.74 &5.4$\times 10^{-9}$ &-540.42 &-10.3\\
$|J=1/2, I=2, F=5/2, m=-1/2 \rangle\rightarrow$ & $|J=3/2, I=2, F=5/2, m=-1/2 \rangle$ &18.34 &3.8$\times 10^{-9}$ &-537.47 &-12.5\\
$|J=1/2, I=2, F=5/2, m=-3/2 \rangle\rightarrow$ & $|J=3/2, I=2, F=5/2, m=1/2 \rangle$ &31.36 &6.0$\times 10^{-9}$ &-530.66 &-11.9\\
$|J=1/2, I=2, F=3/2, m=1/2 \rangle\rightarrow$ & $|J=3/2, I=2, F=5/2, m=5/2 \rangle$ &65.10 &5.2$\times 10^{-9}$ &-300.92 &2.3\\
$|J=1/2, I=2, F=3/2, m=-3/2 \rangle\rightarrow$ & $|J=3/2, I=2, F=7/2, m=-5/2 \rangle$ &14.88 &1.5$\times 10^{-8}$ &-314.32 &4.0\\
$|J=1/2, I=2, F=5/2, m=-1/2 \rangle\rightarrow$ & $|J=3/2, I=2, F=7/2, m=-5/2 \rangle$ &51.30 &3.5$\times 10^{-9}$ &-543.57 &-9.2\\
$|J=1/2, I=2, F=3/2, m=-3/2 \rangle\rightarrow$ & $|J=3/2, I=2, F=7/2, m=-3/2 \rangle$ &48.10 &1.3$\times 10^{-8}$ &-324.81 &6.1\\
$|J=1/2, I=2, F=5/2, m=-1/2 \rangle\rightarrow$ & $|J=3/2, I=2, F=7/2, m=-3/2 \rangle$ &33.18 &2.0$\times 10^{-9}$ &-557.44 &-10.9\\
$|J=1/2, I=2, F=5/2, m=1/2 \rangle\rightarrow$ & $|J=3/2, I=2, F=7/2, m=-3/2 \rangle$ &8.04 &2.2$\times 10^{-9}$ &-568.88 &-10.1\\
$|J=1/2, I=2, F=3/2, m=-1/2 \rangle\rightarrow$ & $|J=3/2, I=2, F=7/2, m=-1/2 \rangle$ &16.21 &1.3$\times 10^{-8}$ &-314.63 &4.2\\
$|J=1/2, I=2, F=5/2, m=-3/2 \rangle\rightarrow$ & $|J=3/2, I=2, F=7/2, m=-1/2 \rangle$ &45.32 &1.8$\times 10^{-9}$ &-545.49 &-13.1\\
$|J=1/2, I=2, F=3/2, m=-1/2 \rangle\rightarrow$ & $|J=3/2, I=2, F=7/2, m=1/2 \rangle$ &54.89 &1.4$\times 10^{-8}$ &-324.78 &2.9\\
$|J=1/2, I=2, F=5/2, m=-3/2 \rangle\rightarrow$ & $|J=3/2, I=2, F=7/2, m=1/2 \rangle$ &33.85 &2.3$\times 10^{-9}$ &-557.16 &-12.0\\
$|J=1/2, I=2, F=5/2, m=-1/2 \rangle\rightarrow$ & $|J=3/2, I=2, F=7/2, m=1/2 \rangle$ &5.95 &1.7$\times 10^{-9}$ &-569.15 &-11.1\\
$|J=1/2, I=2, F=3/2, m=1/2 \rangle\rightarrow$ & $|J=3/2, I=2, F=7/2, m=3/2 \rangle$ &22.93 &1.5$\times 10^{-8}$ &-315.33 &2.9\\
$|J=1/2, I=2, F=3/2, m=3/2 \rangle\rightarrow$ & $|J=3/2, I=2, F=7/2, m=7/2 \rangle$ &41.67 &2.7$\times 10^{-8}$ &-316.98 &1.3\\
$|J=1/2, I=2, F=3/2, m=-3/2 \rangle\rightarrow$ & $|J=5/2, I=2, F=3/2, m=-1/2 \rangle$ &56.95 &4.7$\times 10^{-9}$ &322.93 &10.7\\
$|J=1/2, I=2, F=3/2, m=-1/2 \rangle\rightarrow$ & $|J=5/2, I=2, F=3/2, m=-1/2 \rangle$ &25.65 &8.1$\times 10^{-9}$ &348.66 &9.7\\
$|J=1/2, I=2, F=3/2, m=-3/2 \rangle\rightarrow$ & $|J=5/2, I=2, F=3/2, m=1/2 \rangle$ &16.17 &1.1$\times 10^{-8}$ &352.02 &15.4\\
$|J=1/2, I=2, F=5/2, m=1/2 \rangle\rightarrow$ & $|J=5/2, I=2, F=3/2, m=3/2 \rangle$ &52.04 &2.0$\times 10^{-9}$ &117.76 &-4.6\\
$|J=1/2, I=2, F=5/2, m=3/2 \rangle\rightarrow$ & $|J=5/2, I=2, F=3/2, m=3/2 \rangle$ &16.39 &4.2$\times 10^{-9}$ &101.77 &-5.8\\
$|J=1/2, I=2, F=3/2, m=-1/2 \rangle\rightarrow$ & $|J=5/2, I=2, F=5/2, m=-3/2 \rangle$ &48.87 &3.3$\times 10^{-9}$ &383.90 &9.3\\
$|J=1/2, I=2, F=3/2, m=1/2 \rangle\rightarrow$ & $|J=5/2, I=2, F=5/2, m=-3/2 \rangle$ &23.68 &5.3$\times 10^{-9}$ &403.02 &8.8\\
$|J=1/2, I=2, F=3/2, m=-3/2 \rangle\rightarrow$ & $|J=5/2, I=2, F=5/2, m=-1/2 \rangle$ &44.96 &8.3$\times 10^{-9}$ &382.72 &13.3\\
$|J=1/2, I=2, F=5/2, m=-1/2 \rangle\rightarrow$ & $|J=5/2, I=2, F=5/2, m=-1/2 \rangle$ &6.98 &4.1$\times 10^{-9}$ &152.46 &-2.4\\
$|J=1/2, I=2, F=3/2, m=-3/2 \rangle\rightarrow$ & $|J=5/2, I=2, F=5/2, m=1/2 \rangle$ &25.75 &4.0$\times 10^{-9}$ &400.41 &12.9\\
$|J=1/2, I=2, F=3/2, m=-1/2 \rangle\rightarrow$ & $|J=5/2, I=2, F=5/2, m=1/2 \rangle$ &1.26 &4.8$\times 10^{-9}$ &408.33 &12.7\\
$|J=1/2, I=2, F=5/2, m=3/2 \rangle\rightarrow$ & $|J=5/2, I=2, F=5/2, m=5/2 \rangle$ &60.47 &8.4$\times 10^{-9}$ &162.26 &-1.8\\
$|J=1/2, I=2, F=3/2, m=-1/2 \rangle\rightarrow$ & $|J=5/2, I=2, F=7/2, m=-5/2 \rangle$ &67.42 &3.0$\times 10^{-9}$ &423.69 &7.8\\
$|J=1/2, I=2, F=3/2, m=-3/2 \rangle\rightarrow$ & $|J=5/2, I=2, F=7/2, m=-3/2 \rangle$ &63.13 &2.9$\times 10^{-9}$ &416.02 &12.8\\
$|J=1/2, I=2, F=3/2, m=-1/2 \rangle\rightarrow$ & $|J=5/2, I=2, F=7/2, m=-3/2 \rangle$ &39.26 &3.8$\times 10^{-9}$ &447.47 &10.6\\
$|J=1/2, I=2, F=5/2, m=-3/2 \rangle\rightarrow$ & $|J=5/2, I=2, F=7/2, m=-3/2 \rangle$ &36.74 &5.9$\times 10^{-9}$ &213.99 &-5.3\\
$|J=1/2, I=2, F=3/2, m=-1/2 \rangle\rightarrow$ & $|J=5/2, I=2, F=7/2, m=-1/2 \rangle$ &20.54 &3.3$\times 10^{-9}$ &459.46 &11.7\\
$|J=1/2, I=2, F=3/2, m=-1/2 \rangle\rightarrow$ & $|J=5/2, I=2, F=7/2, m=1/2 \rangle$ &3.88 &2.1$\times 10^{-9}$ &464.18 &11.3\\
$|J=1/2, I=2, F=5/2, m=-1/2 \rangle\rightarrow$ & $|J=5/2, I=2, F=7/2, m=1/2 \rangle$ &67.69 &4.6$\times 10^{-9}$ &222.96 &-2.0\\
$|J=1/2, I=2, F=5/2, m=3/2 \rangle\rightarrow$ & $|J=5/2, I=2, F=7/2, m=5/2 \rangle$ &31.84 &1.5$\times 10^{-9}$ &210.01 &-1.0\\
$|J=1/2, I=2, F=5/2, m=-3/2 \rangle\rightarrow$ & $|J=5/2, I=2, F=9/2, m=-5/2 \rangle$ &9.72 &1.7$\times 10^{-8}$ &259.28 &-3.7\\
$|J=1/2, I=2, F=5/2, m=-3/2 \rangle\rightarrow$ & $|J=5/2, I=2, F=9/2, m=-3/2 \rangle$ &43.36 &1.3$\times 10^{-8}$ &268.40 &-6.2\\
$|J=1/2, I=2, F=5/2, m=-3/2 \rangle\rightarrow$ & $|J=5/2, I=2, F=9/2, m=-1/2 \rangle$ &68.91 &7.8$\times 10^{-9}$ &286.49 &-6.1\\
$|J=1/2, I=2, F=5/2, m=-1/2 \rangle\rightarrow$ & $|J=5/2, I=2, F=9/2, m=-1/2 \rangle$ &13.68 &1.5$\times 10^{-8}$ &259.82 &-4.7\\
$|J=1/2, I=2, F=5/2, m=-1/2 \rangle\rightarrow$ & $|J=5/2, I=2, F=9/2, m=1/2 \rangle$ &47.59 &1.2$\times 10^{-8}$ &269.15 &-3.9\\
$|J=1/2, I=2, F=5/2, m=1/2 \rangle\rightarrow$ & $|J=5/2, I=2, F=9/2, m=3/2 \rangle$ &23.35 &1.5$\times 10^{-8}$ &261.01 &-3.3\\

\label{table:magic}
\end{longtable}

\end{document}